\documentclass{iopart}
\usepackage{graphicx}
\usepackage{epstopdf}
\usepackage{amssymb}
\usepackage{amsfonts}
\usepackage{amsbsy}
\usepackage{multirow}
\usepackage{longtable}
\usepackage{braket}
\usepackage[breaklinks,colorlinks,linkcolor=blue,citecolor=blue,urlcolor=blue]{hyperref}
\usepackage[numbers,square,sort]{natbib} 

\usepackage{doi}

\begin{document}

\title{Path integral Monte Carlo ground state 
approach:
Formalism, implementation, and applications}

\author{Yangqian Yan$^{1,2}$ and D. Blume$^{3,2}$}
\address{$^{1}$Department of Physics, Indiana University Purdue University
Indianapolis (IUPUI), Indianapolis, Indiana 46202, USA}
\address{$^{2}$Department of Physics and Astronomy,
Washington State University,
  Pullman, Washington 99164-2814, USA}
\address{$^3$Homer L. Dodge Department of Physics and Astronomy,
The University of Oklahoma, 440 W. Brooks Street, Norman, 
OK 73019, USA}

\date{\today}

\begin{abstract}
Monte Carlo techniques have played an important role 
in understanding strongly-correlated systems across
many areas of physics, covering a wide range of energy and length scales.
Among the many Monte Carlo methods applicable to
quantum mechanical systems, the path integral Monte Carlo
approach with its variants
has been employed widely. 
Since
semi-classical 
or classical approaches will not be discussed
in this review, path integral based approaches can 
for our purposes be
divided into two categories: approaches
applicable to quantum mechanical systems
at zero temperature and approaches applicable to quantum mechanical
systems at finite
temperature.
While these two
approaches are related to each other, the underlying formulation
and aspects of the algorithm differ.
This paper reviews the path integral Monte Carlo
ground state (PIGS) approach,
which solves the time-independent Schr\"odinger equation.
Specifically, the PIGS approach allows for the determination of
expectation values with respect to eigen states of the
few- or many-body Schr\"odinger equation provided
the system Hamiltonian is known.
The theoretical framework behind the PIGS algorithm, 
implementation details,
and sample applications for fermionic
systems are presented.

\end{abstract}
\maketitle \section{Introduction}
\label{sec_introduction}

Monte Carlo techniques have found many applications,
ranging
from the modeling of the stock market to the
simulation of classical and quantum spin models~\cite{Binder1986}.
This review introduces the path integral 
Monte Carlo ground state (PIGS) method~\cite{pigs99,pigs00,massimo05,rota10},
which allows for the treatment of
quantum mechanical systems with continuous spatial degrees of freedom
at zero temperature. The PIGS method is a variant 
of the finite temperature path integral Monte Carlo
method~\cite{ceperleyrev}.
The key quantity in the finite temperature 
path integral Monte Carlo approach is the 
density matrix
$\hat{\rho}$, $\hat{\rho}= \exp[ -\hat{H}/(k_B T)]$,
where $\hat{H}$ denotes the quantum mechanical system Hamiltonian,
$k_B$ the Boltzmann constant, and $T$ the temperature.
Making the formal replacement
$(k_B T)^{-1}= \imath t / \hbar$,
where $t$ is the time, the density 
matrix turns into the time evolution operator.
Introducing the imaginary time $\tau$, $\tau = \imath t / \hbar$, and 
repeatedly acting
with the ``imaginary time evolution
operator'' $\hat{G}(\Delta \tau)$,
$\hat{G}(\Delta \tau)=\exp ( -\Delta \tau \hat{H})$ (assuming 
$\Delta \tau$ is,
using a metric to be defined later, small) onto an initial state,
the ground state wave function, or more
precisely the lowest energy state that has finite
overlap with the initial state, is projected out. This
projection idea
is the key concept behind the
PIGS approach as well as many 
other imaginary time
propagation schemes~\cite{lester94,press2007numerical}. 
Unlike grid based or basis set expansion based
approaches, the PIGS approach is applicable to systems with
varying degrees of freedom, i.e., few- and many-body systems.
This versatility of the PIGS approach stems from the fact that the
action of the imaginary time evolution operator on the initial
(or propagated) state is evaluated stochastically, i.e.,
by means of Monte Carlo Metropolis sampling.

While the finite temperature path integral Monte
Carlo algorithm, which---as has
already been aluded to---has many
features in common with the PIGS algorithm,
has been reviewed quite extensively~\cite{ceperleyrev,boninsegni05},
the PIGS algorithm has not,
despite its generality,
been reviewed in detail 
in the literature.
The present paper thus serves three main purposes:
(i) It develops,
starting from equations that should be
familiar to an advanced undergraduate student,
the theoretical concepts behind the PIGS
algorithm.
(ii)
It details how the relevant equations
can be evaluated numerically,
provides a good number of implementation details, and
discusses various aspects regarding the algorithm performance.
(iii)
It presents applications
of the PIGS algorithm to 
fermionic systems.

The PIGS approach allows one to solve the time-independent 
non-relativistic Schr\"odinger equation. 
Since the PIGS algorithm does not
provide the full wave function in numerical or analytical
form, the type of expectation values that one would like
to determine needs to be specified {\em{a priori}}
rather than {\em{a posteriori}}. In particular, 
an estimator has to be derived and implemented for each
observable. The PIGS algorithm works, as already mentioned,
through imaginary time propagation. It is imperative to
clarify upfront that the imaginary time propagation
is a numerical tool that facilitates projecting 
out unwanted excited state contributions. An extension
to the real time evolution is, in general, not feasible,
at least not for systems with a large number of degrees of
freedom (for small systems, grid based real time propagation schemes 
do, of course, exist).
A key ingredient of the PIGS algorithm is the stochastic 
evaluation of high-dimensional integrals,
which arise from the discretization of the imaginary time and from the
intrinsic degrees of freedom (particle coordinates)
of the system under study.
The stochastic Monte Carlo based approach to evaluating these integrals
makes the PIGS method applicable to large systems
containing as many as hundreds of particles.
However, as in many Monte Carlo techniques, the
treatment of identical fermions leads to the 
infamous Fermi sign problem. This tutorial applies the PIGS algorithm 
to small fermionic systems with zero-range
interactions. It is shown that the sign problem 
can be ``postponed'' but not be avoided, i.e., application
of the PIGS algorithm to larger fermionic systems 
with zero-range interactions will necessarily fail.

The PIGS algorithm has been applied to systems relevant to physics
and
chemistry.
For example, our understanding of pristine and
doped
bosonic helium clusters of varying size 
has been informed by PIGS 
calculations~\cite{whitley2009spectral}. Unlike
alternative
zero-temperature methods
such as the variational Monte Carlo
method~\cite{vmcref,lester94} and the
diffusion quantum Monte Carlo method with
mixed estimators~\cite{dmcref1,dmcref2}, 
the PIGS approach is known to yield unbiased
results for structural properties like the radial density
and pair distribution function. 
Here the term ``unbiased'' refers to the fact that the 
resulting structural properties are
independent of the initial state, provided the initial state
has finite overlap with the state of interest
and provided the state of interest is the lowest energy 
state with a particular symmetry. Moreover, 
the condensate fraction~\cite{ceperleyrev,moroni2004condensate} and Renyi
entropy~\cite{herdman,hastings} are observables that can be 
calculated relatively 
straightforwardly within the PIGS algorithm
(or 
at least more straightforwardly than within a number
of other approaches). 
The PIGS algorithm has also been, among others, applied 
to 
bulk 
helium
in varying spatial 
dimensions~\cite{moroni2004condensate,Galli2004,vitali08,rossi12,galli07,rossi10,nava13,arrigoni13,carleo09}, 
liquid
helium 
in nano-pores~\cite{Galli20042,nava12}, 
molecular 
para-hydrogen 
in nano-pores~\cite{turnbull05,massimo07}, 
molecular para-hydrogen
clusters~\cite{cuervo2006path,cuervo08,cuervo09}, 
hardsphere bosons~\cite{rossi13,giorgini13}, 
dipolar systems~\cite{macia12,massimo14}, and 
cold atoms loaded into
optical lattices~\cite{pilati12}.

The applications presented in this review deal with 
cold atom systems with infinitely
large two-body $s$-wave scattering length
$a_s$~\cite{stringariFermireview,blumerev12,levinsonarxiv16},
which are---like helium droplets---strongly interacting. However, the
average interparticle spacing 
in cold atom systems tends to be significantly larger
than that in helium droplets. This implies that the sample
applications presented in this review deal with Hamiltonian that are
characterized 
by 
vastly different length scales. To
describe systems for which the average
interparticle distance is large 
compared to the two-body interaction
range, we employ two-body zero-range interactions. 
The use of two-body zero-range interactions removes the
two-body range from the problem. 
If $a_s$ is send to infinity, as in the applications
presented in this review, then
two-component fermions are
characterized 
by the same number of length 
scales as the corresponding non-interacting 
system~\cite{stringariFermireview,blumerev12,heiselberg01,Yvan12}.

For bosons, in contrast, a three-body parameter, which can be
defined in terms of the size
of one of the extremely weakly-bound Efimov trimers,
sets a length scale of the interacting system even if the
range of the two-body interactions is zero.
Since the
use
of two-body zero-range interactions in continuum
Monte Carlo calculations is a fairly novel
development~\cite{ceperley08,krauth14,schmidt14,zr15,schmidt15,giorgini16,giorgini17}, 
the associated implementation details
are discussed in detail. Simulation results
are presented for fermionic
systems. 
Application of the algorithm to bosons requires only
a few changes in the code;
however, due to the existence of a three-body parameter, the
number of time slices, e.g., is much larger than for fermionic systems
with two-body zero-range interactions.

The remainder of this article is organized as follows.
Section~\ref{sec_formalism} introduces, starting from the non-relativistic
Schr\"odinger equation, the key quantum mechanical
equations behind the PIGS
algorithm. 
Section~\ref{sec_pigsalgorithm} discusses a number
of theoretical concepts that are needed to reformulate the
basic quantum mechanical equations in a form amenable to 
computer simulations; many considerations in this
section do not
only apply to the PIGS algorithm but
also to other Monte Carlo algorithms.
Section~\ref{sec_algorithm_intro}
introduces some basic ideas. 
Sections~\ref{sec_trotterformula} and \ref{sec_ppa}
discuss two different
approaches for approximating the short-time propagator,
namely a Trotter formula based approach and an
approach that utilizes the so-called pair product approximation;
these two approaches are compared in 
Sec.~\ref{sec_comparison_of_the_two_approximations}.
The use of two-body zero-range interactions within the pair product approximation
is discussed in Sec.~\ref{sec_propagator_for_tb_zr_interactions}.
Section~\ref{sec_algorithm}
``translates'' the formalism introduced
in Secs.~\ref{sec_formalism} and
\ref{sec_pigsalgorithm} into an algorithm.
Section~\ref{sec_algorithm_sampling} introduces the 
basics of
Monte Carlo sampling of high-dimensional integrals
while Sec.~\ref{metropolis-ch2} reviews formal aspects of 
the Monte Carlo Metropolis sampling.
Section~\ref{sec_moves}
discusses the generation of new configurations, i.e., the moves
employed in the PIGS algorithm; as applicable, differences to the 
path integral Monte Carlo
algorithm are pointed out.
Sections~\ref{sec_algorithm_expectation} 
and \ref{sec_algorithm_error} discuss the 
determination of expectation values and associated error bars, 
respectively.
Last, Sec.~\ref{sec_algorithm_permutation} discusses 
how to treat permutations in the PIGS algorithm; this
discussion is particularly relevant if the system contains two or more
identical fermions.

Section~\ref{sec_application} 
presents a number of applications to harmonically trapped
equal-mass two-component Fermi gases.
The simulation results are discussed from two different
angles. On the one hand, ``technical aspects'' such as
convergence with respect to the propagation
time and the time step
are discussed.
On the other hand, the physical relevance of the simulation 
results presented is highlighted.
Spin-balanced systems with up to $N=10$ particles 
and a non-interacting Fermi gas with a single impurity 
with up to $N=5$ particles are considered.
In both cases, interspecies two-body zero-range interactions
with infinitely large $s$-wave scattering length
are employed.
The construction of different types of trial functions $\psi_T$ 
is discussed 
and the dependence of the simulation results on $\psi_T$ is
elucidated.
PIGS results for the energy,
pair distribution function,  and  contact are presented and compared
to literature results where available.
Last, Sec.~\ref{sec_summary}
provides a summary and an outlook.

\section{Quantum mechanical foundation}
\label{sec_formalism}
We consider $N$ non-relativistic
particles 
described by the time-independent
Hamiltonian $\hat{H}$ at zero temperature.
The Hamiltonian may contain 
single-particle potentials, two-body potentials,
and higher-body potentials.
We work in position space, where the 
potentials are local, i.e., 
we consider potentials that only depend on the position
vectors and not on the momentum vectors as would be the case if,
e.g., spin-orbit
coupling terms were present~\cite{spielmanspinorbit,dalibardspinorbit,huispinorbit}.
The position vector for the $j$-th particle 
with mass $m_j$ is denoted by $\mathbf{r}_j$ and
we collectively denote the position
vectors of all the particles by $\mathbf{R}$,
$\mathbf{R}=\{\mathbf{r}_1, \mathbf{r}_2, \cdots, \mathbf{r}_N\}$.
The stationary
eigen states and corresponding eigen energies are
denoted by $\psi_j(\mathbf{R})$
and $E_j$, where $j=0,1,2,\cdots$.
The $\psi_j(\mathbf{R})$ form a complete set and we are,
throughout this article, interested in 
systems that support at least one $N$-body bound state.
The treatment of scattering states by means of quantum
Monte Carlo approaches is, in general, a challenging 
task~\cite{carlson87,ceperleyscattering,Shumway200519,carlson07}
that is beyond the scope of this paper.
The PIGS algorithm allows one to calculate a subset of
the bound state energies as well as expectation values such as the
pair distribution functions associated with
the
corresponding eigen states. 

The PIGS algorithm is rooted in 
imaginary time propagation, a concept that is used widely
to find the ground state or selected excited states of 
linear and non-linear Schr\"odinger equations~\cite{press2007numerical,itp}.
The concept of imaginary time propagation is also used to solve
non-quantum mechanical wave equations. In what follows, we 
restrict ourselves, for concreteness,
to the linear Schr\"odinger equation. 
To illustrate the key idea behind 
imaginary time propagation algorithms, we assume
that the ground state is non-degenerate, i.e., that 
$E_0 < E_j$ for $j=1,2,\cdots$.
We consider an initial trial function $\psi_T(\mathbf{R})$,
which does not have to be normalized,
that has finite overlap with the ground state 
wave function $\psi_0(\mathbf{R})$.
To analyze what happens when this trial function is propagated in imaginary 
time, we decompose the trial function into the eigen states 
$\psi_j(\mathbf{R})$ 
of the Hamiltonian $\hat{H}$,
\begin{equation}
  \psi_T(\mathbf{R})
  =\sum_{j=0}^{\infty}c_j\psi_j(\mathbf{R}),
  \label{eq_expansion_of_psit}
\end{equation}
where $c_0$ is non-zero by assumption.
Using Eq.~(\ref{eq_expansion_of_psit}),
$\psi_\tau(\mathbf{R})$,
\begin{eqnarray}
  \label{eq_acting_with_exp}
  \psi_{\tau}(\mathbf{R})
  = \exp(-\tau \hat{H})\psi_T(\mathbf{R}),
\end{eqnarray}
can be written as
\begin{eqnarray}
\fl
  \psi_\tau(\mathbf{R})
  =\exp(-\tau E_0)\left\{ c_0
  \psi_0(\mathbf{R})+
\sum_{j=1}^{\infty}c_j 
\exp \left[-\tau (E_j-E_0)\right] \psi_j(\mathbf{R})\right\}.
  \label{eq_expansion_of_psitau}
\end{eqnarray}
Since $E_j$ is, by assumption, greater than $E_0$,
the excited states contained in $\psi_T(\mathbf{R})$
decay out during the imaginary time propagation.
In the $\tau\to\infty$ limit, 
$\psi_{\tau}(\mathbf{R})$
approaches, except for an overall factor,
the eigen state $\psi_0(\mathbf{R})$.
Correspondingly, the energy $E_\tau$,
\begin{equation}
  E_\tau=\frac{\braket{\psi_\tau|\hat{H}|\psi_\tau}}{\braket{\psi_\tau|\psi_\tau}},
  \label{eq_energy}
\end{equation}
approaches the exact ground state energy $E_0$ 
exponentially in the $\tau \to \infty$ limit.
For finite $\tau$,
$E_{\tau}$ provides an upper bound to the exact eigen energy.
This suggests that one can obtain a reliable estimate
of $E_0$ by extrapolating the $E_{\tau}$ for various finite
$\tau$ to the $\tau \to \infty$ limit.
Expectation values of an arbitrary operator $\hat{O}$ can be written
analogously,
 \begin{equation}
  O_\tau=\frac{\braket{\psi_\tau|\hat{O}|\psi_\tau}}{\braket{\psi_\tau|\psi_\tau}},
  \label{eq_general_expectation_value}
\end{equation}
where $O_{\tau}$ denotes the $\tau$-dependent expectation value.
The convergence of $O_{\tau}$ toward the exact
expectation value with respect to $\psi_0(\mathbf{R})$
may not be simply exponential and
needs to be analyzed carefully 
for each operator $\hat{O}$ (see Sec.~\ref{sec_application} 
for examples).

Equations~(\ref{eq_acting_with_exp}),
(\ref{eq_energy}), and (\ref{eq_general_expectation_value}) 
constitute
the starting 
point of the PIGS algorithm (see Sec.~\ref{sec_algorithm}).
Based on these equations, two ingredients or components of
the PIGS algorithm can already be identified.
(i) An initial trial function $\psi_T(\mathbf{R})$
needs to be supplied by the ``simulator''. 
From Eq.~(\ref{eq_expansion_of_psitau}) it is clear that the 
efficiency of the PIGS algorithm depends on the overlap between
$\psi_T(\mathbf{R})$ and $\psi_0(\mathbf{R})$:
If all $c_j$ with $j>0$ vanish, then the imaginary
time propagation is not needed at all.
If the $c_j$ for the states that lie energetically close
to $E_0$ vanish, then the decay of the excited states
is fast, i.e., small $\tau$ should suffice.
The construction of $\psi_T(\mathbf{R})$
is, of course, strongly dependent on the Hamiltonian
under study. Examples are discussed in 
Sec.~\ref{sec_application}.
(ii) Given an initial trial function $\psi_T(\mathbf{R})$,
the action of $\exp(-\tau \hat{H})$
onto $\psi_T(\mathbf{R})$ needs to be evaluated. 
The PIGS 
algorithm as well as many other algorithms
accomplish this by dividing $\tau$ into 
multiple smaller imaginary time steps $\Delta \tau$. 
While non-Monte Carlo based approaches are, typically,
restricted to relatively small system sizes,
the PIGS algorithm as well as other Monte Carlo
algorithms are designed to treat systems for which
$\mathbf{R}$ can be a high-dimensional vector.

The discussion thus far focused on determining the
absolute ground state
of the system. The outlined formalism can be readily adopted 
to the determination of the energetically lowest-lying state
with a given symmetry. For concreteness, let us assume that
the total angular momentum $L$ and the total parity $\Pi$
are good quantum
numbers and that the absolute ground state
has vanishing angular
momentum ($L=0$) and positive parity
($\Pi=+1$). 
If $\psi_T(\mathbf{R})$ is chosen to
have a symmetry other than $(L,\Pi)=(0,+1)$, say $(L',\Pi')$ symmetry, 
then the imaginary time 
propagation projects out the eigen state with $(L',\Pi')$ symmetry
that has the lowest energy. Said differently, the 
imaginary time propagation preserves the symmetry 
of $\psi_T(\mathbf{R})$ and
acts in
a
subspace of the full Hilbert space.

It is instructive to compare the PIGS formalism discussed 
above with another imaginary time propagation based Monte
Carlo technique, namely the diffusion Monte Carlo
technique (for the purpose of the discussion that
follows, the Green's function Monte Carlo
technique behaves identically)~\cite{lester94}.
The diffusion Monte Carlo approach utilizes, in addition to a trial
function, a reference energy $E_{\rm{ref}}$ that is
adjusted continually during the simulation.
If Eq.~(\ref{eq_expansion_of_psitau}) 
is multiplied by $\exp(E_{\rm{ref}} \tau)$,
then the right hand side is, except for an overall
$\mathbf{R}$-independent factor, independent
of $\tau$ for sufficiently large $\tau$ and $E_{\rm{ref}}=E_0$.
This is the key idea behind the diffusion Monte Carlo approach. 
The accumulation of expectation values is started after the excited state
contributions
have decayed out and after $E_{\rm{ref}}$ has been adjusted 
such that $E_{\rm{ref}} \approx E_0$.
While the diffusion Monte Carlo and PIGS approaches share,
as just discussed, similarities, the treatment of particle
permutations differs notably if the system contains
two or more identical fermions. The diffusion Monte Carlo
algorithm does not explicitly apply sequences
of two-particle permutation
operators; the identical particle characteristics (bosons and/or fermions)
of the
$N$-particle wave function are instead encoded via the
trial function, combined with the fixed- or released-node 
approach in the case of identical fermions~\cite{fixnodedmc,ceperley80}. 
The PIGS algorithm, in contrast, explicitly 
anti-symmetrizes the paths if the system contains identical
fermions. If the system contains identical bosons,
explicit symmetrization operations are
not needed provided the ground state of the system where the
bosons are replaced by ``Boltzmann particles'' is the same as that of the
system containing bosons.

\section{PIGS algorithm: General considerations}
\label{sec_pigsalgorithm}

\subsection{Basic concepts}
\label{sec_algorithm_intro}

This section rewrites Eq.~(\ref{eq_energy}) in a form amenable
to evaluation by Monte Carlo techniques.
The actual Monte Carlo implementation is discussed in 
Sec.~\ref{sec_algorithm}.
Using $\ket{\psi_{\tau}}=\exp(-\tau \hat{H})\ket{\psi_T}$ and
$\bra{\psi_{\tau}}=\bra{\psi_T} \exp(-\tau \hat{H})$,
Eq.~(\ref{eq_energy}) reads
\begin{eqnarray}
\label{eq_energy2}
E_{\tau} = \frac{\bra{\psi_T} \exp(-\tau \hat{H}) \hat{H}\exp(-\tau \hat{H})\ket{\psi_T}}
{\bra{\psi_T} \exp(-\tau \hat{H})\exp(-\tau \hat{H})\ket{\psi_T}}.
\end{eqnarray}
The denominator is commonly denoted by $Z(\tau)$,
\begin{eqnarray}
\label{eq_znorm}
Z(\tau)
={\bra{\psi_T} \exp(-2\tau \hat{H})\ket{\psi_T}}
.
\end{eqnarray}
To obtain a prescription for evaluating the operators 
in the integrands, we project $\hat{H}$ and $\exp(-\tau \hat{H})$
onto the position basis. Formally, this amounts to inserting the
identity
\begin{equation}
  \int_{\mathbf{R}}\ket{\mathbf{R}}\bra{\mathbf{R}}d\mathbf{R}=\hat{1},
  \label{chpimc_Ridentity}
\end{equation}
where $\hat{1}$ denotes the unit operator,
multiple times into Eq.~(\ref{eq_energy2}),
\begin{eqnarray}
\fl
  \label{eq_energy_quantum}
  E_\tau=\\
\fl  \frac{\int_{\mathbf{R}}\int_{\mathbf{R}'}\int_{\mathbf{R}''}\int_{\mathbf{R}'''}
F_{\rm{aux}}
  d\mathbf{R}
  d\mathbf{R}'
  d\mathbf{R}''
  d\mathbf{R}'''
}
{
  \int_{\mathbf{R}}\int_{\mathbf{R}'}\int_{\mathbf{R}''}
  \braket{\psi_T|\mathbf{R}}\braket{\mathbf{R}|\exp(-\tau \hat{H})|\mathbf{R}'}
  \braket{\mathbf{R}'|\exp(-\tau \hat{H})|\mathbf{R}''}  \braket{\mathbf{R}''|\psi_T}
  d\mathbf{R}
  d\mathbf{R}'
  d\mathbf{R}''
}
, \nonumber
\end{eqnarray}
where
\begin{eqnarray}
\fl
  F_{\rm{aux}}=
    \braket{\psi_T|\mathbf{R}}\braket{\mathbf{R}|\exp(-\tau \hat{H})|\mathbf{R}'}
  \bra{\mathbf{R}'}
  \hat{H}
  \ket{\mathbf{R}'''}
  \braket{\mathbf{R}'''|\exp(-\tau \hat{H})|\mathbf{R}''}
   \braket{\mathbf{R}''|\psi_T}.
  \end{eqnarray}
We refer to $G(\mathbf{R},\mathbf{R}';\tau)$,
\begin{equation}
  G(\mathbf{R},\mathbf{R}';\tau)=\braket{\mathbf{R}|\exp(-\tau \hat{H})|\mathbf{R}'}, 
  \label{eq_propagator}
\end{equation}
as the imaginary time evolution operator projected onto 
the position basis or, in short, as 
the imaginary time evolution operator or propagator.
Using Eq.~(\ref{eq_propagator}),
we obtain
\begin{eqnarray}
\fl
\label{eq_energy3}
  E_\tau=  \\
\fl  \frac{\int_{\mathbf{R}}\int_{\mathbf{R}'}\int_{\mathbf{R}''}\int_{\mathbf{R}'''}
  \psi_T(\mathbf{R})
  G(\mathbf{R},\mathbf{R}';\tau)
  \bra{\mathbf{R}'}\hat{H}  \ket{\mathbf{R}'''}
  G(\mathbf{R}''',\mathbf{R}'';\tau)
  \psi_T(\mathbf{R}'')
  d\mathbf{R}
  d\mathbf{R}'
  d\mathbf{R}''
  d\mathbf{R}'''
}
{
  \int_{\mathbf{R}}\int_{\mathbf{R}'}\int_{\mathbf{R}''}
  \psi_T(\mathbf{R})
  G(\mathbf{R},\mathbf{R}';\tau)
  G(\mathbf{R}',\mathbf{R}'';\tau)
  \psi_T(\mathbf{R}'')
  d\mathbf{R}
  d\mathbf{R}'
  d\mathbf{R}''}.
\nonumber
\end{eqnarray}

The normalization factor 
$Z(\tau)$,
Eq.~(\ref{eq_znorm}),
plays a key role in the simulations.
For 
example,
if $Z(\tau)$ is known, instead of evaluating
Eq.~(\ref{eq_energy3}), 
one can 
calculate 
the 
energy 
expectation value 
$E_{\tau}$ directly,
\begin{equation}
  E_{\tau}= -\frac{1}{Z(\tau)}\frac{\partial Z(\tau)}{\partial(2\tau)}.
  \label{eq_energy4}
\end{equation}
Equations~(\ref{eq_energy3}) 
and (\ref{eq_energy4}) generate two 
distinct energy estimators (see Sec.~\ref{sec_algorithm_expectation} for details).

In 
the
zero propagation time limit, i.e., for $\tau=0$,
$\hat{G}=\exp(-\tau \hat{H})$ becomes the identity operator.
This implies that the propagator is simply a $\delta$-function in position
space,
\begin{equation}
G(\mathbf{R},\mathbf{R}';0)=\delta(\mathbf{R}-\mathbf{R}').
  \label{chpimc_deltarho}
\end{equation}
To propagate to finite imaginary time, one can solve the Bloch
equation~\cite{ceperleyrev}
\begin{equation}
  \frac{\partial \hat{G}}{\partial \tau}=-\hat{H} \hat{G},
  \label{ch2pimc_bloch}
\end{equation}
which is 
obtained by taking the partial derivative of the propagator with respect
to $\tau$.
Equation~(\ref{ch2pimc_bloch}) can be interpreted 
as a diffusion equation in the 
imaginary time $\tau$.
For the remainder of this section, we write 
the Hamiltonian $\hat{H}$ as a sum of the kinetic energy operator
$\hat{K}$ and the potential energy operator $\hat{V}$.
Moreover, we assume that all particles have the same mass $m$;
this assumption, which can be readily relaxed, simplifies the notation.
If the kinetic energy operator $\hat{K}$ is zero,
the propagator
can be readily solved for.
Similarly, if the potential energy operator $\hat{V}$ is zero,
the propagator can also be solved for.
In this case, the solution $\hat{G}_0$ 
corresponds to free particles diffusing in space
(the subscript ``0'' is used to indicate that the corresponding Hamiltonian contains only 
kinetic energy terms),
i.e., $\hat{G}_0$
is a product of single-particle 
Gaussians,
\begin{equation}
  G_0(\mathbf{R},\mathbf{R}';\tau)=(4\pi \lambda_m \tau)^{-3N/2}\exp\left(-\frac{(\mathbf{R}-\mathbf{R}')^2}{4\lambda_m\tau}\right),
  \label{chpimc_rho0}
\end{equation}
where $\lambda_m$ is equal to $\hbar^2/(2m)$.
Equation~(\ref{chpimc_rho0}) shows that the off-diagonal 
terms (terms for which $\mathbf{R}\neq\mathbf{R}'$) 
of $G_0$, expressed in the position basis, are non-zero.
This shows explicitly that the kinetic energy operator is non-local in
position space.
If $\hat{V}$ and $\hat{K}$ are both non-zero, then the propagator at
finite $\tau$ is known only for a few
selected
problems such as non-interacting particles
in a harmonic trap~\cite{krauth2006statistical} and two
particles with zero-range interactions~\cite{schulman86,lawande88,blinder88,wodkiewicz91,zr15}.
In general, the 
$N$-particle
imaginary time evolution operator
or propagator is unknown. If it was known, the problem
would be 
``trivial''. 

The PIGS algorithm is based
on the idea of writing
the imaginary time evolution operator for large $\tau$ as a product over 
imaginary time evolution operators for small imaginary time steps,
\begin{equation}
  \exp(-\tau \hat{H})=[\exp(-\tau \hat{H}/n)]^n.
  \label{chpimc_split}
\end{equation}
Using Eq.~(\ref{chpimc_split}) in Eq.~(\ref{eq_propagator})
and
inserting 
the unit operator [Eq.~(\ref{chpimc_Ridentity})]
$n-1$ times,
we obtain 
\begin{eqnarray}
  \fl
 G(\mathbf{R}_0,\mathbf{R}_n;\tau)&=
  \bra{\mathbf{R}_0}e^{-\tau \hat{H}/n}
  \underbrace{\int_{\mathbf{R}_1}\ket{\mathbf{R}_1}\bra{\mathbf{R}_1}d\mathbf{R}_1}_{=\hat{1}}
e^{-\tau \hat{H}/n}
\underbrace{\int_{\mathbf{R}_2}\ket{\mathbf{R}_2}\bra{\mathbf{R}_2}d\mathbf{R}_2}_{=\hat{1}} \times \cdots \nonumber \\
&\times\underbrace{\int_{\mathbf{R}_{n-1}}\ket{\mathbf{R}_{n-1}}\bra{\mathbf{R}_{n-1}}d\mathbf{R}_{n-1}}_{=\hat{1}}
e^{-\tau \hat{H}/n}
  \ket{\mathbf{R}_n}
  \label{chpimc_insertrho}
\end{eqnarray}
or
\begin{eqnarray}
  \fl
  G(\mathbf{R}_0,\mathbf{R}_n;\tau)=
  \int_{\mathbf{R}_1}\int_{\mathbf{R}_2} \cdots\int_{\mathbf{R}_{n-1}}&G(\mathbf{R}_0,\mathbf{R}_1;\tau/n)G(\mathbf{R}_1,\mathbf{R}_2;\tau/n)\times \cdots\nonumber\\
  &\times G(\mathbf{R}_{n-1},\mathbf{R}_n;\tau/n)d\mathbf{R}_1d\mathbf{R}_2 \cdots d\mathbf{R}_{n-1}.
  \label{chpimc_rhosplit}
\end{eqnarray}
The problem of evaluating the
propagator at the desired imaginary time 
$\tau$ has been converted to evaluating
$n$ propagators at $\tau/n$ and integrating over $n-1$
(potentially high-dimensional) auxiliary coordinates
$\mathbf{R}_1,\cdots,\mathbf{R}_{n-1}$.
The key points are that one can typically find a 
fairly accurate but approximate
short-time propagator for finite $n$
(see Secs.~\ref{sec_trotterformula}-\ref{sec_propagator_for_tb_zr_interactions})
and that the $n-1$ associated ``auxiliary''
integrations can be performed efficiently 
by Monte Carlo techniques
(see Sec.~\ref{sec_algorithm}).

To simplify the notation, the product 
$G(\mathbf{R},\mathbf{R}';\tau)G(\mathbf{R}',\mathbf{R}'';\tau)$
in the denominator of Eq.~(\ref{eq_energy3}) is rewritten as
$G(\mathbf{R}_0,\mathbf{R}_n;\tau)G(\mathbf{R}_n,\mathbf{R}_{2n};\tau)$.
Each set of coordinates $\mathbf{R}_j$ inserted in Eq.~(\ref{chpimc_insertrho})
is referred to as a ``time slice''.
There are three 
``special''
time slices: 
the initial time slice $\mathbf{R}_{0}$, the middle time slice
$\mathbf{R}_{n}$, and the final time slice $\mathbf{R}_{2n}$.
The initial and middle time slices
are connected by the propagator $G(\mathbf{R}_0,\mathbf{R}_n;\tau)$
and the middle and final time slices are connected
by the propagator 
$G(\mathbf{R}_n,\mathbf{R}_{2n};\tau)$.
Since both propagators are rewritten by inserting $n-1$
auxiliary time slices,
the ``expanded'' partition function contains a total of $2n+1$ time slices.
The position vector ${\mathbf{r}}_{k,j}$ of
the $k$-th particle in the set of coordinates $\mathbf{R}_j$ 
is referred to as a ``bead''.
Thus, a single particle is
represented 
by $2n+1$ beads.
The propagator that 
``connects'' two consecutive time slices is referred to as a ``link''.
The inverse temperature corresponding to this link is $\Delta\tau$, where $\Delta\tau=\tau/n$.
The propagator that 
``connects'' two consecutive beads is referred to as a ``single-particle link''.
In addition, the set of all time slices
$\{\mathbf{R}_0, \cdots, \mathbf{R}_{2n}\}$ is referred to as a configuration.
The definitions are summarized in Table~\ref{chpimc_tabdef}.
\begin{table}
\caption{
PIGS terminology used in this 
article.
  Columns 1-3 show the term, symbol, and explanation,
  respectively.
}
\label{chpimc_tabdef}
\centering
\begin{tabular}{l | c |l}
\hline
\hline
bead  &$\mathbf{r}_{k,j}$ & a single coordinate of the $k$-th particle\\
&& at the $j$-th
imaginary time index\\ \hline
time slice & $\mathbf{R}_j$  & a set of beads at the $j$-th\\ 
&&imaginary time index\\ \hline
configuration &$\{\mathbf{R}_0,\cdots,\mathbf{R}_{2n}\}$ & the set of all time slices \\ \hline
link & $G(\mathbf{R}_j,\mathbf{R}_{j+1};\Delta\tau)$ &  the propagator connecting two\\
&& consecutive time slices\\ \hline
single-particle link & $G(\mathbf{r}_{k,j},\mathbf{r}_{k,j+1};\Delta\tau)$ &  the propagator 
connecting two\\
&&consecutive beads\\
\hline
\hline
\end{tabular}
\end{table}

\begin{figure}
\centering
\includegraphics[angle=0,width=0.4\textwidth]{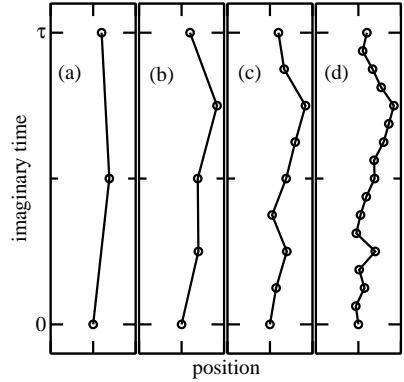}
\caption{
  Schematic world-line representation for a single particle 
with mass $m$
in a one-dimensional
  harmonic trap
with angular trapping frequency $\omega$.
  Panels~(a), (b), (c), and (d) show paths for $3$, 5, 9, and 17 time slices,
  respectively.
 }\label{chpimc_figtrotter}
\end{figure}

Figure~\ref{chpimc_figtrotter} 
shows the world-line representation of a single particle in a
one-dimensional harmonic trap.
World lines move in position space ($x$-axis) and imaginary time ($y$-axis).
Figures~\ref{chpimc_figtrotter}(a), \ref{chpimc_figtrotter}(b), 
\ref{chpimc_figtrotter}(c), and \ref{chpimc_figtrotter}(d) 
show paths for $n=$ 3, 5, 9, and 17 beads,
respectively.
As $n$ increases, the path is resolved in more detail (each link is more accurate) and
observables calculated based on the sampled paths become more accurate.

\begin{figure}
\centering
\includegraphics[angle=0,width=0.4\textwidth]{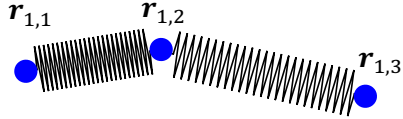}
\caption{
  Isomorphism between path integrals for a single free particle in
  two-dimensional space and classical particles connected by springs.
  In the path integral interpretation, the circles and wiggly lines depict 
  the beads and (single-particle) links of a single particle, respectively.
  In the classical mechanics formulation, the circles and wiggly lines depict
  particles and springs, respectively.
The position vector of the particle is denoted by
$\mathbf{r}_{1,j}$, where
$j$ indicates the imaginary time index
($j=1, 2$, and $3$).
 }\label{chpimc_figspring}
\end{figure}

Figure~\ref{chpimc_figspring} depicts a single particle 
(the Hamiltonian only contains the kinetic energy term) in two-dimensional 
space~\cite{classical}.
Two consecutive beads (circles in Fig.~\ref{chpimc_figspring}) are connected by a
single-particle link (wiggly line in Fig.~\ref{chpimc_figspring}).
The kinetic energy is ``carried'' by the propagators represented by the links.
The expression for the 
propagator in free space reads [Eq.~(\ref{chpimc_rho0}) 
for a single particle with position vector 
$\mathbf{r}_{1,j}=(x_{1,j},y_{1,j})$]
\begin{equation}
  G_0(\mathbf{r}_{1,j},\mathbf{r}_{1,j+1};\tau)=
(4\pi \lambda_m \tau)^{-1}
\exp\left(-\frac{(\mathbf{r}_{1,j}-\mathbf{r}_{1,j+1})^2}
{4\lambda_m\tau}\right).
  \label{chpimc_sr}
\end{equation}
The action $S$~\cite{ceperleyrev},
\begin{equation}
S=-\ln[G_0(\mathbf{r}_{1,j},\mathbf{r}_{1,j+1};\tau)],
  \label{<++>}
\end{equation}
of the single-particle link that connects the beads labeled $\mathbf{r}_{1,j}$
and $\mathbf{r}_{1,j+1}$
reads
\begin{equation}
  S=\ln(4\pi \lambda_m \tau)+
\frac{(\mathbf{r}_{1,j}-\mathbf{r}_{1,j+1})^2}{4\lambda_m\tau}.
  \label{chpimc_ss}
\end{equation}
It can be seen that the action $S$ has the same form 
as that of a ``spring potential'' 
$V_s(\mathbf{r}_{1,j}-\mathbf{r}_{1,j+1})$ 
for two classical particles with position vectors 
$\mathbf{r}_{1,j}$ and $\mathbf{r}_{1,j+1}$ connected via Hooke's law.
The propagator can thus be interpreted as 
being proportional to the Boltzmann factor  $\exp(-\tau V_s)$
of a classical system of springs.
Note that $\mathbf{r}_{1,j}$ and $\mathbf{r}_{1,j+1}$ 
in Eqs.~(\ref{chpimc_sr}) and
(\ref{chpimc_ss}) correspond to 
position vectors of consecutive beads for one single particle
while $\mathbf{r}_{1,j}$ and $\mathbf{r}_{1,j+1}$ 
in the classical isomorphism correspond to position vectors of two different
particles.

In addition to the propagators $G(\mathbf{R}_0,\mathbf{R}_n;\tau)$
and $G(\mathbf{R}_n,\mathbf{R}_{2n};\tau)$,
the partition function contains the trial functions (or ``weights'')
$\psi_T(\mathbf{R}_0)$ and $\psi_T(\mathbf{R}_{2n})$.
For the single two-dimensional particle
in free space, 
$\psi_T(\mathbf{r}_0)$ and $\psi_T(\mathbf{r}_{2n})$
can be 
interpreted 
as 
potentials 
that are felt by the first and last particle of the
chain of classical particles.
Thus, we can interpret the 
PIGS
simulation of a single particle 
as a simulation of a chain of classical particles connected by
springs (or a polymer with nearest neighbor interactions)
and two external forces that act on the particles at the ends of the chain.
The stiffness of the springs is determined by 
$1/\Delta\tau$, i.e., 
by the inverse of the imaginary time step 
associated with the links.

The following two sections introduce two different approaches for
approximating the short-time propagator,
namely the Trotter formula and the pair product approximation.

\subsection{Trotter formula}
\label{sec_trotterformula}
One way to approximate the short-time propagator is to
use the Trotter formula~\cite{trotter1959product}.
For 
sufficiently
small 
time 
steps $\Delta \tau$, the
kinetic energy contribution $\hat{K}$ and the potential 
energy contribution $\hat{V}$ to the propagator can be split,
\begin{equation}
  \exp[-\Delta\tau (\hat{K}+\hat{V})]=\exp(-\Delta\tau \hat{K})\exp(-\Delta\tau \hat{V}) + \mathcal{O}(\Delta\tau^2),
  \label{<++>}
\end{equation}
where the notation $\mathcal{O}(\Delta\tau^2)$ indicates that 
the leading-order error scales, in general, as $\Delta\tau$ to the power of 2.
More specifically,
by Taylor expanding the exponentials, one can prove that the leading-order
error is 
$\Delta\tau^2 [\hat{K},\hat{V}]/2$, 
where $[\hat{K},\hat{V}]$ is the commutator between
$\hat{K}$ and $\hat{V}$, $[\hat{K},\hat{V}]=\hat{K}\hat{V}-\hat{V}\hat{K}$.
In the $\Delta \tau \rightarrow 0$ limit (this
corresponds to the insertion of infinitely many time slices, i.e., the $n \rightarrow \infty$
limit),
the Trotter formula becomes exact.
Since
$n$ cannot be infinitely large
in practice,
one performs 
calculations for different $n$ and
extrapolates
the observables of interest to the infinite $n$ limit.

Importantly, the Trotter formula can be extended to higher orders.
We can readily adopt a $\mathcal{O}(\Delta\tau^3)$ scheme by further splitting the kinetic
energy term or the potential energy term, 
\begin{eqnarray}
  \fl
  \exp[-\Delta\tau (\hat{K}+\hat{V})]&= \exp(-\Delta\tau \hat{K}/2)\exp(-\Delta\tau \hat{V})\exp(-\Delta\tau \hat{K}/2) + \mathcal{O}(\Delta\tau^3)
  \label{chpimc_trotter1}
\end{eqnarray}
or
\begin{eqnarray}
  \fl
  \exp[-\Delta\tau (\hat{K}+\hat{V})]&= \exp(-\Delta\tau \hat{V}/2)\exp(-\Delta\tau \hat{K})\exp(-\Delta\tau \hat{V}/2) + \mathcal{O}(\Delta\tau^3).
  \label{chpimc_trotter2}
\end{eqnarray}
In practice, Eq.~(\ref{chpimc_trotter2}), which is accurate up to second
order [the error is $\mathcal{O}(\Delta\tau^3)$], is easier to use than
Eq.~(\ref{chpimc_trotter1}).
In position space, Eq.~(\ref{chpimc_trotter2}) reads
\begin{eqnarray}
  \fl
  G(\mathbf{R},\mathbf{R}';\Delta\tau)=\exp[-\Delta\tau V(\mathbf{R})/2]\exp[-\Delta\tau V(\mathbf{R}')/2]G_0(\mathbf{R},\mathbf{R}';\Delta\tau)
  +\mathcal{O}(\Delta\tau^3),
  \label{chpimc_trotterfinal}
\end{eqnarray}
where $G_0$ [see Eq.~(\ref{chpimc_rho0})] is the propagator that accounts for the kinetic energy term.

One can reach successively higher accuracy by the repeated use of
the 
Baker-Campell-Hausdorff formula (see, e.g., Ref.~\cite{rossmann2006lie})
\begin{eqnarray}
  e^{\epsilon \hat{A}}e^{\epsilon \hat{B}}=e^{\hat{C}},
  \label{chpimc_bch0}
\end{eqnarray}
where
\begin{eqnarray}
  \hat{C}=
    (\hat{A}+\hat{B})\epsilon
+\frac{1}{2}[\hat{A},\hat{B}]\epsilon^2+\frac{1}{12}
\left( \left[\hat{A},[\hat{A},\hat{B}]
    \right]+ \left[\hat{B},[\hat{B},\hat{A}] \right]
\right)
\epsilon^3\nonumber\\
    -\frac{1}{24} \left[\hat{B}, \left[\hat{A},[\hat{A},\hat{B}] \right] \right]\epsilon^4
+\mathcal{O}(\epsilon^5).
  \label{chpimc_bch}
\end{eqnarray}
Using Eqs.~(\ref{chpimc_bch0}) and (\ref{chpimc_bch}) twice, we obtain~\cite{chin97}
\begin{equation}
  e^{\epsilon \hat{B}}e^{\epsilon \hat{A}}e^{\epsilon \hat{B}}=e^{\hat{D}},
  \label{chpimc_bch2}
\end{equation}
where
\begin{equation}
  \hat{D}=
(\hat{A}+2\hat{B}) \epsilon
-\frac{1}{6}\epsilon^3 \left[\hat{A},[\hat{B},\hat{A}] \right]+\frac{1}{6}\epsilon^3 \left[\hat{B},[\hat{A},\hat{B}] \right]+
\mathcal{O}(\epsilon^5).
  \label{chpimc_bch3}
\end{equation}
Applying Eqs.~(\ref{chpimc_bch2}) and (\ref{chpimc_bch3}) twice to
\begin{eqnarray}
  \fl
\exp \left(-\Delta\tau \frac{\hat{V}}{6} \right)
    \exp \left(-\Delta\tau \frac{\hat{K}}{2} \right) \exp \left(-\Delta\tau \frac{2\tilde{{V}}}{3} \right)
\exp \left(-\Delta\tau \frac{\hat{K}}{2} \right)
\exp \left(-\Delta\tau \frac{\hat{V}}{6} \right),
  \label{<++>}
\end{eqnarray}
we can check that the fourth-order factorization~\cite{chin97}
\begin{eqnarray}
  \fl
  \exp \left[-\Delta\tau(\hat{K}+\hat{V}) \right]=&
\exp \left(-\Delta\tau \frac{\hat{V}}{6} \right)
    \exp \left(-\Delta\tau \frac{\hat{K}}{2} \right) \times \nonumber \\\
&\exp \left(-\Delta\tau \frac{2\tilde{{V}}}{3} \right)
\exp \left(-\Delta\tau \frac{\hat{K}}{2} \right)
\exp \left(-\Delta\tau \frac{\hat{V}}{6} \right) \nonumber \\
&+ \mathcal{O}(\Delta\tau^5),
  \label{chpimc_factorization4}
\end{eqnarray}
where $\tilde{V}$ is 
given by $\hat{V}+\Delta\tau^2[\hat{V},[\hat{K},\hat{V}]]/48$, holds.
The term $[\hat{V},[\hat{K},\hat{V}]]$, 
in position space, can be simplified to 
$(\hbar^2/m)\sum_{i=1}^N|\nabla_i V|^2$,
where
the
gradient $\nabla_i$ in three spatial dimensions 
is given by
\begin{equation}
  \nabla_i=\hat{x}_i\frac{\partial}{\partial x_i}+
\hat{y}_i\frac{\partial}{\partial y_i}+
\hat{z}_i\frac{\partial}{\partial z_i},
  \label{<++>}
\end{equation}
with
$\hat{x}_i$, $\hat{y}_i$, and $\hat{z}_i$ 
denoting
unit vectors 
that point
in the
$x_i$, $y_i$, and $z_i$ directions, respectively.
The
term $|\nabla_i V|^2$ corresponds
to the square of the force on the $i$-th particle.
Care needs to be taken in evaluating the derivatives, since $V$ usually
contains a double sum over two-body potentials or even a triple sum over
three-body
potentials.
In most cases, the evaluation of the force terms cannot be simplified
analytically, implying that the evaluation of double commutators involves
double or triple sums over 
the total number of particles.
This makes the numerical evaluation comparatively
expensive.
Note that 
the exponentials in Eq.~(\ref{chpimc_factorization4}) 
that contain the potential energy can 
have different numerical factors.
In addition, 
Eq.~(\ref{chpimc_factorization4}) 
is not the only form of the fourth-order
factorization~\cite{chin97,suzuki95}.

\begin{figure}
\centering
\includegraphics[angle=0,width=0.4\textwidth]{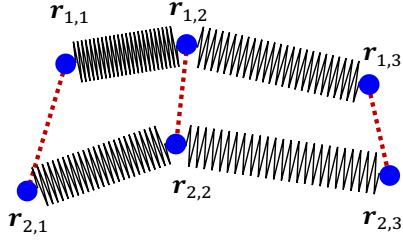}
\caption{
  Isomorphism between path integrals for two interacting particles in
  two-dimensional space and classical particles connected by springs.
In the path integral interpretation,
the circles and wiggly lines depict the beads and links of a single particle,
respectively,
and the
dotted lines depict 
the 
two-body 
interaction
between beads with the same imaginary time index.
In the
classical mechanics 
interpretation,
the circles and wiggly lines depict particles and springs,
respectively,
and the
dotted
lines depict 
the 
two-body 
interaction 
between 
selected classical
particles.
The 
position vector
of the $k$-th particle 
at the $j$-th imaginary time index $j$ is
denoted by
$\mathbf{r}_{k,j}$. 
 }\label{chpimc_figspring2}
\end{figure}

Using the Trotter formula, the isomorphism 
between
a
single 
quantum
particle in free space and the classical spring system can be extended
to multiple 
quantum
particles with interactions.
Figure~\ref{chpimc_figspring2} depicts two interacting particles in two-dimensional 
space 
(it is assumed that
the particles do not feel a single-particle potential).
In this case, the propagator for the link that connects the beads labeled 
by
$\mathbf{r}_{1,j}$, $\mathbf{r}_{1,j+1}$, 
$\mathbf{r}_{2,j}$, and $\mathbf{r}_{2,j+1}$ reads
[see Eq.~(\ref{chpimc_trotter2})]
\begin{eqnarray}
  \fl
  G(\{\mathbf{r}_{1,j},\mathbf{r}_{2,j}\},\{\mathbf{r}_{1,j+1},\mathbf{r}_{2,j+1}\};\Delta\tau) 
  =&e^{-\Delta\tau V_{\rm{2b}}(\mathbf{r}_{1,j}-\mathbf{r}_{2,j})/2}
e^{-\Delta\tau
  V_{\rm{2b}}(\mathbf{r}_{1,j+1}-\mathbf{r}_{2,j+1})/2}\nonumber\\
&\times G_0(\mathbf{r}_{1,j},\mathbf{r}_{1,j+1};\Delta\tau)G_0(\mathbf{r}_{2,j},\mathbf{r}_{2,j+1};\Delta\tau),
  \label{chpimc_2sr}
\end{eqnarray}
where $V_{\rm{2b}}$ denotes the two-body interaction potential between particles
1 and 2 and $G_0(\mathbf{r}_{k,j},\mathbf{r}_{k,j+1};\Delta\tau)$ the single-particle
propagator of the $k$-th particle [see Eq.~(\ref{chpimc_sr})].
As in Fig.~\ref{chpimc_figspring}, two consecutive beads for the same particle
(e.g., the circles labeled by $\mathbf{r}_{1,j}$ and $\mathbf{r}_{1,j+1}$
in Fig.~\ref{chpimc_figspring2}) are connected by 
single-particle links
(wiggly lines in Fig.~\ref{chpimc_figspring2})
that 
represent the propagators $G_0$. 
Since the two-body interaction 
potential
(dotted lines in Fig.~\ref{chpimc_figspring2}) 
is diagonal in position space 
[see the exponentials on the right hand side of Eq.~(\ref{chpimc_2sr})],
it connects beads of different particles with the same index, i.e., it
connects $\mathbf{r}_{1,1}$ and $\mathbf{r}_{2,1}$, $\mathbf{r}_{1,2}$ and 
$\mathbf{r}_{2,2}$, and $\mathbf{r}_{1,3}$ and $\mathbf{r}_{2,3}$ (or,
in general, it connects
$\mathbf{r}_{k,j}$ and $\mathbf{r}_{k+1,j}$).
Each particle in the 
PIGS 
simulation corresponds to 
$2n+1$ 
classical particles connected by springs.
Classical
particles 
associated with 
different chains interact only
if they have the same bead index.

\subsection{Pair product approximation}
\label{sec_ppa}
To introduce the pair product approximation, we assume for simplicity
that the potential energy operator $\hat{V}$ can be written as a sum
of two-body terms,
\begin{equation}
    V(\mathbf{R})=\sum_{k=1}^{N-1}\sum_{l=k+1}^{N}V_{\rm{2b}}(\mathbf{r}_{k}-\mathbf{r}_l),
  \label{}
\end{equation}
i.e., we assume for now that single-particle and three- and higher-body
forces are absent.
Under these assumptions,
the short-time propagator can be evaluated using the pair
product approximation~\cite{ceperleyrev}.
It is convenient to define the two-body kinetic energy operator
$\hat{K}_{kl}$ for the $k$-th and $l$-th particle in position space,
\begin{equation}
  \hat{K}_{kl}= -\frac{\hbar^2}{m}\nabla_{\mathbf{r}_{k}-\mathbf{r}_l}^2
    .
  \label{<++>}
\end{equation}
The relative non-interacting and interacting two-body 
Hamiltonian are $\hat{K}_{kl}$
and $\hat{K}_{kl}+\hat{V}_{\rm{2b}}(\mathbf{r}_{k}-\mathbf{r}_l)$, respectively.
The pair product approximation considers two-body correlations explicitly,
but not
higher-body correlations, and writes
the many-body propagator as a product over single-particle propagators and two-body propagators,
\begin{eqnarray}
  G(\mathbf{R},\mathbf{R}';\Delta\tau)
  \approx
  G_0(\mathbf{R},\mathbf{R}';\Delta\tau) 
    \left(\prod_{k<l}^N\bar{G}^{\rm{rel}}(\mathbf{r}_{k}-\mathbf{r}_l,\mathbf{r}'_{k}-\mathbf{r}_l';\Delta\tau)
    \right)
    ,
  \label{chpimc_rhotot}
\end{eqnarray}
where $\bar{G}^{\rm{rel}}$,
\begin{equation}
  \fl
    \bar{G}^{\rm{rel}}(\mathbf{r}_{k}-\mathbf{r}_l,\mathbf{r}'_{k}-\mathbf{r}_l';\Delta\tau)=
\frac{\braket{\mathbf{r}_{k}-\mathbf{r}_l|
\exp[-\Delta\tau(\hat{K}_{kl}+\hat{V}_{\rm{2b}}(\mathbf{r}_{k}-\mathbf{r}_l))]|
\mathbf{r}_{k}'-\mathbf{r}_l'}}{\braket{\mathbf{r}_{k}-\mathbf{r}_l|\exp(-\Delta\tau \hat{K}_{kl})|\mathbf{r}_{k}'-\mathbf{r}_l'}},
  \label{eq_gbarrel}
\end{equation}
is the reduced pair propagator.
The denominator of the reduced pair propagator coincides with the 
known relative non-interacting two-body 
propagator.
Thus the only ``non-trivial input'' is the relative propagator of the
interacting two-body system.
One can readily see that the pair product approximation is exact for two
particles for any propagation time
because the center-of-mass and relative degrees of
freedom separate in this case.
In some cases such as for the two-body zero-range interaction potential, the
exact reduced pair propagator is known
analytically~\cite{schulman86,blinder88,lawande88,wodkiewicz91}.
In other cases such as for the two-body hardcore potential, 
an approximate reduced pair propagator is known analytically 
in closed form~\cite{barker79,caoberne}.
If the reduced propagator is not known analytically, one can perform a partial wave decomposition and obtain
a numerical representation of the reduced two-body propagator~\cite{ceperleyrev}.

In dilute gases or weakly-bound droplets, 
the interparticle spacing is typically 
so large that two-body collisions dominate over three- and
higher-body collisions.
The leading-order error of the pair product approximation is determined by the
importance of three-body correlations.
For two-component equal-mass Fermi gases with two-body zero-range interactions,
three-body correlations are suppressed by the Pauli exclusion principle.
For this system, we found that
the pair product approximation provides an extremely good description of the
propagator.
Specifically, we obtain accurate simulation results for a small number of
beads (see Sec.~\ref{sec_application} for details).
For bosons, in contrast, three-body correlations can be significant.
As a consequence, the pair product approximation is not as efficient as for
two-component 
fermions and simulations typically 
employ a large
number of beads (``large'' in this context means about two orders of
magnitude more number of beads as in the simulations for 
fermions~\cite{zr15}).

To illustrate the pair product approximation, we cannot use the classical
isomorphism because the kinetic and potential energy contributions are
mixed. One needs to evaluate the single-particle propagator, which can
be represented by springs as in Figs.~\ref{chpimc_figspring} and
\ref{chpimc_figspring2}.
However, one also needs to evaluate 
the reduced two-body propagator, which connects
two consecutive beads of one particle's path 
with the same consecutive beads of another particle's path.
These ``four-bead connections'' do not have a simple classical analog.

\subsection{Comparison of the two approximations}
\label{sec_comparison_of_the_two_approximations}

This section discusses the advantages and disadvantages of
approximating the short-time propagator with the help of
the Trotter formula
and the pair product approximation.

In the Trotter formula based scheme, 
the kinetic and the potential energy terms are treated separately.
Inserting the identity 
$\int_{\mathbf{R}}\ket{\mathbf{R}}\bra{\mathbf{R}}d\mathbf{R}=\hat{1}$,
Eq.~(\ref{chpimc_Ridentity}), 
multiple times into
Eqs.~(\ref{chpimc_trotter2}) 
or (\ref{chpimc_factorization4}),
it can be seen that the potential energy is diagonal in position space.
This means that one can
directly evaluate the potential energy term at each time slice.
The kinetic energy term contains off-diagonal terms and needs to be evaluated
at each link instead of at each time slice.
Nevertheless, since the kinetic energy term corresponds to a simple Gaussian, 
the sampling 
of the kinetic energy piece of the propagator can, in general,
be performed
efficiently (see Sec.~\ref{wiggle-ch2} for details).

Even though the Trotter formula can formally be generalized to expressions
that are accurate to order
$\Delta\tau^5, \Delta\tau^6,\cdots$~\cite{chin05,chin10},
many of these expressions are of limited use in practice
because they contain either commutators 
that involve rather complicated expressions or 
terms
that correspond
to negative imaginary time, which are not
normalizable.
There exists a multi-product expansion for the propagator~\cite{chin10};
however, applications thereof are still rare~\cite{chin10app}.
Thus present-day algorithms
mostly employ Trotter formula based decompositions that are accurate to order 
$\Delta\tau^4$.

In the pair product approximation, the two-body reduced propagator contains 
kinetic
energy and potential energy contributions. This means that the reduced
two-body propagator
has
to be evaluated at each link.
Because the reduced two-body propagator is, in general, not a simple
Gaussian,
the sampling is 
typically
less efficient than in the case where the Trotter formula is used.
Furthermore, as discussed in Sec.~\ref{sec_algorithm_permutation},
the evaluation of the permutations is computationally more involved.

Our discussion of the pair
product approximation assumed that the potential
energy can be written as a sum over two-body terms.
If the Hamiltonian contains one-, three-, or higher-body potential energy terms,
one can include them by combining the Trotter formula and the pair product
approximation.
To this end, one first splits the propagator into two terms using the
Trotter formula.
The first term contains the kinetic energy operator and the two-body 
interactions while the second term contains all other
potential terms.
One then 
applies the pair product approximation to the first term.
In this approach, it is most convenient to use the 
second-order Trotter formula for two reasons.
First, if a higher-order Trotter formula was used, one
would need to evaluate the commutator between the one-,
three- and higher-body potential terms and the two-body potential terms.
For the two-body zero-range interactions considered in 
Sec.~\ref{sec_application}
this is a 
rather challenging task.
Second, both the second-order Trotter formula and the pair
product approximation yield errors for the energy
that scale quadratically with the time step $\Delta \tau$.
While the error in the pair product approximation tends to be smaller
than that associated with the pair product approximation, 
ultimately it is the scaling with $\Delta \tau$ that determines
the accuracy and use of
the fourth-order Trotter formula typically
leads only to a small overall improvement.

From our perspective,
the pair product approximation has one key advantage:
It can deal with 
a class of two-body potentials that the Trotter formula based scheme cannot
deal with (at least no such treatment is known to us).
For example, 
the two-body hardcore and zero-range potentials contain infinities and can thus not be
treated by the Trotter formula based scheme.
However, the infinities 
can, 
as discussed in the next section
exemplarily for the two-body zero-range potential,
be dealt with analytically in the pair product approximation.

\subsection{Propagator for two-body zero-range interactions}
\label{sec_propagator_for_tb_zr_interactions}
As alluded to in 
the
previous section,
two-body 
zero-range interactions can be
incorporated into 
continuum Monte Carlo simulations through the pair product approximation~\cite{zr15,ceperley08,schmidt14,schmidt15},
which employs the relative
propagator $\bar{G}^{\rm{rel}}$
[see Eqs.~(\ref{chpimc_rhotot}) and (\ref{eq_gbarrel})].
In what follows, we limit our discussion to three spatial dimensions.
To determine $\bar{G}^{\rm{rel}}$, one considers the relative
Hamiltonian $H^{\rm{rel}}$
for two particles interacting through the regularized zero-range Fermi-Huang 
pseudopotential $V_{\rm{F}}(r)$~\cite{zerorangepotential} in free space,
  \begin{equation}
    H^{\rm{rel}}=-\frac{\hbar^2}{2\mu} \nabla_{\mathbf{r}}^2+ V_{\rm{F}}(r),
    \label{eq_hrel_twobodyzr}
\end{equation}
where
\begin{equation}
V_{\rm{F}}(r)=
\frac{2\pi\hbar^2 a_s}{\mu}\delta^{(3)}(\mathbf{r})\frac{\partial}{\partial r}r.
\label{eq_fermihuangregularized}
  \end{equation}
   Here, $\mu$
denotes the two-body reduced mass, $\mathbf{r}$
the interparticle distance vector, and $a_s$ the $s$-wave scattering length.
The regularization operator $(\partial/\partial r)r$ in
Eq.~(\ref{eq_fermihuangregularized}) ensures that the Hamiltonian
is well-behaved. Without this operator,
the two-body coupling constant would have to be renormalized.
With the regulator, however, the coupling strength is uniquely
determined and given by $2 \pi \hbar^2 a_s/ \mu$.

The reduced (or normalized) relative propagator corresponding to the 
Hamiltonian given in Eq.~(\ref{eq_hrel_twobodyzr}) reads~\cite{lawande88,wodkiewicz91}
  \begin{equation}
    \fl
    \bar{G}^{\rm{rel}}(\mathbf{r},\mathbf{r}';\tau)= 1+\frac{\hbar^2
    \tau}{\mu r r'}\exp\left(-\frac{\mu r r'(1+\cos\theta)}{\hbar^2 \tau}\right)
    \left(1+\frac{\hbar}{a_s}\sqrt{\frac{\pi\tau}{2\mu}}{\rm{erfc}}(v)\exp(v^2)\right),
    \label{3dfiniteg}
  \end{equation}
where $\cos\theta= \mathbf{r}\cdot\mathbf{r}'/(r r')$ and
$v=[r+r'-\tau \hbar^2/(\mu a_s)]/\sqrt{2\tau\hbar^2/\mu}$.
For $|a_s|=\infty$, the length scale $a_s$ 
drops out of the expression for the propagator
and 
Eq.~(\ref{3dfiniteg})
simplifies to
\begin{equation}
  \bar{G}^{\rm{rel}}(\mathbf{r},\mathbf{r}';\tau)= 1+\frac{\hbar^2
  \tau}{\mu r r'}\exp\left(-\frac{\mu r r'(1+\cos\theta)}{\hbar^2 \tau}\right)
  .
  \label{3dfreedensitymatrix}
\end{equation}
Importantly, the reduced relative propagator diverges when $r$ or
$r'$ approach zero.
These divergencies have implications for the Monte Carlo sampling
of the paths. As discussed in detail in Sec.~\ref{sec_pdm},
moves have to be designed carefully such that detailed balance is
fulfilled.
For example, while $\bar{G}^{\rm{rel}}$ diverges for $r$ and $r' \rightarrow 0$,
the probability to find two particles
at vanishing interparticle distance does not diverge.
The treatment of systems with two-body hardcore interactions is
similar in spirit to that detailed here for two-body zero-range 
interactions.
  
Adding the spherically symmetric 
harmonic confining potential
$V_{\rm{trap}}(r)= \mu \omega^2 r^2 /2$ for
the relative degrees of freedom to the Hamiltonian
$H^{\rm{rel}}$ given in Eq.~(\ref{eq_hrel_twobodyzr})
and assuming that $a_s$ is infinitely large, the reduced relative
propagator $\bar{G}^{\rm{rel}}(\mathbf{r},\mathbf{r}';\tau)$
reads
\begin{eqnarray}
  \bar{G}^{\rm{rel}}(\mathbf{r},\mathbf{r}';\tau)= 1+
  \frac{2 a_{\rm{ho}}^2}{ r r'}
\mbox{sinh}(\tau \hbar \omega)
  \exp\left(-\frac{r r'(1+\cos\theta)}{2 a_{\rm{ho}}^2 {\rm{sinh}}(\tau \hbar \omega)}\right)  ,
  \label{3dHOdensitymatrix}
  \end{eqnarray}
where $a_{\rm{ho}}=\sqrt{\hbar/(m \omega)}$.
In the limit that the angular trapping frequency $\omega$ goes to
zero, Eq.~(\ref{3dHOdensitymatrix})
reduces to Eq.~(\ref{3dfreedensitymatrix}).
Expression~(\ref{3dHOdensitymatrix}) is used in Sec.~\ref{sec_application}
to treat harmonically trapped two-component Fermi gases
with two-body zero-range interactions
at unitarity using the pair product approximation.

\section{Monte Carlo techniques and the PIGS algorithm}
\label{sec_algorithm}

Throughout this section we assume that the trial function
$\psi_T(\mathbf{R})$ is given 
and that its value can be determined for any set of coordinates $\mathbf{R}$.
The functional form of $\psi_T(\mathbf{R})$ depends sensitively
on the system under study. The choice of
$\psi_T(\mathbf{R})$ and the dependence of the PIGS results
on $\psi_T(\mathbf{R})$ will be discussed
in Sec.~\ref{sec_application} for 
harmonically trapped two-component Fermi gases.

\subsection{General sampling scheme: Importance sampling}
\label{sec_algorithm_sampling}
Equation~(\ref{chpimc_rhosplit}) writes the long-time propagator as
a high-dimensional integral over a product of 
short-time propagators. This implies that 
the evaluation of the normalization factor 
$Z(\tau)$ is a high-dimensional integral.
This section discusses the Monte Carlo sampling of this high-dimensional
integral over $\{\mathbf{R}_0,\cdots,\mathbf{R}_{2n}\}$ 
(there are $3 \times (2n+1) \times N$
independent coordinates if we are 
considering three spatial dimensions).
To proceed,
we write $Z(\tau)$ explicitly in terms of the short-time propagator,
\begin{equation}
  Z(\tau)=
\int_{\mathbf{R}_0}\cdots\int_{\mathbf{R}_{2n}}\pi(\mathbf{R}_0,\cdots,\mathbf{R}_{2n}) d \mathbf{R}_0  \cdots  d \mathbf{R}_{2n},
  \label{<++>}
\end{equation}
where
\begin{eqnarray}
\pi(\mathbf{R}_0,\cdots,\mathbf{R}_{2n})=&
\psi_T(\mathbf{R}_0)
G(\mathbf{R}_0,\mathbf{R}_{1};\Delta\tau)
G(\mathbf{R}_1,\mathbf{R}_{2};\Delta\tau)\nonumber\\
&
\times
\cdots
\times
G(\mathbf{R}_{2n-1},\mathbf{R}_{2n};\Delta\tau)
\psi_T(\mathbf{R}_{2n}).
  \label{chpimc_pix}
\end{eqnarray}
To simplify the notation,
we denote the configuration $\{\mathbf{R}_0,\cdots,\mathbf{R}_{2n}\}$ by
$\mathbf{x}$ and the probability distribution
$\pi(\mathbf{R}_0,\cdots,\mathbf{R}_{2n})$ by $\pi(\mathbf{x})$.
The notation of these and other quantities is summarized in
Tables~\ref{chpimc_tabdef} and \ref{tabchpimc_MC}.
\begin{table}
  \centering
\caption{Definitions of Monte Carlo sampling terminology used in this article.
  Columns 1-3 show the symbol, name, and related equation number,
  respectively. The configuration $\mathbf{x}$ is defined as
  $\mathbf{x}=\{\mathbf{R}_0,\cdots,\mathbf{R}_{2n}\}$.
}
\label{tabchpimc_MC}
\centering
\begin{tabular}{c | c |c}
\hline
\hline
$\pi(\mathbf{x})$ & probability distribution &Eq.~(\ref{chpimc_pix})\\
$p(\mathbf{x})$ & probability density function &Eq.~(\ref{chpimc_pdf})\\
$w(\mathbf{x})$ & weight function (observable specific)
&Eqs.~(\ref{chpimc_generalobservable})-(\ref{chpimc_observablerewrite3});
Sec.~\ref{sec_algorithm_expectation}\\
$P(\mathbf{x}\to\mathbf{x}')$ & transition probability &Eq.~(\ref{chpimc_detailedbalance})\\
$\cal{G}(\mathbf{x}\to\mathbf{x}')$ & proposal distribution (selected by simulator)
& around Eqs.~(\ref{chpimc_detailedbalance})-(\ref{chpimc_accept})\\
$\cal{A}(\mathbf{x}\to\mathbf{x}')$ & acceptance distribution
&
Eq.~(\ref{chpimc_accept})\\
$\psi_T(\mathbf{R})$ & trial function (selected by simulator) & 
Eq.~(\ref{eq_expansion_of_psit}) \\
\hline
\hline
\end{tabular}
\end{table}
The expectation value $\braket{{O}}$ 
of an arbitrary observable ${O}$ can be written as
\begin{equation}
  \braket{{O}}=
  \frac{\int_{\mathbf{x}}w(\mathbf{x})\pi(\mathbf{x}) d\mathbf{x}}{\int_{\mathbf{x}}\pi(\mathbf{x}) d\mathbf{x}},
  \label{chpimc_generalobservable}
\end{equation}
where the integration goes over $3 \times (2n+1) \times N$
coordinates and the weight function $w(\mathbf{x})$ needs to
be determined, as will be discussed in Sec.~\ref{sec_algorithm_expectation}, 
for each observable.
To see the structure 
of $\braket{{O}}$ more clearly, we rewrite Eq.~(\ref{chpimc_generalobservable})
as
\begin{equation}
  \braket{{O}}=\int_{\mathbf{x}} w(\mathbf{x})p(\mathbf{x})  d\mathbf{x},
  \label{chpimc_observablerewrite3}
\end{equation}
where
the probability density function $p(\mathbf{x})$ is defined as
\begin{equation}
  p(\mathbf{x})=\frac{\pi(\mathbf{x})}{\int_{\mathbf{x}'}\pi(\mathbf{x}')
  d\mathbf{x}'}.
  \label{chpimc_pdf}
\end{equation}
In contrast to the probability 
distribution $\pi(\mathbf{x})$, the probability density function $p(\mathbf{x})$
is normalized;
$w(\mathbf{x})$ and 
$p(\mathbf{x})$ represent the weight contributed to the observable by 
the configuration $\mathbf{x}$ and the normalized probability to be in
the configuration $\mathbf{x}$, respectively.
Equation~(\ref{chpimc_observablerewrite3}) provides the basis of importance 
sampling:
Configurations are not blindly distributed uniformly in space but instead
are distributed according to
$p(\mathbf{x})$.
The advantage of importance sampling is that most computer time is used 
to sample configurations that are physically relevant and little time to sample configurations that do not contribute significantly to $\braket{{O}}$. 

The general idea of the PIGS algorithm is to generate configurations
$\mathbf{x}$ according to $p(\mathbf{x)}$ and to 
use the generated configurations to accumulate 
the weight functions $w(\mathbf{x})$ for a set of observables.
Thus, it is crucial to have correct and 
efficient sampling schemes that explore the full configuration
space with a relatively high acceptance ratio
and without 
getting stuck around a local minimum.
Section~\ref{metropolis-ch2} reviews the basics of selected Monte Carlo methods,
which are then used in the subsequent sections.

\subsection{Some background on Monte Carlo methods}
\label{metropolis-ch2}
This section discusses 
how to update or generate configurations using the Metropolis algorithm.
A Markov process is uniquely defined by the transition probability 
$P(\mathbf{x}\to\mathbf{x}')$ to go from configuration $\mathbf{x}$ to
configuration $\mathbf{x}'$.
The Metropolis algorithm 
satisfies the detailed balance condition~\cite{lester94}
\begin{equation}
  \pi(\mathbf{x})P(\mathbf{x}\to \mathbf{x}')=\pi(\mathbf{x}')P(\mathbf{x}'\to \mathbf{x}),
  \label{chpimc_detailedbalance}
\end{equation}
which states that 
the flow of probability from $\mathbf{x}$ to $\mathbf{x}'$ is equal to
that from $\mathbf{x}'$ to $\mathbf{x}$.
This means that there is no net flow of probability.
The Metropolis algorithm needs to ensure ergodicity of the Markov process.
If the process is ergodic, the Markov chain (i) returns to any 
previously generated configuration $\mathbf{x}$ 
after a sufficiently long simulation time
and (ii) is not periodic (a Markov chain of 
$\{\mathbf{x}, \mathbf{x}', \mathbf{x}, \mathbf{x}', \cdots\}$, e.g., is periodic).
The ergodicity ensures that the probability distribution $\pi(\mathbf{x})$
gets sampled fully.
For example, as discussed in Ref.~\cite{b416}, if we use the traditional 
scheme of treating the permutations~\cite{ceperleyrev}, for
a two-component Fermi gas with zero-range interactions, 
the Markov process ends up with a configuration in which all particles
sit on top of each other and the configuration 
never returns to the original configuration 
$\mathbf{x}$. This means that
ergodicity is violated and that the Markov process does not
generate samples according to $p(\mathbf{x})$. This renders the sampled
configurations meaningless.
We note, however, that while the detailed balance condition together with the
ergodicity guarantees that the equilibrium distribution coincides with
the desired
probability distribution $\pi(\mathbf{x})$,
there exist other Monte Carlo methods that do not satisfy the detailed
balance condition but yield an equilibrium distribution that coincides 
with the desired probability distribution 
$\pi(\mathbf{x})$~\cite{krauth09}.

The Metropolis algorithm consists of two steps~\cite{lester94}: (i)
the generation of a proposed configuration 
(or move) and (ii) the acceptance or rejection of
the proposed configuration (or move).
The combination of (i) and (ii) leads to a new configuration.
Starting from the configuration $\mathbf{x}$,
we propose a new configuration $\mathbf{x}'$ according to a proposal 
distribution $\cal{G}(\mathbf{x}\to \mathbf{x}')$ and accept 
(the new configuration would be $\mathbf{x}'$) or reject the new configuration 
(the new configuration would be $\mathbf{x}$) according to the 
acceptance distribution 
${\cal{A}}(\mathbf{x}\to \mathbf{x}')$.
This implies that $P(\mathbf{x} \rightarrow \mathbf{x}')$
is given by $\cal{G}(\mathbf{x}\to \mathbf{x}') {\cal{A}}(\mathbf{x}\to \mathbf{x}')$.
The Metropolis algorithm chooses ${\cal{A}}(\mathbf{x}\to\mathbf{x}')$ such
that~\cite{lester94}
\begin{equation}
  {\cal{A}}(\mathbf{x}\to \mathbf{x}')=
\min\left(1,\frac{\pi(\mathbf{x}')\cal{G}(\mathbf{x}'\to \mathbf{x})}{\pi(\mathbf{x})\cal{G}(\mathbf{x}\to \mathbf{x}')}\right).
  \label{chpimc_accept}
\end{equation}
We verify that the detailed balance condition 
[Eq.~(\ref{chpimc_detailedbalance})] is satisfied in the following.
If $\pi(\mathbf{x}')\cal{G}(\mathbf{x}'\to \mathbf{x})$ is smaller 
than $\pi(\mathbf{x})\cal{G}(\mathbf{x}\to \mathbf{x}')$, we obtain
from Eq.~(\ref{chpimc_accept}) that
\begin{equation}
  P(\mathbf{x}\to\mathbf{x}')
  =\frac{\pi(\mathbf{x}')\cal{G}(\mathbf{x}'\to\mathbf{x})}{\pi(\mathbf{x})}
  \label{chpimc_detailedb1}
\end{equation}
and
\begin{equation}
  P(\mathbf{x}'\to\mathbf{x})
  =\cal{G}(\mathbf{x}'\to\mathbf{x}).
  \label{chpimc_detailedb2}
\end{equation}
Plugging the right-hand sides of
Eqs.~(\ref{chpimc_detailedb1}) and (\ref{chpimc_detailedb2}) into
Eq.~(\ref{chpimc_detailedbalance}), we confirm that
Eq.~(\ref{chpimc_detailedbalance}) holds.
If $\pi(\mathbf{x}')\cal{G}(\mathbf{x}'\to \mathbf{x})$ is larger or equal to 
$\pi(\mathbf{x})\cal{G}(\mathbf{x}\to \mathbf{x}')$, it
can be checked similarly
that
Eq.~(\ref{chpimc_detailedbalance}) holds.
Thus, we have shown that detailed balance is fulfilled.

A key task is to design proposal distributions  ${\cal{G}}(\mathbf{x} \to \mathbf{x}')$
that ensure the complete and, ideally, efficient exploration
of the entire configuration space.
In most cases, efficient simulation schemes are achieved if
more than one proposal distribution (and hence type of move) is
utilized.
As discussed more in the next section, the proposal distribution
might be designed based on the knowledge of the non-interacting
system (see, e.g., Sec.~\ref{wiggle-ch2}) or based on knowledge of
certain limiting behaviors of the interacting system
(see, e.g., Sec.~\ref{sec_pdm}).

In practice, the acceptance ratio $A$, 
i.e., 
one minus
the fraction of rejected moves, should be monitored
(note, $A$ is a number and not an 
$\mathbf{x}$- and $\mathbf{x}'$-dependent function).
The acceptance ratio $A$ for
Metropolis sampling is different from the acceptance ratio encountered 
in the rejection sampling.
In the rejection sampling, a rejected configuration does not lead to a new
configuration. In the Metropolis sampling, in contrast, a rejected
configuration does lead to a new configuration.
When a configuration is rejected, the old
configuration becomes the new configuration.
For most of the updates (i.e., the generation of proposed new configurations), the acceptance ratio should not be too large and not be too small. 
A high acceptance ratio typically implies that the deviation between the old and new configurations is, on average, small.
This 
means that the configuration space is explored comparatively slowly.
A low acceptance ratio, in contrast, means that the Markov chain 
contains many identical configurations;
again, typically 
this means that the configuration space is explored comparatively
slowly.
Both cases can result in large correlations of the sample and should be avoided.
As a rule of thumb, the acceptance ratio should lie roughly between 30\% and
50\%~\cite{berg2004markov}.

\subsection{Moves}
\label{sec_moves}
The previous section outlined the basics of the
Metropolis algorithm. 
This section discusses the PIGS moves that are used
to update the configurations.
For all moves, the proposed new configuration
$\mathbf{x}'$ is chosen based on the proposal 
distribution ${\cal{G}}(\mathbf{x} \to \mathbf{x}')$ and 
accepted/rejected based on the acceptance distribution 
${\cal{A}}(\mathbf{x} \to \mathbf{x}')$.
Once ${\cal{G}}(\mathbf{x} \to \mathbf{x}')$ is specified,
${\cal{A}}(\mathbf{x} \to \mathbf{x}')$ follows from Eq.~(\ref{chpimc_accept}).
This section discusses three different moves.
The ``naive move'' and the ``wiggle move''
are ``all purpose'' moves, which have
proven to be useful for nearly all systems.
In some cases, the use of these two types of moves alone
does not lead to an efficient (or even correct) exploration of the
configuration space.
For systems with two-body zero-range interactions, e.g.,
the ``pair distance move'' is needed.
In general, the simulator decides on the frequency with which
the individual moves are used. The optimal ratio can be found empirically
from the performance of the simulation itself or
through the implementation of some sort of learning algorithm.

The list of moves discussed below does not include a ``permutation move'',
i.e., the stochastic sampling 
of the permutations is not discussed. The
reason for this is twofold.
If the system contains identical bosons, the ground
state wave function is typically identical to that of
Boltzmann particles, eliminating the need for an explicit 
symmetrization.
If the system contains identical fermions,
we employ the on-the-fly anti-symmetrization scheme
discussed in Sec.~\ref{sec_algorithm_permutation},
which is particularly useful if two-body zero-range interactions
are employed.

The moves employed in the PIGS algorithm have much in common with the 
moves employed in the finite-temperature path
integral Monte Carlo algorithm.
As already discussed, one difference is that the PIGS algorithm contains
the trial function $\psi_T$ while the finite-temperature path
integral Monte Carlo algorithm does not.
Quite generally, whether a move depends on the trial function
$\psi_T$ or not depends on whether or not
the beginning time slice $\mathbf{R}_0$
or the ending time slice $\mathbf{R}_{2n}$
are being updated.
In the implementations of the moves discussed below,
the wiggle move 
does not depend on $\psi_T$,
and
the naive move and the pair move may depend
on $\psi_T$ (it depends on whether or not the
randomly selected time slice to be updated is the $0$-th
or $2n$-th time slice).
It should be noted, though, that the moves can, in principle, be implemented 
in a variety of ways,
i.e., slightly different proposal distributions
${\cal{G}}(\mathbf{x} \to \mathbf{x}')$
might be used in different implementations and 
be referred to by the same move name.

\subsubsection{Naive move}
\label{naive-ch2}
The simplest move (the naive move) consists of shifting the position
vector $\mathbf{r}_{\rm{old}}$ of
a single bead by $\delta\mathbf{r}$, where $\delta\mathbf{r}$ is drawn 
uniformly from the interval $[-\Delta\mathbf{r},\Delta\mathbf{r}]$.
The basic idea behind this move is that the propagator is a smooth
function of $\mathbf{x}$ and that a small change in $\mathbf{x}$ does not 
introduce a
huge change in the probability distribution $\pi(\mathbf{x})$.
The size $2 \Delta\mathbf{r}$ of the interval 
(if we are simulating a three-dimensional system,
then the interval corresponds to a cube)
can be adjusted such that the
acceptance ratio of the proposed new position vector is around 50\%.
The proposal distribution $\cal{G}(\mathbf{x}\to\mathbf{x}')$ is 
equal to a constant if
the new bead lies in the 
interval 
$[\mathbf{r}_{\rm{old}}-\Delta\mathbf{r},\mathbf{r}_{\rm{old}}+\Delta\mathbf{r}]$;
otherwise, ${\cal{G}}(\mathbf{x}\to\mathbf{x}')$ is equal to $0$.
Following Eq.~(\ref{chpimc_accept}), the move is accepted according to
\begin{equation}
  {\cal{A}}(\mathbf{x}\to\mathbf{x}')=
\min\left(1,\frac{\pi(\mathbf{x}')}{\pi(\mathbf{x})}\right).
  \label{eq_calanaivemove}
\end{equation}
Importantly, one cannot choose an ``unbalanced'' interval like 
$[-\epsilon\Delta\mathbf{r},\Delta\mathbf{r}]$, where $\epsilon<1$,
since the detailed balance condition, Eq.~(\ref{chpimc_detailedbalance}), is
not satisfied in this case. 
The reason is that it is possible to go 
from $\mathbf{r}$ to $\mathbf{r}+\Delta\mathbf{r}$ in one move but 
that it
is impossible to go from $\mathbf{r}+\Delta\mathbf{r}$ to $\mathbf{r}$ in
one move.

\begin{figure}
\centering
\includegraphics[angle=0,width=0.4\textwidth]{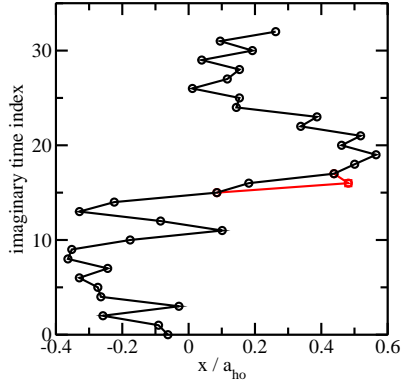}
\caption{
  Illustration of the naive move for a single particle in a one-dimensional
  harmonic trap for $n=32$ beads.
  The black circles depict the old bead positions.
  The red square shows the proposed bead position for the 16-th time slice
  index.
  It can be seen that only two links (namely the link involving the 15-th and
  16-th beads and that involving the 16-th and 17-th beads) are changed.
 }\label{chpimc_fignaive}
\end{figure}

The algorithm for the naive move can be summarized as follows.
(i) Let the current configuration 
be $\mathbf{x}=\{\mathbf{R}_{0},\cdots,\mathbf{R}_{2n}\}$. 
Randomly select a particle index $k$ and a time slice index $j$, where
$j$ can take any value from $0$ to $2n$.
Set $\mathbf{r}_{\rm{old}}=\mathbf{r}_{k,j}$ 
and calculate the old probability distribution 
$\pi_{\rm{old}}=\pi(\mathbf{R}_{0},\cdots,\mathbf{R}_{2n})$.
(ii) Generate a new position
$\mathbf{r}_{\rm{new}}=\mathbf{r}_{\rm{old}}+\delta{\mathbf{r}}$,
where $\delta{\mathbf{r}}$ is
drawn uniformly from the interval $[-\Delta\mathbf{r},\Delta\mathbf{r}]$.
Define 
$\mathbf{R}_j^{\rm{new}}=\{\mathbf{r}_{1,j},\cdots,\mathbf{r}_{k-1,j},\mathbf{r}_{\rm{new}},\mathbf{r}_{k+1,j},\cdots,\mathbf{r}_{N,j}\}$ 
and calculate the new probability distribution
$\pi_{\rm{new}}=\pi(\mathbf{R}_{0},\cdots\mathbf{R}_{j-1},\mathbf{R}_j^{\rm{new}},\mathbf{R}_{j+1},\cdots,\mathbf{R}_{2n})$.
(iii) Calculate the ratio $\pi_{\rm{new}}/\pi_{\rm{old}}$. If this ratio
is larger
than a random number drawn uniformly from 0 to 1, accept the move and set
$\mathbf{r}_{k,j}=\mathbf{r}_{\rm{new}}$; otherwise,
reject the move and set 
$\mathbf{r}_{k,j}=\mathbf{r}_{\rm{old}}$
(i.e., do not change $\mathbf{r}_{k,j}$).
Figure~\ref{chpimc_fignaive} illustrates the naive move for a single particle
in a one-dimensional harmonic trap.
It can be seen that the proposed move involves only one bead.

Although the naive move attemps to change only one bead at a time, whether the
proposed move gets accepted or rejected depends,
in principle, on all the beads, i.e., the
coordinates of all particles at all time slices, since the
acceptance/rejection depends on the ratio $\pi_{\rm{new}}/\pi_{\rm{old}}$.
However, because the probability distribution $\pi(\mathbf{x})$ 
is a product over propagators 
and the trial function $\psi_T$ evaluated at the two ends
[Eq.~(\ref{chpimc_pix})],
the only terms that contribute to the ratio $\pi_{\rm{new}}/\pi_{\rm{old}}$
are the propagators $G(\mathbf{R}_{j-1},\mathbf{R}_{j};\Delta\tau)$ and
$G(\mathbf{R}_{j},\mathbf{R}_{j+1};\Delta\tau)$ and,
if $j$ equals $0$ or $2n$, the trial function
 $\psi_T(\mathbf{R}_0)$ or
$\psi_T(\mathbf{R}_{2n})$.
If one uses the second-order Trotter formula 
[Eq.~(\ref{chpimc_trotterfinal})],
which treats the potential and
kinetic energy terms separately, the terms that contribute to
$G(\mathbf{R}_{j-1},\mathbf{R}_j;\Delta\tau)$ and $G(\mathbf{R}_{j},\mathbf{R}_{j+1};\Delta\tau)$ are
the potential energy term $\exp[-\Delta\tau V(\mathbf{R}_j)]$ 
and the kinetic energy 
terms $G_0(\mathbf{r}_{k,j-1},\mathbf{r}_{k,j};\Delta\tau)$ 
and $G_0(\mathbf{r}_{k,j},\mathbf{r}_{k,j+1};\Delta\tau)$.

The caveat of the naive move is that the correlation length is typically
large. In the best case scenario (i.e., in the case where all beads of all particles are
considered exactly once and all proposed moves are accepted), 
$(2n+1) \times N$
moves are needed to generate a configuration in
which every bead differs from the starting configuration.
Thus, we calculate observables for every 
$(\alpha\times (2n+1)\times N)$-th 
configuration, where $\alpha$ is a constant greater than 1 that is 
adjusted to ensure that the observables are calculated from 
configurations with small correlations.
In practice, we find that $\alpha$ lies between 2 and 20
for the applications considered in this review.

\subsubsection{Wiggle move}
\label{wiggle-ch2}
The wiggle move uses the non-interacting propagator to design the
proposal distribution
${\cal{G}}(\mathbf{x} \to \mathbf{x}')$.
Since the non-interacting propagator is a product of simple Gaussians
for particles in free space [Eq.~(\ref{chpimc_rho0})] and 
for particles confined in a harmonic
trap~\cite{krauth2006statistical,zr15},
one can generate configurations efficiently with 100~\% acceptance ratio using 
the Box-Muller transformation~\cite{box1958} or with finite
acceptance ratio 
using the Marsaglia 
polar method
(the acceptance ratio is around 80~\%)~\cite{marsaglia64}
or the Ziggurat algorithm (the acceptance ratio is around 
98~\%)~\cite{JSSv005i08,zigguratflaw06}.
If the difference between the propagator of the system to be simulated and
the non-interacting propagator is small, 
the acceptance ratio for a move generated based on the
propagator of the non-interacting system
is high. Despite the large acceptance ratio, the
correlation between consecutive configurations is, in general, small.
In the non-interacting limit, the acceptance ratio is exactly 1.

Depending on the number of beads changed simultaneously, the wiggle move is  
a single-bead move or a multi-bead move.
The single-bead version of the wiggle move
randomly selects a particle index $k$ and a time slice index $j$
($j>0$ and $j<2n$).
Since the beads to be moved exclude the time slices $0$ and $2n$,
the wiggle move does not involve the trial function $\psi_T$.
We denote the
new proposed position
vector by 
$\mathbf{r}_{k,j}^{\rm{new}}$
(how to choose $\mathbf{r}_{k,j}^{\rm{new}}$ is discussed below)
and
define $\mathbf{R}_j^{\rm{new}}=
(\mathbf{r}_{1,j},\cdots,\mathbf{r}_{k-1,j},\mathbf{r}_{k,j}^{\rm{new}},\mathbf{r}_{k+1,j},\cdots,\mathbf{r}_{N,j})$.
The old and proposed configurations read
\begin{equation}
\mathbf{x}=\{\mathbf{R}_{0},\cdots,\mathbf{R}_{2n}\}
  \label{<++>}
\end{equation}
 and 
 \begin{equation}
 \mathbf{x}'=\{\mathbf{R}_{0},\cdots\mathbf{R}_{j-1},\mathbf{R}_j^{\rm{new}},\mathbf{R}_{j+1},\cdots,\mathbf{R}_{2n}\},
   \label{<++>}
 \end{equation}
respectively.
We choose ${\cal{G}}(\mathbf{x} \to \mathbf{x}')$ 
according to the propagator $G_0$ [Eq.~(\ref{chpimc_rho0})]
of the non-interacting system without confinement
(the proposal distribution based on the propagator of the
non-interacting harmonic oscillator Hamiltonian
can be treated similarly),
\begin{equation}
  {\cal{G}}(\mathbf{x} \to \mathbf{x}') = \exp\left(-
  \frac{(\mathbf{r}_{k,j}^{\rm{new}}-\mathbf{r}_{k,j-1})^2+
(\mathbf{r}_{k,j}^{\rm{new}}-\mathbf{r}_{k,j+1})^2}
  {4\lambda_m\Delta\tau}\right)
  \label{}
\end{equation}
or, rearranging the exponent,
\begin{equation}
  {\cal{G}}(\mathbf{x} \to \mathbf{x}') = \exp\left(-
  \frac{[\mathbf{r}_{k,j}^{\rm{new}}-(\mathbf{r}_{k,j-1}+\mathbf{r}_{k,j+1})/2]^2}
  {2\lambda_m\Delta\tau}\right).
  \label{chpimc_wigglemovegaussian}
\end{equation}
The right hand side of Eq.~(\ref{chpimc_wigglemovegaussian}) is equal to a
Gaussian whose mean value is
given by the midpoint of the $(j-1)$-th
and the $(j+1)$-th bead of the $k$-th particle and whose variance is 
$\lambda_m\Delta\tau$.
Thus, $\mathbf{r}_{k,j}^{\rm{new}}$ can be generated using the Box-Muller
transformation, the Marsaglia polar method, or the Ziggurat algorithm
discussed above.
If we use the second-order Trotter formula,
the acceptance probability ${\cal{A}}(\mathbf{x} \to \mathbf{x}')$
[Eq.~(\ref{chpimc_accept})]
takes a fairly simple form since a large number of terms
(those not involving time slice $j$
and those not involving particle $k$)
in the ratio $\pi(\mathbf{x'})/\pi(\mathbf{x})$ can be cancelled.
If the pair product approximation is used, 
the evaluation of ${\cal{A}}(\mathbf{x} \to \mathbf{x}')$
is more involved since fewer
terms in the ratio $\pi(\mathbf{x'})/\pi(\mathbf{x})$ can be cancelled
due to the fact that the kinetic energy and the potential energy terms are ``linked'' in
the pair product approximation.

The single-bead
version of the wiggle move can be generalized to multiple consecutive beads.
Since the multi-bead move leads to a deformation of a segment of the path,
the move is called ``wiggle move''.
In what follows, our discussion is guided by
Ref.~\cite{boninsegni05}.
Instead of a single bead of the path
we propose to change a path segment consisting of multiple beads 
according to a proposal distribution
${\cal{G}}(\mathbf{x} \to \mathbf{x}')$
that generalizes the expression given in
Eq.~(\ref{chpimc_wigglemovegaussian}).
We denote the time slice indices of the two ends that are unchanged 
by $j$ and $j+s$, where $s$ is an integer power of 2;
the condition for $s$ allows one, as will become clear below, to organize the move into
``levels''.
The corresponding 
position vectors are $\mathbf{r}_{k,j}$ and $\mathbf{r}_{k,j+s}$,
where $s>0$.
Note that the wiggle move does not explicitly involve the trial function $\psi_T$
since the path segment to be changed has to be continuous. This means that the zeroth bead can be
at the beginning of the segment but nowhere else and that the $2n$-th
bead can be at the end of the segment but nowhere else.
In the finite-temperature path integral Monte Carlo approach, 
in contrast,
any path segment can be considered, provided the path is closed.

We now outline the multi-bead move, both without  and with ``staging''.
The algorithm without staging is less efficient but can be employed in
connection with the Trotter formula and the pair product approximation.
The staging version can only be used in connection with the Trotter formula.
Both multi-bead move versions generate a proposed new path segment
$\{\mathbf{r}_{k,j+1}^{\rm{new}},\cdots,\mathbf{r}_{k,j+s-1}^{\rm{new}}\}$
that is completely independent of the old path segment
$\{\mathbf{r}_{k,j+1},\cdots,\mathbf{r}_{k,j+s-1}\}$.
Here, $j+s$ has to be smaller than or equal to $2n$.

To motivate the strategy of the multi-bead wiggle move, we write
the pieces of the non-interacting propagator $G_0$ that
depend on the particle index $k$ and the
time slice indices $j$ to $j+s$ out explicitly,
\begin{eqnarray}
  \fl
  G_0
\sim
&\underbrace{\exp\left[-\frac{(\mathbf{r}_{k,j}-\mathbf{r}_{k,j+s})^2}{4s\lambda_m\Delta\tau}
\right]}_{\rm{constant}}\times\nonumber\\
&\underbrace{\exp\left[-\frac{(\mathbf{r}_{k,j+s/2}-\bar{\mathbf{r}}_{k,j,j+s})^2}{s\lambda_m\Delta\tau}\right]}_{\rm{zeroth level}}\times\nonumber\\
&\underbrace{\exp\left[-\frac{(\mathbf{r}_{k,j+s/4}-\bar{\mathbf{r}}_{k,j,j+s/2})^2}{s\lambda_m\Delta\tau/2}\right]\exp\left[-\frac{(\mathbf{r}_{k,j+3s/4}-\bar{\mathbf{r}}_{k,j+s/2,j+s})^2}{s\lambda_m\Delta\tau/2}\right]}_{\rm{first level}}\times\nonumber\\
&\dots,
  \label{chpimc_multilevel}
\end{eqnarray}
where
$\bar{\mathbf{r}}_{k,\alpha,\beta}=(\mathbf{r}_{k,\alpha}+\mathbf{r}_{k,\beta})/2$.
If $s$ is equal to $2^l$, Eq.~(\ref{chpimc_multilevel}) contains $l$ levels
(the zeroth level is counted as one level but the constant term is not).
Equation~(\ref{chpimc_multilevel}) suggests that the sampling can be done
level by level.
For example, for a path segment consisting of three time slices ($s=4$
and $j=0$), the beginning bead is
$\mathbf{r}_{k,0}$ and the ending bead is $\mathbf{r}_{k,4}$.
Thus, there exist two levels in total.
First, the new midpoint bead $\mathbf{r}_{k,2}^{\rm{new}}$
is proposed according to the $0$-th level (partial) proposal distribution
${\cal{G}}_{0-{\rm{th}}}(\mathbf{x} \to \mathbf{x}')$, i.e.,
$\mathbf{r}_{k,2}^{\rm{new}}$ 
is generated 
by sampling a three-dimensional Gaussian distribution with variance
$s \lambda_m \Delta \tau/2$.
Second, the new midpoint beads $\mathbf{r}_{k,1}^{\rm{new}}$ 
and $\mathbf{r}_{k,3}^{\rm{new}}$
are proposed according to the
$1$-st level (partial) proposal distribution
${\cal{G}}_{1-{\rm{st}}}(\mathbf{x} \to \mathbf{x}')$, i.e.,
$\mathbf{r}_{k,1}^{\rm{new}}$  and $\mathbf{r}_{k,3}^{\rm{new}}$ 
are generated 
by sampling three-dimensional Gaussian distributions with variance
$s \lambda_m \Delta \tau/4$.

In general, the $u$-th level (partial) proposal distribution
${\cal{G}}_{u-{\rm{th}}}(\mathbf{x} \to \mathbf{x}')$
reads
\begin{eqnarray}
  \fl
  {\cal{G}}_{u-{\rm{th}}}(\mathbf{x}\to\mathbf{x}') = \nonumber \\
\fl  \prod_{v=1}^{2^u}
  \exp\left(-
  \frac{[\mathbf{r}_{k,j+s(2v-1)/2^{u+1}}^{\rm{new}}-
(\mathbf{r}_{k,j+s (v-1)/2^{u}}+\mathbf{r}_{k,j+s v/2^{u}})/2]^2}
  {\lambda_m s \Delta\tau /2^{u}} \right),
  \label{chpimc_wigglemultibead}
\end{eqnarray}
which implies that $\mathbf{r}_{k,j+s(2v-1)/2^{u+1}}^{\rm{new}}$
can be generated by sampling a three-dimensional Gaussian 
with variance
$\lambda_m s \Delta\tau /2^{u+1}$.
Since the $u$-th level
proposal distribution
${\cal{G}}_{u-{\rm{th}}}(\mathbf{x}\to\mathbf{x}')$ 
depends only on 
the position vectors of the $(u-1)$-th level,
the new path segment 
$\{\mathbf{r}_{k,j+1}^{\rm{new}},\cdots,\mathbf{r}_{k,j+s-1}^{\rm{new}} \}$
can, indeed, be generated level by level.
The product of ${\cal{G}}_{u-{\rm{th}}}(\mathbf{x}\to\mathbf{x}')$
over all levels (i.e., $\prod_{u=0}^{l-1} {\cal{G}}_{u-{\rm{th}}}$) yields 
the ``full'' proposal distribution
${\cal{G}}(\mathbf{x} \to \mathbf{x}')$.
Denoting the time slices that involve the newly
proposed beads by $\mathbf{R}_v^{\rm{new}}$,
where $v$ ranges from $j+1$ to $j+s-1$, and 
using the second-order Trotter formula,
the acceptance distribution ${\cal{A}}(\mathbf{x}\to\mathbf{x}')$ 
becomes
\begin{equation}
  {\cal{A}}(\mathbf{x}\to\mathbf{x}')=
  \min\left(1,\prod_{v=j+1}^{j+s-1}
\frac{\exp[-\Delta\tau V(\mathbf{R}_{v}^{\rm{new}})]}
{\exp[-\Delta\tau V(\mathbf{R}_{v})]}\right).
  \label{chpimc_wiggletestsimple}
\end{equation}

The staging algorithm allows one to reject the multi-bead wiggle move
in advance, i.e., before the entire new path segment has been
generated, if ``bad bead positions'' are drawn~\cite{boninsegni05}.
The ``in advance rejection'' is checked for at each level $u$. Let us assume
that we are considering level $u$ with 
the new midpoint beads
$\mathbf{r}_{k,j+s(2v-1)/2^{u+1}}^{\rm{new}}$,
where $v$ ranges from 1 to $2^u$.
Using the second-order Trotter formula,
the move is accepted or rejected based on 
\begin{equation}
  {\cal{A}}_{u-{\rm{th}}}(\mathbf{x} \rightarrow \mathbf{x}')=
  \min\left(1,\prod_{v=1}^{2^u}\frac{\exp[-\Delta\tau V(\mathbf{R}_{j+s(2v-1)/2^{u+1}}^{\rm{new}})]}{\exp[-\Delta\tau V(\mathbf{R}_{j+s(2v-1)/2^{u+1}})]}\right).
  \label{chpimc_wiggletest}
\end{equation}
If ${\cal{A}}_{u-{\rm{th}}}$ is smaller than a random number drawn uniformly
from the interval [0,1], the move is rejected at the $u$-th level and 
the new configuration is set equal to the
old configuration; otherwise, the move is accepted.
If the move is accepted at the $u$-th level, we go to the $(u+1)$-th
level and repeat the procedure.
If the final level is reached and the new proposed beads are accepted, 
then the entire path segment consisting of 
the proposed new beads 
$\mathbf{r}_{k,j+1}^{\rm{new}},\cdots,\mathbf{r}_{k,j+s-1}^{\rm{new}}$
is accepted and a configuration with a new path segment has been generated.

The outcome of the ``multi-bead sampling + staging'' algorithm is equivalent to 
that of the 
multi-bead algorithm without staging,
which proposes all the beads of the path segment considered first and then
accepts or rejects at the very end.
The 
staged (or in-advance) rejection
speeds up the algorithm.
Importantly, the staging algorithm only works if the Trotter
formula is used.
If the pair product approximation is used, the rejection needs to be done at the very end because
the propagator for consecutive time slices cannot be reorganized into
different levels.

\begin{figure}
\centering
\includegraphics[angle=0,width=0.4\textwidth]{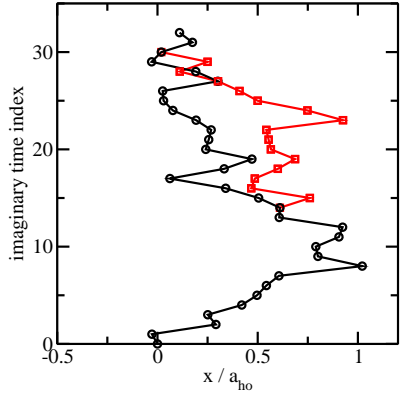}
\caption{
  Illustration of the wiggle move for a single particle in a one-dimensional
  harmonic trap.
  The black circles and red squares depict the old and proposed new configurations,
  respectively.
  It is assumed that the construction of the new path segment was
  continued after the construction of the first midpoint bead, the next
  two midpoint beads, and so on.
 }\label{chpimc_figwiggle}
\end{figure}

Figure~\ref{chpimc_figwiggle} 
illustrates the wiggle move for a single particle
in a one-dimensional harmonic trap ($j=14$ and $s=16$).
The proposed new path is constructed as follows:
A new midpoint bead
with index 22 is proposed and tested according to Eq.~(\ref{chpimc_wiggletest}):
If rejected (i.e., if the random number generated is smaller than 
${\mathcal{A}}_{0-{\rm{th}}}$), 
the move is aborted in advance and the new configuration is set to the old
configuration;
if not rejected, the construction of the new path segment is continued (this
is what is assumed in making Fig.~\ref{chpimc_figwiggle}).
In the latter case, two new midpoint beads with index 18 and 26
are proposed and tested simultaneously according to
Eq.~(\ref{chpimc_wiggletest}).
If rejected, the move is aborted in advance and the new configuration is set
to the old
configuration;
if not rejected, four new midpoint beads with index 16, 20, 24, and 28
are proposed and tested simultaneously according to
Eq.~(\ref{chpimc_wiggletest}).
If rejected, the move is aborted in advance and the new configuration is set
to the old
configuration;
if not rejected, eight new midpoint beads with index 15, 17, 19, 21, 23, 25, and 29
are tested simultaneously according to
Eq.~(\ref{chpimc_wiggletest}).
If rejected, the move is aborted with the new configuration being the old
configuration;
if not rejected, the move is accepted in its entirety and 
the path segment that involves the beads with index 
15 to 29 is changed to the new position vectors.

\subsubsection{Pair distance move}
\label{sec_pdm}
The pair distance move is employed in systems with two-body zero-range interactions.
As discussed earlier, two-body zero-range interactions can be treated
using the pair product approximation but not, to the best of our
knowledge, using the Trotter formula based decomposition of the
propagator.
The use of the pair distance move is 
especially important if the two-body $s$-wave scattering length diverges.
The key motivation is that two particles can, if zero-range interactions are present, 
be close to each other or even on top of each other.
Traditional moves such as the wiggle move and the naive
move, however, do not generate configurations in which particles sit on top of each other.
The reason is that the scaled pair distribution function $4\pi r^2P_{12}(r)$
for non-interacting particles or for uniformly distributed
particles is zero at $r=0$, implying that 
configurations with vanishing pair distance are not generated by the
traditional moves.
The pair distance move involves two particles with the
same time slice index $j$, where $j=0,\cdots,2n$.
The proposed move keeps
the center of mass of the selected pair unchanged but modifies its relative distance
vector.

The pair distance move 
can be implemented as follows.
(i) Randomly choose the indices $k$, $l$ and $j$
of the single-particle beads $\mathbf{r}_{k,j}$ and
$\mathbf{r}_{l,j}$ involved in the move
and set 
$\mathbf{r}_{\rm{old}}=\mathbf{r}_{k,j}-\mathbf{r}_{l,j}$
and
$r_{\rm{old}}=|\mathbf{r}_{\rm{old}}|$.
Store the old center-of-mass vector $\mathbf{b}$ of the selected pair,
$\mathbf{b}=(\mathbf{r}_{k,j}+\mathbf{r}_{l,j})/2$.
(ii) Generate a new relative distance vector 
$\mathbf{r}_{\rm{new}}=\mathbf{r}_{\rm{old}}+ \delta r \hat{\mathbf{r}}_{\rm{old}}$,
where $\delta r$ is obtained by choosing a value
uniformly from the pre-set interval
$[ -\Delta r, \Delta r]$.
This prescription
implies that $\delta r$ can be negative and that
$\mathbf{r}_{\rm{new}}$, in turn, lies along the  directions
$\mathbf{r}_{\rm{old}}$ or $-\mathbf{r}_{\rm{old}}$.
(iii) Calculate the
ratio $w$, $w=
(r_{\rm{old}}+\delta r)^2 \pi_{\rm{new}}/
[(r_{\rm{old}})^2
\pi_{\rm{old}}
]$.
If this ratio is larger than a random number drawn uniformly from 0 to 1, accept the move
and set
$\mathbf{r}_{k,j}=\mathbf{b}+\mathbf{r}_{\rm{new}}/2$
and
$\mathbf{r}_{l,j}=\mathbf{b}-\mathbf{r}_{\rm{new}}/2$;
otherwise, reject the move (in this case,
$\mathbf{r}_{k,j}$ and $\mathbf{r}_{l,j}$
remain unchanged).
The value of $\Delta r$ is adjusted such that
approximately 50~\% of the proposed moves are accepted.

The acceptance/rejection step
involves the quantity $w$. 
If $j$ is not equal to $0$ or $2n$,
$w$ reduces to
\begin{eqnarray}
w=
\frac{
(r_{\rm{old}}+\delta r)^2
G(\mathbf{R}_{j-1},\mathbf{R}_j^{\rm{new}};\Delta \tau)
G(\mathbf{R}_{j}^{\rm{new}},\mathbf{R}_{j+1};\Delta \tau)
}
{
(r_{\rm{old}})^2
G(\mathbf{R}_{j-1},\mathbf{R}_j^{\rm{old}};\Delta \tau)
G(\mathbf{R}_{j}^{\rm{old}},\mathbf{R}_{j+1};\Delta \tau)
}.
\label{eq_pair_weight1}
\end{eqnarray}
If $j$ is equal to $0$, one finds
\begin{eqnarray}
w=
\frac{
(r_{\rm{old}}+\delta r)^2
\psi_T(\mathbf{R}_0^{\rm{new}})
G(\mathbf{R}_{0}^{\rm{new}},\mathbf{R}_{1};\Delta \tau)
}
{
(r_{\rm{old}})^2
\psi_T(\mathbf{R}_0^{\rm{old}})
G(\mathbf{R}_{0}^{\rm{old}},\mathbf{R}_{1};\Delta \tau)
}.
\label{eq_pair_weight2}
\end{eqnarray}
The expression for $j=2n$ is similar to that for $j=0$.
In Eqs.~(\ref{eq_pair_weight1})-(\ref{eq_pair_weight2}),
$\mathbf{R}_j^{\rm{new}}$ is defined
as
\begin{eqnarray}
  \fl
\mathbf{R}_j^{\rm{new}}= \nonumber \\
\fl
\{ \mathbf{r}_1,\cdots, \mathbf{r}_{k-1,j},\mathbf{b}+\mathbf{r}_{\rm{new}}/2,
\mathbf{r}_{k+1,j},\cdots,
\mathbf{r}_{l-1,j},\mathbf{b}-\mathbf{r}_{\rm{new}}/2,\mathbf{r}_{l+1,j},\cdots,
\mathbf{r}_{N,j}
\}.
\end{eqnarray}
As already alluded to, the
propagator $G$ entering in the expressions for $w$ 
is expressed
using the pair product approximation,
i.e., each of the $G$'s is replaced by the expression given in
Eq.~(\ref{chpimc_rhotot}).
Careful inspection of the resulting expression for $w$
shows that a number of terms in the numerator can be canceled
by corresponding terms in the denominator. However, since the
reduced pair propagator $\bar{G}_{\rm{rel}}$
``connects'' the $k$-th particle with all other
particles   and the $l$-th particle with all other
particles, two products over a particle dummy index survive 
for each of the $G$'s, making the pair distance move 
computationally more expensive than the naive move
implemented using the Trotter decomposition
(in fact, the argument just given 
explains why the pair product approximation is,
generally speaking, computationally more demanding than
Trotter formula based schemes).
The ratio $(r_{\rm{old}}+\delta r)^2/(r_{\rm{old}})^2$, which
is included in the acceptance step, ensures that the 
small interparticle distance behavior is described properly.  
The reader is referred to Ref.~\cite{zr15} for more details.

\subsection{Expectation values}
\label{sec_algorithm_expectation}

The previous section outlined how to generate new configurations.
Assuming that no symmetrization or anti-symmetrization is needed
(see Sec.~\ref{sec_algorithm_permutation} for details)
and that a suitable trial function $\psi_T$ is known, the missing
piece for completing the PIGS algorithm is the determination of the weight function
$w(\mathbf{x})$ [see Eq.~(\ref{chpimc_observablerewrite3})].
This section discusses the derivation of the form of the weight function $w(\mathbf{x})$ for selected
observables; the steps outlined below can be generalized to other observables.
Explicit expressions for $w(\mathbf{x})$ 
can be derived for many observables using either 
quantum estimator relations such as Eq.~(\ref{eq_energy3})
or thermodynamic type relations
such as Eq.~(\ref{eq_energy4}).
The determination of the superfluid or condensate fractions
is more involved and not considered in this review.

\subsubsection{Example: Energy estimator}
Using the thermodynamic type relation, Eq.~(\ref{eq_energy4}), 
and plugging in one of the approximate expressions for the propagator,
an explicit expression for the
weight function $w(\mathbf{x})$ can be derived.
As an example, we consider the thermodynamic energy estimator for the second-order
Trotter formula
for particles without permutations.
Using Eq.~(\ref{chpimc_rhosplit}) and Eq.~(\ref{chpimc_trotter2})
without the ${\cal{O}}(\Delta \tau^3)$ term
in 
Eq.~(\ref{eq_znorm}),
the 
normalization factor $Z(\tau)$ reads
\begin{eqnarray}
  \fl
  Z(\tau)=
  \int_{\mathbf{R}_0} \cdots \int_{\mathbf{R}_{2n}}
  \psi_T(\mathbf{R}_0)
  \psi_T(\mathbf{R}_{2n})
  G_0(\mathbf{R}_0,\mathbf{R}_1;\Delta\tau) \times \cdots 
  \times G_0(\mathbf{R}_{2n-1},\mathbf{R}_{2n};\Delta\tau)
\times
  \nonumber\\
  \exp\left[-\Delta\tau\left(\frac{1}{2}V(\mathbf{R}_0)+\frac{1}{2}V(\mathbf{R}_{2n})+ \sum_{j=1}^{2n-1}V(\mathbf{R}_j)\right)\right]
  d \mathbf{R}_0\cdots d\mathbf{R}_{2n}.
  \label{chpimc_energyz}
\end{eqnarray}
Using Eq.~(\ref{chpimc_energyz}) in Eq.~(\ref{eq_energy4}),
recalling that $\tau$ is equal to $n \Delta \tau$,
and denoting the energy estimator $E_{\tau}$---calculated 
for a finite number of time slices by 
$\langle E_T \rangle$---,
we obtain
\begin{eqnarray}
  \fl
  \braket{E_T}
  =-\frac{1}{Z(\tau)}\int_{\mathbf{R}_0} \cdots \int_{\mathbf{R}_{2n}}
  \frac{1}{2n}
  \psi_T(\mathbf{R}_0)
  \psi_T(\mathbf{R}_{2n})
\times
  \nonumber \\
  \frac{\partial}{\partial \Delta\tau} \Bigg\{ G_0(\mathbf{R}_0,\mathbf{R}_1;\Delta\tau) \times \cdots \times
  G_0(\mathbf{R}_{2n-1},\mathbf{R}_{2n};\Delta\tau) \times
  \nonumber \\
  \exp\left[-\Delta\tau\left(\frac{1}{2}V(\mathbf{R}_0)+\frac{1}{2}V(\mathbf{R}_{2n})+ \sum_{j=1}^{2n-1}V(\mathbf{R}_j)\right)\right] \Bigg\} \nonumber \\
   d \mathbf{R}_0 \cdots d\mathbf{R}_{2n}.
  \label{chpimc_energyz2}
\end{eqnarray}
The goal is now to rewrite the right-hand side of Eq.~(\ref{chpimc_energyz2})
such that we can read off $w(\mathbf{x})$ by comparing with 
Eq.~(\ref{chpimc_observablerewrite3}).
Combining Eq.~(\ref{chpimc_trotterfinal}) 
[without the ${\cal{O}}(\Delta \tau^3)$ term] and Eq.~(\ref{chpimc_pix}),
we recognize that Eq.~(\ref{chpimc_energyz2}) can be rewritten in terms of
$\pi(\mathbf{x})$,
\begin{equation}
  \braket{E_T}=-\frac{1}{Z(\tau)}\int_{\mathbf{x}}
  \frac{1}{2n}\frac{\partial \pi(\mathbf{x})}{\partial \Delta\tau}
  d\mathbf{x}.
  \label{chpimc_energyz2s}
\end{equation}
The probability distribution $\pi(\mathbf{x})$ depends on $\Delta \tau$
through the $2n$ propagators $G_0$.
Applying the chain rule to evaluate the derivative with respect to $\Delta \tau$, 
we obtain
\begin{eqnarray}
  \fl
  \label{chpimc_energyz3}
  \braket{E_T}=\frac{1}{Z(\tau)} \times \\
\fl
  \int_{\mathbf{x}}\frac{1}{2n}\left[\sum_{j=0}^{2n-1}\left(\frac{3N}{2\Delta\tau}-\frac{(\mathbf{R}_j-\mathbf{R}_{j+1})^2}{4\lambda_m\Delta\tau^2}\right)+\frac{1}{2}V(\mathbf{R}_{0})+\frac{1}{2}V(\mathbf{R}_{2n})+\sum_{j=1}^{2n-1}V(\mathbf{R}_j)\right] \times \nonumber \\
\fl \pi(\mathbf{x})d \mathbf{x}.
\nonumber
\end{eqnarray}
Comparing Eq.~(\ref{chpimc_energyz3}) with Eq.~(\ref{chpimc_observablerewrite3}),
one reads off
\begin{equation}
  \fl
  w(\mathbf{x})=\frac{1}{2n}\left[\sum_{j=0}^{2n-1}\left(\frac{3N}{2\Delta\tau}-\frac{(\mathbf{R}_j-\mathbf{R}_{j+1})^2}{4\lambda_m\Delta\tau^2}\right)+\frac{1}{2}V(\mathbf{R}_{0})+\frac{1}{2}V(\mathbf{R}_{2n})+\sum_{j=1}^{2n-1}V(\mathbf{R}_j)\right].
  \label{chpimc_energyf1}
\end{equation}
The right-hand side of 
Eq.~(\ref{chpimc_energyf1}) can be evaluated straightforwardly, provided the  
configuration $\mathbf{x}$ is known.

Using the quantum estimator relation, Eq.~(\ref{eq_energy3}), 
an alternative energy estimator can be derived.
Here, we derive the quantum energy estimator for particles without permutations using the second-order
Trotter formula as an example.
Our goal is to rewrite Eq.~(\ref{eq_energy3}) such that we can read off the form of $w(\mathbf{x})$
by comparing with Eq.~(\ref{chpimc_observablerewrite3}).
To this end, we derive an auxiliary identity 
[see Eq.~(\ref{eq_helpaux2})] 
that we use below
to rewrite the integrand of the numerator of Eq.~(\ref{eq_energy3}).

Using
the position representation of the Hamiltonian $\hat{H}$~\cite{sakurai2011modern},
\begin{equation}
  \braket{\mathbf{R}|\hat{H}|\mathbf{R}'}=H_\mathbf{R}\delta(\mathbf{R}-\mathbf{R}'),
  \label{chpimc_hmatrix}
\end{equation}
where
\begin{equation}
  H_{\mathbf{R}}
  =-\lambda_m\nabla_{\mathbf{R}}^2+V(\mathbf{R})
  \label{chpimc_hmatrix2},
\end{equation}
one finds
\begin{eqnarray}
  \int_{\mathbf{R}'} \langle \mathbf{R} | \hat{H} | \mathbf{R}' \rangle
  G(\mathbf{R}',\mathbf{R}'';\tau) d \mathbf{R}'
  = \nonumber \\
  \int_{\mathbf{R}'}
  \left\{ [-\lambda_m\nabla_{\mathbf{R}}^2+V(\mathbf{R})]\delta(\mathbf{R}-\mathbf{R}') \right\}
    G(\mathbf{R}',\mathbf{R}'';\tau) d \mathbf{R}'
\end{eqnarray}
or, integrating by parts twice,
\begin{eqnarray}
  \int_{\mathbf{R}'} \langle \mathbf{R} | \hat{H} | \mathbf{R}' \rangle
  G(\mathbf{R}',\mathbf{R}'';\tau) d \mathbf{R}'
  = \nonumber \\
  \int_{\mathbf{R}'}
  \delta(\mathbf{R}-\mathbf{R}') \left\{ [-\lambda_m\nabla_{\mathbf{R}}^2+V(\mathbf{R})]
  G(\mathbf{R}',\mathbf{R}'';\tau) \right\} d \mathbf{R}'.
  \label{eq_helpaux}
\end{eqnarray}
Performing the integration over $\mathbf{R}'$ on the right-hand side of Eq.~(\ref{eq_helpaux}),
we have
\begin{eqnarray}
\fl
  \int_{\mathbf{R}'} \langle \mathbf{R} | \hat{H} | \mathbf{R}' \rangle
  G(\mathbf{R}',\mathbf{R}'';\tau) d \mathbf{R}'
  = 
 [-\lambda_m \nabla_{\mathbf{R}}^2+V(\mathbf{R})]
 G(\mathbf{R},\mathbf{R}'';\tau) .
 \label{eq_helpaux2}
\end{eqnarray}

We denote the quantum estimator $E_{\tau}$ [Eq.~(\ref{eq_energy3})],
evaluated using a finite number of time slices, by $\langle E_H \rangle$.
Inserting the closure relation [Eq.~(\ref{chpimc_Ridentity})] 
$2n-2$ 
times into Eq.~(\ref{eq_energy3})
  and using Eq.~(\ref{eq_helpaux2}) with $\tau$ replaced by $\Delta \tau$,
 we obtain
\begin{eqnarray}
  \fl
  \braket{E_H}
  =\frac{1}{Z(\tau)} \times \nonumber \\
\fl
\int_{\mathbf{R}_0}\cdots\int_{\mathbf{R}_{2n}}
   \psi_T(\mathbf{R}_0) G(\mathbf{R}_0,\mathbf{R}_1;\Delta\tau)G(\mathbf{R}_1,\mathbf{R}_2;\Delta\tau)
\times \cdots \times G(\mathbf{R}_{n-2},\mathbf{R}_{n-1};\Delta\tau)
  \times\nonumber \\
\fl  \left\{[-\lambda_m\nabla_{\mathbf{R}_n}^2+V(\mathbf{R}_n)]G(\mathbf{R}_{n-1},\mathbf{R}_{n};\Delta\tau)\right\}G(\mathbf{R}_{n},\mathbf{R}_{n+1};\Delta\tau)
  \times\nonumber\\
\fl  G(\mathbf{R}_{n+1},\mathbf{R}_{n+2};\Delta\tau)
  \times \cdots \times G(\mathbf{R}_{2n-1},\mathbf{R}_{2n};\Delta\tau)
  \psi_T(\mathbf{R}_{2n})d \mathbf{R}_0 \cdots d\mathbf{R}_{2n}.
  \label{chpimc_energyz4}
\end{eqnarray}
Applying the second-order Trotter formula [Eq.~(\ref{chpimc_trotterfinal}) without the ${\cal{O}}(\Delta \tau^3)$
term]
to Eq.~(\ref{chpimc_energyz4}), we obtain
\begin{eqnarray}
  \fl
  \braket{E_H}=
  \frac{1}{Z(\tau)}
\int_{\mathbf{x}} \bigg[\frac{3N}{2\Delta\tau}-\frac{(\mathbf{R}_n-\mathbf{R}_{n-1})^2}{4\lambda_m\Delta\tau^2}+V(\mathbf{R}_n)
-(\mathbf{R}_{n}-\mathbf{R}_{n-1})\cdot\nabla_{\mathbf{R}_n}V(\mathbf{R}_n)
  \nonumber\\
  +\lambda_m \Delta\tau \nabla_{\mathbf{R}_n}^2V(\mathbf{R}_n)-\lambda_m\Delta\tau^2(\nabla_{\mathbf{R}_n}V(\mathbf{R}_n))^2\bigg]\pi(\mathbf{x}) d\mathbf{x}.
  \label{chpimc_energyz5}
\end{eqnarray}
Because $\hat{H}$ commutes with the propagator, $\hat{H}$ can be
applied to any time slice
(in the derivation above, $\hat{H}$ was applied to the $n$th time slice).
This implies that one can average 
over all time slices to improve the accuracy (i.e., to take more
``measurements'' 
for each configuration). 
Averaging over all possible time slice indices, we obtain
\begin{eqnarray}
  \fl
  \braket{E_H}=
  \frac{1}{Z(\tau)}
  \int_{\mathbf{x}}\frac{1}{2n}\Bigg\{\sum_{j=1}^{2n}
    \bigg[\frac{3N}{2\Delta\tau}-\frac{(\mathbf{R}_j-\mathbf{R}_{j-1})^2}{4\lambda_m\Delta\tau^2}+V(\mathbf{R}_j)
    -(\mathbf{R}_{j}-\mathbf{R}_{j-1})\cdot\nabla_{\mathbf{R}_j}V(\mathbf{R}_j)
  \nonumber\\
  +\lambda_m \Delta\tau \nabla_{\mathbf{R}_j}^2V(\mathbf{R}_j)-\lambda_m\Delta\tau^2(\nabla_{\mathbf{R}_j}V(\mathbf{R}_j))^2 \bigg]  \Bigg\}
  \pi(\mathbf{x}) d\mathbf{x}.
  \label{chpimc_energyz6}
\end{eqnarray}
Comparing Eq.~(\ref{chpimc_energyz6}) with Eq.~(\ref{chpimc_observablerewrite3}),
we obtain
\begin{eqnarray}
  \fl
  w(\mathbf{x})=\frac{1}{2n}\bigg\{\sum_{j=1}^{2n} \bigg[\frac{3N}{2\Delta\tau}-\frac{(\mathbf{R}_j-\mathbf{R}_{j-1})^2}{4\lambda_m\Delta\tau^2}+V(\mathbf{R}_j)
  -(\mathbf{R}_{j}-\mathbf{R}_{j-1})\cdot\nabla_{\mathbf{R}_j}V(\mathbf{R}_j)
  \nonumber\\
  +\lambda_m \Delta\tau \nabla_{\mathbf{R}_j}^2V(\mathbf{R}_j)-\lambda_m\Delta\tau^2(\nabla_{\mathbf{R}_j}V(\mathbf{R}_j))^2\bigg]
  \bigg\}.
  \label{chpimc_energyf2}
\end{eqnarray}
In Eq.~(\ref{chpimc_energyf2}),
the head ($0$-th time slice) and tail ($2n$-th time slice)
are not treated on
equal footing because of the partial derivative.
The expression can be made
``symmetric'' by averaging over additional
terms for which the derivative yields a term that contains the
factor $\mathbf{R}_{j+1} - \mathbf{R}_{j}$.
Compared to Eq.~(\ref{chpimc_energyf1}), Eq.~(\ref{chpimc_energyf2}) contains 
three extra terms in the sum. 
In the $n\to\infty$ limit,
both estimators approach
the true expectation value. However, for finite $n$,
$\langle E_T \rangle$ and $\langle E_H \rangle$  generally give different estimates of the energy.
To obtain accurate results, one needs to extrapolate the finite $\Delta\tau$
calculations to the zero time step limit.
The difference between
the two estimators for a single $\Delta\tau$ may be used
as a rough estimate of the systematic error~\cite{ceperleyrev}.

\subsubsection{Example: Structural properties}
The energy estimator is special in that the information carried by all
$2n+1$
time slices can be used [see
  the sum over $j$ in Eqs.~(\ref{chpimc_energyf1}) and (\ref{chpimc_energyf2})].
The reason is that the Hamiltonian operator commutes with the propagator.
For other estimators, only the information carried by the middle 
or $n$-th time slice
and the associated propagators can, in general, be used.
This section exemplarily discusses the determination of structural properties
within the PIGS framework.

Quite generally, the operator $\hat{D}$ corresponding to a structural 
observable can be written as a function $f(\mathbf{R})$ times a $\delta$-function. 
For example, for
the scaled pair distribution function $4\pi r^2P_{12}(r)$ for particles 
one and two, $f(\mathbf{R})$ is equal to one
and the $\delta$-function is equal to 
$\delta(r_{\rm{ref}}- |\mathbf{r}_{12}|)$,
where $\mathbf{r}_{12}$ is the distance vector between particles one
and two.
Replacing $\langle \mathbf{R}'| \hat{H} | \mathbf{R}''' \rangle$
on the right-hand side of Eq.~(\ref{eq_energy3})
by $\delta(r_{\rm{ref}}- |\mathbf{r}_{12}|) \delta(\mathbf{R}'-\mathbf{R}''')$,
one finds 
$w(\mathbf{x})=\delta(r_{\rm{ref}} - | \mathbf{r}_{1,n} - \mathbf{r}_{2,n}|)$.
Similarly to the stochastic evaluation of structural properties for
a given many-body (zero-temperature) wave function, the $\delta$-function in the 
operator yielding the pair distribution function amounts to
sorting the configurations into small intervals or bins and counting the 
number of configurations that fall into each of the intervals.

In practice, to obtain the scaled pair distribution function $4\pi r^2 P_{12}(r)$ for 
particles one and two,
we discretize the pair distance $r_{12}$, 
  $r_{12}=|\mathbf{r}_{1}-\mathbf{r}_{2}|$, into a series
of $k_{\rm{max}}$ bins $[k \delta r, (k+1)
  \delta r]$, where $k$ range from 0 to $k_{\rm{max}}-1$.
  During the simulation, the pair distance is calculated for many
  configurations and sorted into the bins, i.e., a histogram of 
  the pair distances is collected.
  For each configuration considered (note, we may skip configurations 
  to ensure that the samples collected have neglegible correlations), 
  the pair distance $r_{12}$ is calculated for the middle time slice $\mathbf{R}_n$.
  The bin number $l$ of the histogram is calculated by evaluating
  $l={\rm{Floor}}(r_{12}/\delta r)$, where Floor$(x)$ gives the 
  largest integer smaller or equal to $x$, and the histogram value $v_l$
  of the $l$-th bin is increased by one.
  At the end, the histogram defined by the $v_l$ is 
  normalized by dividing by the total number $B_t$ of
  pair distances considered and the bin size $\delta r$. 
  The histogram created is a discretized version of the scaled pair
  distribution function $4\pi r^2 P_{12}(r)$.
  The approach outlined 
  yields the correct normalization even if some pair distances generated
  during the simulation are larger than $k_{\rm{max}} \delta r$.
  We typically monitor how many distances
  cannot be sorted into the histogram by comparing $\sum_l v_l$ with $B_t$.
  If the fraction is too large, then 
  the ``cutoff'' $k_{\rm{max}}\delta r$ needs to be increased.

Because the process involved in calculating 
different structural properties
such as the pair
distribution function and triple distribution function 
is the same,
the data structure used to accumulate different 
distribution functions and the accumulation process can be 
described by a single class in object
oriented programming languages.
This avoids duplication of the code.
In the code, the desired estimator (the ``object'') such as the pair
distribution function estimator and the triple distribution 
function estimator can be constructed
according to the
same class and can be initialized with observable specific parameters such as the bin size, the bin
number, and the number of particles.
To accumulate the weight and finalize the
results, the same virtual methods can be called for different estimators.
The actual implementation of these virtual methods
may or may not be
the same for different estimators. 
For example, 
the scaled pair and scaled triple distribution functions can share the same
implementation since both are described by an operator of the form
$\delta(r_{\rm{ref}}-r)$, where $r_{\rm{ref}}$ is either the pair distance
or the three-body hyperradius (see, e.g., 
Refs.~\cite{statistics14,nbody15}
for the resulting distribution
function).
The (unscaled) pair distribution function, in contrast, is described by an
operator of the form 
$\delta(r_{\rm{ref}}-r)/(4\pi r^2)$ 
and
has to be implemented separately.

\subsection{Error analysis}
\label{sec_algorithm_error}
The expectation value $\braket{f(\mathbf{x})}$ of a function $f(\mathbf{x})$ 
with respect to the probability density function $p(\mathbf{x})$ 
is defined as
\begin{equation}
  \braket{f(\mathbf{x})}=\int_{\mathbf{x}}f(\mathbf{x})p(\mathbf{x})d\mathbf{x}.
  \label{<++>}
\end{equation}
In the PIGS algorithm, we generate a finite series $X$ of
configurations $\mathbf{x}_j$,
\begin{equation}
X=\{\mathbf{x}_1, \mathbf{x}_2,\cdots, \mathbf{x}_M\},
  \label{<++>}
\end{equation}
according to the probability distribution $\pi(\mathbf{x})$.
The expectation value $\braket{{O}}$ of an operator
can then be
estimated by the mean value $\bar{{O}}$ of the series $X$,
\begin{equation}
  \bar{{O}}=
  \frac{1}{M} \sum_{j=1}^M w(\mathbf{x}_j).
  \label{chpimc_meanpimc}
\end{equation}
We refer to $\bar{{O}}$ as $\braket{{O}}_X$.
In the limit $M\to\infty$, the mean value $\bar{{O}}$
approaches the expectation value $\braket{{O}}$
[see Eq.~(\ref{chpimc_observablerewrite3})].

According to the central limit theorem, 
the mean value $\bar{{O}}$ of the series $X$ approaches the expectation value $\braket{{O}}$
in a predictive manner.
The ``standard'' central limit theorem states that the mean of a sufficiently large
number of random samples, drawn from a 
distribution with a well-defined
mean value and variance, is approximately normally distributed~\cite{krauth2006statistical,feller1971introduction}.
Since the Markov chain generates a series of data
that is correlated for small ``lag'' and uncorrelated
for large ``lag'', the standard central limit theorem 
cannot be applied directly.
However, it has been shown that the central limit theorem can be extended
to Markov-chain generated data~\cite{jones2004}.
Thus, we divide the series $X$, obtained from
the PIGS samples $w(\mathbf{x}_j)$, 
into $L$ blocks, each with $l=M/L$ configurations.
Defining the block averages $S_k$,
\begin{equation}
S_k= \frac{1}{l} \sum_{j=(k-1)\times l+1}^{k \times l}w(\mathbf{x}_j),
  \label{chpimc_meanpimcsk}
\end{equation}
we construct the series $\{S_1, \cdots, S_L\}$.
Provided $l$ is sufficiently large
(in our applications, the value of $l$ ranges from 10 to $10^4$), 
the block averages $S_k$ are
normally distributed and
the variance $\sigma^2$ of the block averages 
can be estimated from the sample variance $\braket{\sigma^2}_X$,
\begin{equation}
\braket{\sigma^2}_X= \frac{1}{L-1} \sum_{j=1}^L (S_j-\bar{S}),
  \label{<++>}
\end{equation}
where 
\begin{equation}
\bar{S}= \frac{1}{L} \sum_{j=1}^L S_j.
  \label{chpimc_meanpimcfinal}
\end{equation}
Note that $\bar{S}$ is equal to $\bar{{O}}$ [this can be seen by comparing Eqs.~(\ref{chpimc_meanpimc})
  and (\ref{chpimc_meanpimcsk})].
In our simulations,
we estimate the error of the expectation
value $\braket{{O}}$ using $ \braket{\sigma_{\bar{{O}}}}_X$,
\begin{equation}
  \braket{\sigma_{\bar{{O}}}}_X=
  \sqrt{\frac{\braket{\sigma^2}_X}{L}},
  \label{chpimc_errorpimc}
\end{equation}
 i.e., we report the mean $\bar{{O}}$
with error $\braket{\sigma_{\bar{{O}}}}_X$.

In our simulations, the number $l$ of configurations
per block is determined such that the block averages $S_k$
are normally distributed.
Alternatively, the value of $l$ (and, assuming
$M$ is fixed, that of $L$) 
can be determined by calculating the autocorrelation
length~\cite{Binder1986,berg2004markov}.
The latter approach is more commonly used and is somewhat 
simpler to implement. The two approaches should yield
comparable results.

Considering $Q$ simulations, each yielding a series $X_j$ and correspondingly 
$\sqrt{\braket{\sigma^2}_{X_j}}$ (assuming finite $L$),
the estimate of the standard deviation is biased because the mean value of a square root
function is not equal to the square root of the mean, i.e.,
\begin{equation}
  \frac{\sum_{j=1}^{Q}\sqrt{\braket{\sigma^2}_{X_j}}}{Q}\neq
  \sqrt{\frac{\sum_{j=1}^Q\braket{\sigma^2}_{X_j}}{Q}}.
  \label{<++>}
\end{equation}
Since the bias becomes negligible for sufficiently large $L$,
there is no need to correct for the
bias.
In our simulations, $L$ is typically 80 or larger.
Because the elements $S_j$ in $\{S_1,\cdots,S_L\}$ are normally distributed, the variance
$\braket{\sigma^2}_X$ is approximately a constant for sufficiently large $L$
and the error $\braket{\sigma_{\bar{{O}}}}_X$ 
scales, according to Eq.~(\ref{chpimc_errorpimc}), as $1/\sqrt{L}$. 
Thus, to improve the accuracy of an observable by an 
order of magnitude, the computational time
needs to be increased by two orders of magnitude.

\begin{figure}
\centering
\includegraphics[angle=0,width=0.4\textwidth]{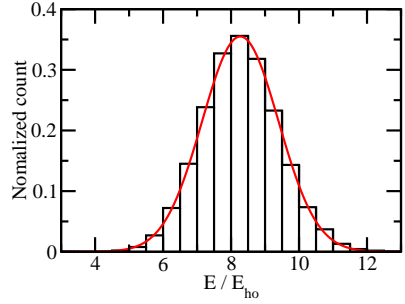}
\caption{
  Normalized histogram of the energy  $E/E_{\rm{ho}}$,
calculated using the thermal
estimator, for the harmonically trapped (3,3) system 
at unitarity.
The histogram is constructed using 38,400 block averages
generated using 480 processors
(see the text for details).
The red solid line shows the normal distribution with 
mean value of $E=8.273E_{\rm{ho}}$ and standard deviation
of $1.124E_{\rm{ho}}$ (the error of the mean is 
$1.124E_{\rm{ho}}/\sqrt{38,400}=0.006 E_{\rm{ho}}$).
 }\label{chpimc_realdata}
\end{figure}

To check whether the final distribution is approximately normal, one can
make a histogram of the observable under study.
Figure~\ref{chpimc_realdata} shows the normalized histogram for 
the energy
of the (3,3) system 
at unitarity.
The notation (3,3) refers to
three spin-up fermions and three spin-down fermions under external harmonic
confinement with angular frequency $\omega$ (the
oscillator energy of
$\hbar \omega$ is denoted by $E_{\rm{ho}}$).
The physics of this small fermionic system is discussed in
more detail in Sec.~\ref{sec_application}.
The example considered here
uses $\tau = 0.5 (E_{\rm{ho}})^{-1}$, 
$\Delta \tau=0.125 (E_{\rm{ho}})^{-1}$, and $\psi_T$ given in Eq.~(\ref{eq_psitrial_correlations});
the resulting extrapolated $\Delta \tau=0$
energy is reported in Table~\ref{tab1}.
The simulation is done on $480$ 
processors with each processor producing 80
block averages.
This yields a total of 38,400 block averages.
Even though these block averages are not obtained from a single
Markov chain but from 480 independent Markov chains,
we calculate the mean and error of the mean using,
respectively,
Eqs.~(\ref{chpimc_meanpimcfinal}) and (\ref{chpimc_errorpimc}) with $L = 38,400$.
The resulting sample mean is 
$8.273E_{\rm{ho}}$ with an error or
uncertainty
of
$0.006 E_{\rm{ho}}$.
Using the calculated mean and standard deviation, 
the solid line in Fig.~\ref{chpimc_realdata}
shows the corresponding normal distribution. It can be seen that the solid
line provides a faithful description of the histogram, indicating that the
underlying samples are indeed normally distributed.

The presented analysis requires a sufficiently large number of block averages.
In practice, it may not be feasible or advisable to calculate many block averages.
In such a case, 
one can check if the error scales as $\sqrt{1/L}$
with the number of blocks $L$.
Reducing the number of blocks by a factor of two,
one should observe that, if the block averages are 
normally distributed, the error increases
roughly by a factor of $\sqrt{2}$.
This check can be performed for as few as 5 or 10 blocks and provides,
in many cases, enough information to reliably assign error bars.

To check explicitly whether the samples are independent, one
needs to perform autocorrelation (or serial correlation) tests~\cite{lester94}.
Given a series of numbers $\{x_1, \cdots, x_L\}$, the lag $k$ 
correlation coefficient $r_k$, which measures the correlation of the series 
of numbers $\{x_1, \cdots, x_{L-k}\}$ and 
$\{x_{1+k}, \cdots, x_L\}$, 
is given by~\cite{salas1980applied} 
\begin{equation}
  r_k=\frac{\sum_{j=1}^{L-k}(x_j-\bar{x})(x_{j+k}-\bar{x})}{\sum_{j=1}^{L}(x_j-\bar{x})^2},
  \label{chpimc_rk}
\end{equation}
where $\bar{x}$ denotes the average of the numbers 
$\{x_1,\cdots,x_L\}$.
If the samples are truly uncorrelated,
the $r_k$ approximately follow
a normal distribution for sufficiently large $L-k$ and 
the variance of $r_k$
is approximately equal to $1/L$.
Furthermore, the probability that $r_k$ falls into the interval
\begin{equation}
  \left[\frac{-1-1.96\sqrt{L-k-1}}{L-k}, \frac{-1+1.96\sqrt{L-k-1}}{L-k} \right]
  \label{chpimc_confidenceinterval}
\end{equation}
is 95\%~\cite{salas1980applied}.
Based on hypothesis testing theory~\cite{gonick1993cartoon},
it is claimed, with 95\% confidence, 
that a sample is correlated  if $r_k$ (based on a single test for one $k$)
falls outside the interval given in
Eq.~(\ref{chpimc_confidenceinterval}).

\begin{figure}
\centering
\includegraphics[angle=0,width=0.4\textwidth]{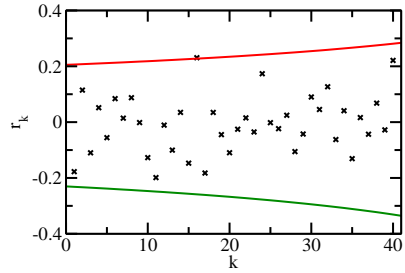}
\caption{
  The crosses show the correlation coefficient $r_k$ as a function of the lag $k$ 
($k=1-40$)
for 
  the sample and observable considered in
  Fig.~\ref{chpimc_realdata}.
  A series of 80 block averages, obtained from a single Markov chain
  (and processor), is analyzed.
  The upper and lower solid lines show the 95\% confidence interval defined in
    Eq.~(\ref{chpimc_confidenceinterval}).
 }\label{chpimc_figautoreal}
\end{figure}

Figure~\ref{chpimc_figautoreal}
shows the correlation coefficient $r_k$ [Eq.~(\ref{chpimc_rk}) with $L=80$] 
for 80 block averages generated on a single processor
(i.e., obtained from a single Markov chain)
as a function of the lag $k$ for the system and observable considered
in Fig.~\ref{chpimc_realdata}.
Since the correlation coefficients 
for $k \ge 1$ all lie within the confidence band,
it is said that the data pass the autocorrelation test.
For the data shown in
Fig.~\ref{chpimc_realdata},
similar correlation coefficient plots
are obtained for each of the 80 block averages generated by the other 479
processors.
This verifies that the samples are truly independent.

\subsection{Permutations:
On-the-fly anti-symmetrization scheme}
\label{sec_algorithm_permutation}
To account for the particle statistics, one needs to ensure the proper
behavior of the propagator under particle
permutations.
The Hilbert space for identical bosons or identical fermions is
restricted compared to that of Boltzmann particles described by the same
Hamiltonian.
The discussion so far,
including the short-time approximations for the propagator introduced in
Sec.~\ref{sec_algorithm_intro}, applies to Boltzmann particles.

In the ``standard'' approach of generating paths for
systems containing identical particles,
the symmetrizer and anti-symmetrizer
are evaluated stochastically
(the corresponding move is referred to as
``permutation move'')~\cite{ceperleyrev,boninsegni05}.
This implies that Bose and Fermi systems are
simulated by the same paths.
Expectation values, in contrast, are
accumulated by including ``weight factors''
(plus and minus signs) that account
for the particle statistics. This standard approach can be thought of
as an analog of a post-symmetrization scheme, where one first generates
configurations that represent the entire Hilbert space and then
projects out those configurations that have the proper symmetry.

Here, we introduce an alternative ``on-the-fly''
symmetrization/anti-symmetrization
scheme that explicitly enforces the proper symmetry at each
imaginary time index.
This scheme is
particularly useful for fermionic systems with zero-range interactions.
Without 
a three-body regulator
and without this on-the-fly
anti-symmetrization scheme,
two-component Fermi gases would undergo Thomas 
collapse~\cite{thomas35}.
A downside of the scheme discussed below is that the
computational effort scales factorially with the number of
identical particles; as a consequence,
the scheme becomes prohibitively expensive with increasing
number of particles.
While the on-the-fly symmetrization scheme
can be applied to systems that contain identical bosons, the
explicit symmetrization
is typically not needed in this case since the ground state
wave function of the system in which the bosons are replaced
by Boltzmann particles is identical to that of
the system with bosons.
Thus, the discussion in this section is most
useful for fermions.

We start with a general discussion that will be useful
for our on-the-fly anti-symmetrization
scheme and then discuss on-the-fly anti-symmetrization
scheme and PIGS
specific aspects.
To this end, we introduce 
the symmetrizer $\hat{\mathcal{P}}$~\cite{ceperleyrev}.
For single-component Bose and Fermi
systems ($N$ identical particles), $\hat{\mathcal{P}}$ can
be written as~\cite{basicTextBaym}
\begin{equation}
\hat{\mathcal{P}}=
\frac{1}{N!}\sum_{\sigma}(\pm1)^{N_{\rm{I}}(\sigma)} \hat{P}_{\sigma},
  \label{chpimc_symm}
\end{equation}
where $\sigma$ denotes
the permutation of particle indices, $N_{\rm{I}}(\sigma)$ the
number of inversions in $\sigma$~\cite{notestrmathe}, 
and $\hat{P}_{\sigma}$ the corresponding
permutation operator.
The plus (minus) sign in Eq.~(\ref{chpimc_symm})
applies to identical bosons (fermions).
For example, the symmetrizers 
(to unify the notation, we use the term symmetrizer for bosons and fermions)
for two and three identical fermions are $\hat{\mathcal{P}}=\hat{\mathcal{A}}_2$
and $\hat{\mathcal{P}}=\hat{\mathcal{A}}_3$,
$\hat{\mathcal{A}}_2=(1-\hat{P}_{12})/2$ and 
$\hat{\mathcal{A}}_3=
(1-\hat{P}_{12}-\hat{P}_{13}-\hat{P}_{23}+\hat{P}_{123}+\hat{P}_{132})/6$,
respectively.
Here, $\hat{P}_{ijk\cdots l}$ 
replaces the identity of particle $i$ (i.e., its entire ``information'' 
including spatial coordinates, spin degrees of freedom, etc.)
with that of particle $j$, that of particle $j$ with that of particle $k$, $\cdots$,
and that of particle $l$ with that of particle $i$.
The symmetrizer $\hat{\mathcal{P}}$ commutes with $\hat{P}_{ij}$ if the $i$-th and $j$-th particles are
identical.
In the previous examples, $\hat{\mathcal{A}}_2$ commutes with $\hat{P}_{12}$ and
$\hat{\mathcal{A}}_3$ commutes with $\hat{P}_{12}$, $\hat{P}_{13}$, and $\hat{P}_{23}$.
The definition of the symmetrizer $\hat{\mathcal{P}}$ can be generalized to multi-component Bose and Fermi systems as well as
Bose-Fermi mixtures.
In these cases, the total symmetrizer is written as a product of symmetrizers
for each component.
For example, the symmetrizer for the mixture of two identical bosons (particles 1 and 2)
and two identical fermions (particles 3 and 4) reads 
$(1+\hat{P}_{12})(1-\hat{P}_{34})/4$.

The symmetrizer $\hat{\mathcal{P}}$ also 
commutes with the Hamiltonian $\hat{H}$ and the
propagator $\hat{G}$.
$\hat{\mathcal{P}}$ serves the purpose of projecting out the wave functions
that satisfy the proper exchange symmetry,
i.e., it divides the Hilbert space into two parts: (i)
If $\psi_s$ is an eigen state with the proper symmetry, then one has
$\hat{\mathcal{P}} \psi_s=\psi_s$.
(ii)
If, in contrast, $\psi_{\rm{ns}}$ is an eigen state that does not have the proper
exchange symmetry, then we have
$\hat{\mathcal{P}} \psi_{\rm{ns}}=0$.
We note that the eigen values of the symmetrizer $\hat{\mathcal{P}}$ are
0 and 1 while those of the two-particle permute operator $\hat{P}_{12}$
are $-1$ and 1.
As we will show in the following, the fact that
the eigen values of $\hat{\mathcal{P}}$ are either 0 or 1 implies
\begin{equation}
  \hat{\mathcal{P}}^2=\hat{\mathcal{P}}.
  \label{chpimc_p2}
\end{equation}

To prove Eq.~(\ref{chpimc_p2}), we introduce 
a unitary matrix $\mathcal{U}$ that diagonalizes the Hermitian 
symmetrizer $\hat{\mathcal{P}}$, i.e., 
$\mathcal{U}$ is constructed such that
$\hat{\mathcal{D}}=\mathcal{U}\hat{\mathcal{P}}\mathcal{U}^{-1}$ is
diagonal.
Because $\hat{\mathcal{P}}$ and $\hat{\mathcal{D}}$ are related through a 
unitary transformation, $\hat{\mathcal{D}}$ and
$\hat{\mathcal{P}}$ share the same eigen values.
Since the eigen values of $\hat{\mathcal{D}}$ are 
either 0 or 1, $\hat{\mathcal{D}}$ is diagonal with diagonal elements 0 or 1.
This implies
that $\hat{\mathcal{D}}^2$ is equal to $\hat{\mathcal{D}}$.
We now rewrite $\hat{\mathcal{P}}^2$ 
using $\hat{\mathcal{P}}=\mathcal{U}^{-1}\hat{\mathcal{D}}\mathcal{U}$,
\begin{eqnarray}
  \hat{\mathcal{P}}^2=(\mathcal{U}^{-1}\hat{\mathcal{D}}\mathcal{U})
(\mathcal{U}^{-1}\hat{\mathcal{D}}\mathcal{U}).
\end{eqnarray}
Using $\mathcal{U}^{-1} \mathcal{U}=1$, we have
\begin{eqnarray}
\hat{\mathcal{P}}^2
  =\mathcal{U}^{-1}\hat{\mathcal{D}}\hat{\mathcal{D}}\mathcal{U}.
\end{eqnarray}
Since $\hat{\mathcal{D}}^2$ is equal to $\hat{\mathcal{D}}$ (see above),
one finds $\hat{{\mathcal{P}}}^2={\mathcal{U}}^{-1} \hat{\mathcal{D}} {\mathcal{U}}$
and thus ${\hat{\mathcal{P}}}^2=\hat{\mathcal{P}}$, which is what we set
out to prove.

The symmetrized propagator can, as we prove below, be written as
$\hat{G}_{\rm{unsymm}}\hat{\mathcal{P}}$,
where $\hat{G}_{\rm{unsymm}}$ is the 
unsymmetrized propagator, i.e., the propagator for the corresponding system with Boltzmann particles.
In position space, the symmetrized propagator 
$G(\mathbf{R},\mathbf{R}',\hat{\mathcal{P}};\tau)$
can be rewritten
as~\cite{ceperleyrev}
\begin{equation}
G(\mathbf{R},\mathbf{R}',\hat{\mathcal{P}};\tau)=  
\braket{\mathbf{R}|\exp(-\tau \hat{H})\hat{\mathcal{P}}|\mathbf{R}'}
  \label{chpimc_symmrho1}
\end{equation}
or
as a sum over the unsymmetrized propagators,
\begin{equation}
  G(\mathbf{R},\mathbf{R}',\hat{\mathcal{P}};\tau)
\propto
\sum_\sigma {\rm{sgn}}(\sigma)G(\mathbf{R},\hat{\mathcal{P}}_{\sigma}\mathbf{R}';\tau),
  \label{eq_gsymm}
\end{equation}
where ${\rm{sgn}}(\sigma)$ is the sign for the permutation $\sigma$ (for
single-component fermions, ${\rm{sgn}}(\sigma)=(-1)^{N_{\rm{I}}(\sigma)}$).
In Eq.~(\ref{eq_gsymm}), we use the proportionality symbol
since the ``normalization factor'' depends on the number of identical
particles in the system (for $N$ identical particles, the
proportionality symbol becomes an equal sign if the right-hand-side is
multiplied by $(N!)^{-1}$). 

We now prove that $\hat{G}_{\rm{unsymm}} \hat{\mathcal{P}}$
is, indeed, the symmetrized propagator.
In Schr{\"o}dinger quantum mechanics, the symmetrized 
propagator in position space reads
\begin{equation}
G(\mathbf{R},\mathbf{R}',\hat{\mathcal{P}};\tau)=
\braket{\mathbf{R}|\exp(-\tau \hat{H}) \sum_j\ket{\psi_{{\rm{symm}},j}}\bra{\psi_{{\rm{symm}},j}}\mathbf{R}'},
  \label{chpimc_rhosymm2}
\end{equation}
where $\{\psi_{{\rm{symm}},j}\}$ is the complete set of symmetrized 
eigen states, e.g., for $N$ identical
bosons or $N$ identical fermions.
The complete set of unsymmetrized eigen states of $\hat{H}$, i.e.,
the set of eigen states for Boltzmann particles is denoted by $\{\psi_{{\rm{unsymm}},j}\}$.
Recall that the $\hat{\mathcal{P}}$ operator can be 
diagonalized using the unitary matrix
$\mathcal{U}$.
$\mathcal{U}$ ``reorganizes'' the eigen states $\psi_{{\rm{unsymm}},j}$ such that the new eigen
states are also eigen states of $\hat{\mathcal{P}}$.
The resulting eigen states $\psi_{r,j}$,
\begin{equation}
\psi_{r,j}=\sum_l\mathcal{U}_{jl}\psi_{{\rm{unsymm}},l},
  \label{chpimc_newbasis}
\end{equation}
either have the proper symmetry, i.e.,
$\hat{\mathcal{P}}\psi_{r,j}$ is equal to $\psi_{r,j}$ (in this case, the eigen value of
$\hat{\mathcal{P}}$ is 1) or $\hat{\mathcal{P}} \psi_{r,j}$ gives
zero
(in this case, the eigen value of $\hat{\mathcal{P}}$ is 0).
The subset of eigen states 
$\{\psi_{r,j}\}$, for which $\hat{\mathcal{P}}\psi_{r,j}$
is equal to $\psi_{r,j}$, coincides 
with the complete set of symmetrized eigen states.
This process of constructing a set of properly symmetrized eigen states from a
complete set of unsymmetrized eigen states is known as post-symmetrization.
For later reference, we write down the auxiliary identity
\begin{equation}
  \mathcal{U}\sum_l\ket{\psi_{{\rm{unsymm}},l}}\bra{\psi_{{\rm{unsymm}},l}}\mathcal{U}^{-1}
  =\sum_l\ket{\psi_{r,l}}\bra{\psi_{r,l}},
  \label{chpimc_auxi}
\end{equation} 
which can be obtained using the matrix form of Eq.~(\ref{chpimc_newbasis}),
\begin{equation}
  \{\psi_{r,1},\psi_{r,2},\dots\}^T=
\mathcal{U}\{\psi_{{\rm{unsymm}},1},\psi_{{\rm{unsymm}},2},\dots\}^T.
  \label{<++>}
\end{equation}

We now manipulate the right-hand side
of Eq.~(\ref{chpimc_symmrho1}) so that it can be readily 
related to Eq.~(\ref{chpimc_rhosymm2}).
Starting with the right-hand side of Eq.~(\ref{chpimc_symmrho1}) and 
using Eq.~(\ref{chpimc_p2}),
we find
\begin{eqnarray}
  \braket{\mathbf{R}|\exp(-\tau \hat{H})\hat{\mathcal{P}}|\mathbf{R}'}&=
  \braket{\mathbf{R}|\exp(-\tau \hat{H})\hat{\mathcal{P}}\hat{\mathcal{P}}|\mathbf{R}'}.
\end{eqnarray}
Inserting $\mathcal{U}\mathcal{U}^{-1}=\hat{1}$ and then
\begin{equation}
  \sum_l\ket{\psi_{{\rm{unsymm}},l}}\bra{\psi_{{\rm{unsymm}},l}}=\hat{1},
  \label{<++>}
\end{equation}
we find
\begin{eqnarray}
  \fl
  \braket{\mathbf{R}|\exp(-\tau \hat{H})\hat{\mathcal{P}}|\mathbf{R}'}&=
  \braket{\mathbf{R}|\exp(-\tau \hat{H})\hat{\mathcal{P}}\mathcal{U}\sum_l\ket{\psi_{{\rm{unsymm}},l}}\bra{\psi_{{\rm{unsymm}},l}}\mathcal{U}^{-1}\hat{\mathcal{P}}|\mathbf{R}'}.
  \label{chpimc_mids}
\end{eqnarray}
Using Eq.~(\ref{chpimc_auxi}) in Eq.~(\ref{chpimc_mids}), we find
\begin{eqnarray}
  \braket{\mathbf{R}|\exp(-\tau \hat{H})\hat{\mathcal{P}}|\mathbf{R}'}&=
  \braket{\mathbf{R}|\exp(-\tau \hat{H})\hat{\mathcal{P}}\sum_l\ket{\psi_{r,l}}\bra{\psi_{r,l}}\hat{\mathcal{P}}|\mathbf{R}'}.
\end{eqnarray}
Finally, 
noting that $\hat{\mathcal{P}}\sum_l\ket{\psi_{r,l}}\bra{\psi_{r,l}}\hat{\mathcal{P}}$
is equal to
$\sum_l\ket{\psi_{{\rm{symm}},l}}\bra{\psi_{{\rm{symm}},l}}$, we arrive
at
\begin{eqnarray}
  \fl
  \braket{\mathbf{R}|\exp(-\tau \hat{H})\hat{\mathcal{P}}|\mathbf{R}'}&=
  \braket{\mathbf{R}|\exp(-\tau \hat{H})\sum_l\ket{\psi_{{\rm{symm}},l}}\bra{\psi_{{\rm{symm}},l}}\mathbf{R}'},
  \label{chpimc_symmrhod}
\end{eqnarray}
i.e., 
we have proven that
$\hat{G}_{{\rm{unsymm}}}\hat{\mathcal{P}}$ [the left-hand side
of
Eq.~(\ref{chpimc_symmrhod}) in position space]
is identical to the symmetrized propagator
in position space in Schr\"odinger quantum mechanics
[Eq.~(\ref{chpimc_rhosymm2})].

Replacing the propagator 
$\braket{\mathbf{R}|\exp(-\tau \hat{H})|\mathbf{R}'}$ by 
$\braket{\mathbf{R}|\exp(-\tau \hat{H})\hat{\mathcal{P}}|\mathbf{R}'}$
in all expressions involving the propagator (such as the normalization
factor $Z(\tau)$, the probability
distribution $\pi(\mathbf{x})$, and the weight function $w(\mathbf{x})$),
we have all elements of the PIGS algorithm for bosons and fermions;
as indicated at the beginning of this section, the trial function
$\psi_T$ will be discussed in Sec.~\ref{sec_application}
for specific examples.
The symmetrized probability distribution $\pi_{{\rm{symm}}}(\mathbf{x})$, e.g.,
reads
\begin{eqnarray}
  \fl
  \pi_{{\rm{symm}}}(\mathbf{R}_0,\cdots,\mathbf{R}_{2n})=
\psi_T(\mathbf{R}_0)
G(\mathbf{R}_0,\mathbf{R}_{1},\hat{\mathcal{P}};\Delta\tau)
G(\mathbf{R}_1,\mathbf{R}_{2},\hat{\mathcal{P}};\Delta\tau)
\nonumber\\
\times
\cdots \times
G(\mathbf{R}_{2n-1},\mathbf{R}_{2n},\hat{\mathcal{P}};\Delta\tau)
\psi_T(\mathbf{R}_{2n}),
\label{eq_examplesymmetrizer}
\end{eqnarray}
where each symmetrized propagator is a sum over unsymmetrized propagators with
permuted configurations. This implies that the complexity 
of the symmetrized PIGS algorithm is 
increased by up to a factor of $N!$ compared to the non-symmetrized
PIGS algorithm.

Since the symmetrizer is applied at each link 
[see Eq.~(\ref{eq_examplesymmetrizer})],
the number of operations needed to evaluate
the symmetrized propagator
scales as
$2n N!$ for $N$ identical particles.
This implies that the on-the-fly
symmetrization scheme becomes
inefficient if the number of time slices is too large;
Sec.~\ref{sec_application} demonstrates that reliable results
for two-component Fermi gases can be obtained for a series of $n$ as small
as $1$, $2$ and $4$.
If the Trotter formula is used, some terms can typically be pulled 
out of the sum over the permutations,
reducing the computational effort somewhat.
If the pair product approximation is
used, analogous simplifications are, in general, not
possible.
However, as mentioned at the beginning of this section,
the on-the-fly symmetrization scheme is particularly useful
for fermionic systems with zero-range interactions,
which cannot be treated using the Trotter formula.

The change of the probability distribution from $\pi(\mathbf{x})$
to $\pi_{\rm{symm}}(\mathbf{x})$ needs to be accounted for in the
implementation of the moves and the collection of the
expectation values.
For example, the acceptance function for the naive move
is equal to $\pi_{\rm{symm}}(\mathbf{x})$,
implying that ${\cal{A}}(\mathbf{x} \rightarrow \mathbf{x}')$
in Eq.~(\ref{eq_calanaivemove}) is given by
$\min (1, \pi_{\rm{symm}}(\mathbf{x}')/\pi_{\rm{symm}}(\mathbf{x}))$.
The pair distance move has to be modified 
analogously.
To obtain the
acceptance function for the wiggle move,
the derivation outlined in Sec.~\ref{wiggle-ch2} needs
to be carried out using the symmetrized propagator.
Similarly, the estimators need to be changed accordingly.
For example, to account for the permutations in 
the thermodynamic energy estimator $\langle E_T \rangle$,
$\pi(\mathbf{x})$ and $Z(\tau)$ in Eq.~(\ref{chpimc_energyz2})
have to be replaced by the corresponding symmetrized quantities.
If we wrote out, in analogy to Eq.~(\ref{chpimc_energyf1}), the fully
symmetrized
expression for the weight function
$w(\mathbf{x})$
using the second-order Trotter formula, it would be rather 
lengthy since
the symmetrized propagator contains a sum over permutations
at each time slice.

The ground state of
fermionic systems usually corresponds to an excited state of the
corresponding system with Boltzmann statistics.
This implies that explicit anti-symmetrization 
is necessary to propagate the trial 
function to the ground state with the correct particle statistics.
The anti-symmetrization introduces a ``sign problem''
since
the probability distribution can be positive or negative.
The fact that the probability distribution can take
either sign can be dealt with using the ideas of
Refs.~\cite{statistics14,FermiSignProblem}.

The probability distribution for a given configuration can take either
sign.
Integrating all the positive (negative) portions, we obtain $Z_+$ ($Z_-$).
The normalization factor $Z(\tau)$ is the sum of $Z_+$ and $Z_-$, 
$Z(\tau)=|Z_+|-|Z_-|$.
When accumulating observables, the sign needs to be kept track of.
In general, the expectation value of an observable $O$
can be written as
\begin{equation}
  \braket{{O}}=\frac{|Z_+|}{|Z_+|-|Z_-|} \braket{{O}_+} -\frac{|Z_-|}{|Z_+|-|Z_-|}
\braket{{O}_-}.
  \label{eqpmo}
\end{equation}
For the thermal energy estimator, e.g., this can be worked out
explicitly.
Using $Z(\tau)=|Z_+|-|Z_-|$, we find
\begin{eqnarray}
  \braket{{E_T}}
=-(|Z_+|-|Z_-|)^{-1}\frac{\partial(|Z_+|-|Z_-|)}{\partial\tau}
\end{eqnarray}
or
\begin{eqnarray}
  \braket{{E_T}}
=(|Z_+|-|Z_-|)^{-1}\left(|Z_+| \braket{{E_{T,+}}} -|Z_-| \braket{{E_{T,-}}}\right),
  \label{<++>}
\end{eqnarray}
where 
\begin{equation}
  \braket{{E_{T,\pm}}}=
-(|Z_{\pm}|)^{-1}\frac{\partial(|Z_{\pm}|)}{\partial\tau}.
  \label{<++>}
\end{equation}
In an actual calculation, 
the contributions
$\langle O_+ \rangle$ and $\langle O_- \rangle$
to the estimator are first calculated separately
and then weighted according 
to their relative magnitudes. 
Alternatively, Eq.~(\ref{eqpmo}) can be written as
\begin{equation}
  \braket{{O}}= S^{-1}
\left(\frac{|Z_+|}{|Z_+|+|Z_-|}\braket{{O}_+} +\frac{|Z_-|}{|Z_+|+|Z_-|}
\braket{{-O}_-}\right),
  \label{eqpmo2}
\end{equation}
where
\begin{eqnarray}
\label{eq_sfactor}
S = \frac{|Z_+|-|Z_-|}{|Z_+|+|Z_-|}.
\end{eqnarray} 
This suggests that one can think of the simulation as yielding
a purely positive
normalization factor $|Z_+|+|Z_-|$;
however, if one does so, 
a minus sign needs to be included in the observable
if the probability distribution for the chosen configuration is
negative.
The final result is obtained if the expression is multiplied
by the factor $S$.

In Eqs.~(\ref{eqpmo}) and (\ref{eqpmo2}), 
the term $|Z_+|-|Z_-|$ appears in the denominator.
If $|Z_+|$ becomes closer to $|Z_-|$ with increasing propagation
time, then the simulation becomes increasingly more challenging since 
the statistical noise needs to be
smaller than the difference between $|Z_+|$ and $|Z_-|$.

Finally, one may wonder if there are any constraints on the trial  function
$\psi_T$. For the fixed-node 
diffusion Monte Carlo simulations, the trial function $\psi_T$ needs to be
an eigen state of the symmetrizer.
This is not the case 
for PIGS simulations because the symmetrized propagator projects
out the corresponding wave function.

\section{Application to fermionic systems}
\label{sec_application}

\subsection{General considerations}
\label{sec_application_fermi_general}
This section discusses applications of the PIGS algorithm to 
harmonically-trapped equal-mass two-component Fermi gases 
consisting of $n_1$ spin-up and $n_2$ spin-down
particles ($N=n_1+n_2$) in three-dimensional space.
We refer to these systems as $(n_1,n_2)$.
The model Hamiltonian $\hat{H}$ reads
\begin{eqnarray}
\hat{H} = \hat{H}_{\rm{free-space}} + \hat{V}_{\rm{trap}},
\label{eq_hamfermi0}
\end{eqnarray}
where
\begin{eqnarray}
\hat{H}_{\rm{free-space}}=
  \sum_{j=1}^{N} \frac{-\hbar^2}{2m}\nabla_j^2+
\sum_{j=1}^{N-1} \sum_{k>j}^N V_{\rm{F}}(r_{jk}).
\label{eq_hamfermi}
\end{eqnarray}
The interspecies two-body zero-range potential
$V_{\rm{F}}(r_{jk})$ is given in Eq.~(\ref{eq_fermihuangregularized})
and the confining potential $\hat{V}_{\rm{trap}}$
with angular trapping frequency $\omega$ reads
\begin{eqnarray}
\hat{V}_{\rm{trap}} = 
  \frac{1}{2}m
  \omega^2 \sum_{j=1}^{N} \mathbf{r}_{j}^2.
\end{eqnarray} 
Throughout this section, we assume that the
interspecies two-body interaction is characterized by an infinitely large
$s$-wave scattering length $a_s$, i.e., we consider systems at unitarity.
Like fermions are assumed to be non-interacting, i.e.,
no intraspecies interactions are considered.
This assumption is realized in cold atom systems provided one 
operates at magnetic field strengths away from $p$- and 
higher-partial-wave resonances.
As discussed in
Secs.~\ref{sec_ppa}-\ref{sec_propagator_for_tb_zr_interactions}, 
two-body zero-range
interactions are most conveniently treated using the pair
product approximation. In the calculations presented below,
the reduced relative propagator
$\bar{G}^{\rm{rel}}(\mathbf{r},\mathbf{r}';\tau)$
[Eq.~(\ref{3dHOdensitymatrix})],
which accounts for the two-body
zero-range interactions,
the kinetic energy, and the relative two-body
confining potential, is being used.

The Hamiltonian $\hat{H}$ is, for infinitely large $a_s$,
characterized by one (meaningful) length scale, the harmonic
oscillator length $a_{\rm{ho}}$, $a_{\rm{ho}}=\sqrt{\hbar/(m \omega)}$.
The harmonic oscillator length also characterizes the non-interacting
system.
The range of the interaction potential, which is zero, and the
$s$-wave scattering length, which is infinitely large, do not
define meaningful length scales.
Moreover, for two-component fermions with equal masses,
the three-body system in free space is unbound, implying that 
the three-body system does not introduce a new (finite) length scale;
in particular, Efimov physics is absent~\cite{petrov03,skor57}.
In what follows, we express lengths in units of $a_{\rm{ho}}$ and 
energies in units of the harmonic oscillator energy $E_{\rm{ho}}$, 
$E_{\rm{ho}}=\hbar \omega$.

Two-component Fermi gases with vanishing interaction
range and infinitely large interspecies $s$-wave scattering length
have been and continue to be a paradigmatic 
strongly-correlated 
system,
for which few analytical results are known and which are 
challenging to treat numerically. 
The PIGS applications presented in this 
section have not been published before. 
The examples are chosen for their pedagogical value and for their
relevance with regards to obtaining a more complete understanding of small
two-component Fermi gases.
Two types of systems are considered, spin-balanced systems
($n_1=n_2=N/2$; see Sec.~\ref{sec_application_fermibalanced}) and spin-imbalanced
systems ($n_1=N-1$ and $n_2=1$; see Sec.~\ref{sec_application_fermiimbalanced}).

\subsection{Spin-balanced Fermi gas}
\label{sec_application_fermibalanced}

The ground state
of spin-balanced two-component Fermi gases
has $(L,\Pi)=(0,+1)$
symmetry, i.e., vanishing
total orbital angular momentum $L$ and
positive parity $\Pi$.
Intuitively, this can be understood by realizing that 
each spin-up fermion is paired with a spin-down fermion.
In reality, the pairing respects the identical particle
characteristics, i.e., 
each spin-up fermion is paired 
with $(1/n_2)$-th of each spin-down fermion 
and
each spin-down fermion is paired 
with $(1/n_1)$-th of each spin-up fermion. The particle statistics is
enforced along the paths
by explicitly applying the symmetrizer to each of the 
$2n$ propagators [see, e.g., Eq.~(\ref{eq_examplesymmetrizer}) for the
symmetrized probability distribution].
The applications below use 
trial functions $\psi_T$ that have the proper particle
symmetry build in. In general, one could employ
any trial function that has finite overlap with the eigen state
to be determined. In practice, however, it seems best to build as
much ``prior knowledge'' as possible into the trial function.

In the following, we discuss the construction of the trial function
$\psi_T(\mathbf{R})$ and the dependence of the energy on the 
variational parameters entering into $\psi_T$.
In addition, the convergence of the energy with respect to 
the total imaginary 
propagation time $\tau$ is analyzed. As can be seen from 
Eq.~(\ref{eq_expansion_of_psitau})
and the surrounding discussion, $\tau$ should, in principle, be taken to infinity
to allow for the excited state contributions to fully die out. In practice,
this is not feasable since the noise or error that arises
due to the anti-symmetrization (the sign error) increases
with increasing $\tau$. Thus, the task is to find a regime of
$\tau$ values, for which the excited state contributions can be 
neglected and the sign error is sufficiently small.
For each fixed $\tau$ simulation,
the convergence of the results with respect to $\Delta \tau$ 
or, equivalently, the number of time slices needs to be checked.
Typically, for each fixed $\tau$, the results for several $n$ are,
in a first step, 
extrapolated to the infinite $n$ limit. In a second
step, the $n\rightarrow \infty$ results for several $\tau$ are
considered to determine for which $\tau$ excited state contributions can be neglected.
In considering larger $\tau$, it has to be checked that 
the sign error is sufficiently small.
Last, the calculations should, ideally, 
be repeated for different $\psi_T$ to
ensure that the trial function does not introduce a bias.
Having an overview of the general PIGS procedure,
we now discuss the construction of the trial function $\psi_T$.

Quite generally, the construction of the
trial function is guided by physical considerations.
For example, one may parameterize the trial function $\psi_T$
in terms of a set of variational parameters $\boldsymbol{\alpha}$,
which are optimized by minimizing the expectation value
of the Hamiltonian, calculated using $\psi_T$, with respect
to the variational parameters $\boldsymbol{\alpha}$.
A good trial function $\psi_T$ is associated with a small
energy variance. In fact, if the variance is zero, the
trial function coincides with one of the exact eigen states
of the model Hamiltonian.
While the outlined optimization strategy 
has been 
applied quite fruitfully to a number of
systems, it cannot---in general---be used for the 
model Hamiltonian and trial functions considered in this article
since 
both the kinetic
energy and the potential energy expectation values diverge
for Hamiltonian with zero-range interactions.
For
an exact eigen state, the divergencies cancel, yielding a finite 
energy expectation value. 
For a trial function that does not fulfill the boundary conditions
imposed by the two-body zero-range interactions (see below),
the infinities
prevent one from estimating the energy expectation value reliably.
As a consequence, we optimize the ``variational parameters'' 
contained in $\psi_T$ using the PIGS approach itself
or using analytical arguments.

To motivate the functional form of our trial function $\psi_T$,
we consider the behavior of the many-body
eigen function $\psi$ when a spin-up fermion (particle $j$)
and a spin-down fermion (particle $k$) approach each other
while the other $3N-3$ coordinates,
collectively denoted by $\mathbf{Y}$,
where
$\mathbf{Y}=\{(\mathbf{r}_j+\mathbf{r}_k)/2,
\mathbf{r}_1,\cdots,\mathbf{r}_{j-1},\mathbf{r}_{j+1},\cdots,
\mathbf{r}_{k-1},\mathbf{r}_{k+1},\cdots,\mathbf{r}_N
\}$,
are kept fixed~\cite{stringariFermireview},
\begin{eqnarray}
  \psi |_{r_{jk}\to 0}\propto
  \left( \frac{1}{r_{jk}} -\frac{1}{a_s} \right) B(\mathbf{Y}).
  \label{eq_psiSRBP}
\end{eqnarray}
Here, $B$ is a function that is independent of the distance vector
$\mathbf{r}_{jk}$.
The Bethe-Peierls boundary condition,
Eq.~(\ref{eq_psiSRBP}), is a 
direct consequence of the interspecies 
two-body
zero-range interactions and holds for any up-down pair distance $r_{jk}$.
If the scattering length $a_s$ diverges, as assumed throughout 
this section,
the boundary condition reduces to 
\begin{equation}
  \psi |_{r_{jk}\to 0}\propto
  \frac{1}{r_{jk}}B(\mathbf{Y}).
  \label{eq_psiSRBP2}
\end{equation}
The trial function is constructed 
such that $\psi_T$ 
(i)  approximately fulfills the boundary condition,
Eq.~(\ref{eq_psiSRBP2});
(ii) approximately describes a state containing $N/2$ up-down pairs;
(iii) changes sign under the exchange of any two
identical fermions; 
(iv) approximately accounts for the external harmonic confinement;
and
(v) has $(L,\Pi)=(0,+1)$ symmetry.

Guideline (ii) suggests a 
term of the form
$\prod_{j=1}^{N/2}(r_{j,N/2+j})^{-1}$.
This term fulfills the boundary condition for the pairs
containing the first and $(N/2+1)$-st particle, the second and 
$(N/2+2)$-nd particle, and so on but not the boundary condition 
for pairs containing, e.g., the first and $(N/2+2)$-nd particle.
Application of the symmetrizer $\hat{\mathcal{P}}$
makes the short-distance behavior ``less ideal''.
To see this, let us exemplarily consider the $(n_1,n_2)=(2,2)$
system and act with $\hat{\mathcal{P}}$ onto $(r_{13} r_{24})^{-1}$,
\begin{eqnarray}
\hat{\mathcal{P}} \left( \frac{1}{r_{13} r_{24}} \right)
= \frac{1}{2 r_{13} r_{24}} - 
\frac{1}{2 r_{23} r_{14}}.
\label{eq_fourparticleexample}
\end{eqnarray} 
Rewriting 
the second term on the right-hand side
of Eq.~(\ref{eq_fourparticleexample})
in terms of the independent Jacobi vectors
$\mathbf{r}_{13}$, $\mathbf{r}_{24}$,
and $\mathbf{r}_{13,24}$, where
$\mathbf{r}_{13,24}=\mathbf{r}_2+\mathbf{r}_4-(\mathbf{r}_{1}+\mathbf{r}_3)$
(these coordinates correspond to one of the so-called 
H-trees~\cite{avery1989hyperspherical}),
we obtain
\begin{equation}
\hat{\mathcal{P}} \left( \frac{1}{r_{13} r_{24}} \right)=
  \frac{1}{2r_{13} r_{24}} -\frac{2}{|\mathbf{r}_{13,24}+\mathbf{r}_{13}-\mathbf{r}_{24}||\mathbf{r}_{13,24}-\mathbf{r}_{13}+\mathbf{r}_{24}|}.
  \label{eq_fourparticleexample2}
\end{equation}
It can now be seen that the right-hand side of Eq.~(\ref{eq_fourparticleexample2})
cannot be brought into the form
of Eq.~(\ref{eq_psiSRBP2}), implying that application of the symmetrizer
leads to a functional form for which neither the ``paired''
nor the ``unpaired'' interspecies distances obey
the Bethe-Peierls boundary condition.

Attempting to find a compromise between guidelines (i) and (ii),
we write $\psi_T(\mathbf{R})$ as 
\begin{eqnarray}
\psi_T(\mathbf{R})=
f_{\rm{trap}}(\mathbf{R})\Phi_{\alpha}(\mathbf{R}),
\label{eq_psitrial_ff}
\end{eqnarray}
where $f_{\rm{trap}}(\mathbf{R})$
accounts for the external confinement,
\begin{eqnarray}
f_{\rm{trap}}(\mathbf{R})=
\exp \left(-\sum_{j=1}^{N}
\frac{\mathbf{r}_j^2}{2a_{\rm{ho}}^2}\right),
\label{eq_psitrial_trap}
\end{eqnarray}
and
$\Phi_{\alpha}(\mathbf{R})$
for the correlations,
\begin{eqnarray}
  \fl
\Phi_{\alpha}(\mathbf{R})=
(a_{\rm{ho}})^{-\alpha N/2(N/2-1)-N}\hat{\mathcal{P}} \left( \frac{
    \prod_{j=1}^{N/2}\prod_{k=1,k\neq j}^{N/2} \left(r_{j,N/2+k} \right)^{\alpha}
    }{\prod_{j=1}^{N/2}r_{j,N/2+j}} \right).
\label{eq_psitrial_correlations}
\end{eqnarray}
It can be readily checked that the right hand side of 
Eq.~(\ref{eq_psitrial_correlations})
has units of $length^{-3N/2}$, as required for a $3N$-particle
wave function.
Since $\Phi_{\alpha}(\mathbf{R})$ depends only on scalars, i.e., interparticle distances,
$\psi_T(\mathbf{R})$ has the desired $(L,\Pi)=(0,+1)$ symmetry.
In Eq.~(\ref{eq_psitrial_ff}), $\alpha$ is an adjustable parameter. 
For $\alpha=0$, $\Phi_{\alpha}(\mathbf{R})$ fulfills
the Bethe-Peierls boundary condition for all up-down pairs,
provided the symmetrizer $\hat{\mathcal{P}}$ is dropped.
A finite value of $\alpha$ reduces the probability of
spin up-spin down particles 
that are ``not paired'' via the product in the denominator to be close
to each other.
We pursue
two avenues to determine the optimal $\alpha$. We use results
from the literature to fix $\alpha$,
and we determine the optimal $\alpha$ by analyzing our PIGS results.

To determine $\alpha$ using literature results, we 
rewrite the time-independent 
Schr\"odinger equation in terms
of the hyperspherical coordinates $R$ and $\mathbf{\Omega}$,
where $R$ denotes the hyperradius,
\begin{eqnarray}
R^2 = \sum_{k=1}^N \mathbf{r}_k^2,
\end{eqnarray}
and $\mathbf{\Omega}$ the $3N-1$ hyperangles.
Note that the hyperradius $R$, which
is simply given by $|\mathbf{R}|$,
is defined without separating off the
center-of-mass degrees of freedom.
For our purposes, the exact definition 
of the hyperangles is not important. 
The key ingredient for our train of thought is
that the hyperradial and hyperangular degrees
of freedom decouple when the $s$-wave scattering length
is infinitely large~\cite{castin06a}.
The eigen value of the hyperangular equation 
[we denote the hyperangular function by $\phi_{\nu}(\mathbf{\Omega})$]
is typically written in
terms of $s_{\nu}$, which---in turn---determines
the total energy of the system,
$E_{\nu q}=(2q+s_{\nu}+1) E_{\rm{ho}}$,
where $E_{\nu q}$ includes the center-of-mass energy of
$3E_{\rm{ho}}/2$, $5E_{\rm{ho}}/2, \cdots$, 
and $q$ is the hyperradial quantum number, which takes the
values
$q=0,1,\cdots$.
Writing the total wave function $\psi$
as $R^{-(3N-1)/2} F_{\nu q}(R) \phi_{\nu}(\mathbf{\Omega})$,
the hyperradial Schr\"odinger-like equation reads
\begin{equation}
  \left[ -\frac{\hbar^2}{2m}\frac{\partial^2}{\partial R^2}+V_{\rm{eff}}(R)+\frac{1}{2}
    m\omega R^2\right]F_{\nu q}(R)=E_{\nu q} F_{\nu q}(R),
  \label{<++>}
\end{equation}
where \begin{equation}
  V_{\rm{eff}}(R)=\frac{\hbar^2(s_\nu^2-1/4)}{2m R^2}.
  \label{<++>}
\end{equation}
Solving the differential equation,
one finds
\begin{equation}
  F_{\nu q}(R)=\exp \left( -\frac{R^2}{2 (a_{\rm{ho}})^2} \right) 
\frac{R^{s_{\nu}+1/2}}{(a_{\rm{ho}})^{s_{\nu}+1}}L_q^{(s_{\nu})}
\left((R^2/(a_{\rm{ho}})^2 \right),
  \label{<++>}
\end{equation}
where $L_q^{(s_{\nu})}$ denotes the associated
Laguerre polynomial and $F_{\nu q}(R)$ is normalized according to
$\int_{0}^{\infty}|F_{\nu q}(R)|^2 d R=1$.
For the ground state ($q=0$, $\nu=0$, and no center-of-mass excitations),
the hyperradial solution becomes
\begin{equation}
  F_{0 0}(R)=\exp \left( -\frac{R^2}{2 (a_{\rm{ho}})^2} \right) 
\frac{R^{s_{0}+1/2}}{(a_{\rm{ho}})^{s_0+1}}.
  \label{eq_hyperradial00}
\end{equation}
Comparing the power of $a_{\rm{ho}}$ on the right-hand side
of Eq.~(\ref{eq_psitrial_correlations}) with
the power of $a_{\rm{ho}}$ in Eq.~(\ref{eq_hyperradial00}),
we deduce
\begin{equation}
  \alpha N/2 (N/2-1) +N =s_0+1
  \label{eq_alpha1}
\end{equation}
or
\begin{equation}
\alpha
=\frac{E_{00}/E_{\rm{ho}}-N}{N/2(N/2-1)}.
  \label{eq_alpha2}
\end{equation}
Using this reasoning, $\alpha$ can be estimated
using the ground state energy of the $(N/2,N/2)$
Fermi system.
Selected $\alpha$ values, obtained using the energies reported
in Ref.~\cite{yin15}, 
are shown in column 2 of Table~\ref{tab1}.
Our first set of PIGS calculations use the value of $\alpha$
reported in Table~\ref{tab1}. In a second set of calculations,
$\alpha$ is varied and optimized using the PIGS results themselves.
We emphasize that the trial function constructed above provides
a fairly descent description of the hyperradial degree of freedom.
The hyperangular degrees of freedom are, however, less 
well described; we return to this point below.

\begin{table}[!tbp]
  \centering
  \caption{Spin-balanced two-component Fermi gas ($N/2\le5$) with
zero-range interactions at unitarity.
Column~1 lists the $(N/2,N/2)$ system considered.
Column 2 reports the value of $\alpha$, obtained from
Eq.~(\ref{eq_alpha2}) using the energies reported in Ref.~\cite{yin15}.
Columns 3-6 show the propagation time $\tau$,
the scheme used to extrapolate the
energy to $\Delta \tau=0$,
the $n$ used (the number of time slices is $2n+1$),
and the resulting extrapolated $\Delta \tau=0$ PIGS energy $E_{\rm{PIGS}}$ 
with error bars, 
respectively. The abbreviation ``extrap.'' in the header of
column~4 stands for ``extrapolation'' and ``4-th'' and ``2-nd'' are
to be read as ``4-th order'' and ``2-nd order'', respectively.
For comparison, columns 7 and 8 show energies from the literature,
obtained using the explicitly correlated Gaussian (ECG) approach~\cite{yin15}
(the energies are denoted by $E_{\rm{ECG}}$)
and the diffusion Monte Carlo method~\cite{calson14}
(the energies are denoted by $E_{\rm{DMC}}$),
respectively;
the energies $E_{\rm{ECG}}$ and $E_{\rm{DMC}}$ are obtained by
extrapolating a series of finite-range energies to the
zero-range limit. 
}
\begin{tabular} 
{c c c c c c c c}
    \hline
    \hline
     & $\alpha$ & $\tau E_{\rm{ho}}$ & extrap. & $n$ used & $E_{\rm{PIGS}}/E_{\rm{ho}}$ & $E_{\rm{ECG}}/E_{\rm{ho}}$ & $E_{\rm{DMC}}/E_{\rm{ho}}$        \\
    \hline   
    $(2,2)$ &0.505& 1& 4-th & 4, 5, 6, 8&   5.0069(29) & 5.0091(4) & 5.028(2) \\
    $(3,3)$ &0.390& 0.5& 2-nd & 3, 4& 8.353(14) & 8.337(4) & 8.377(3) \\
    $(4,4)$ &0.335& 0.5& 2-nd & 2, 4& 11.99(7) & 12.03(3) & 12.04(1) \\
    $(5,5)$ &0.306& 0.25& 2-nd & 1, 2 & 16.12(6) & 16.12(6) & 16.10(1) \\
    \hline
    \hline
  \end{tabular}
  \label{tab1}
\end{table}

\begin{figure}
\centering
\includegraphics[angle=0,width=0.4\textwidth]{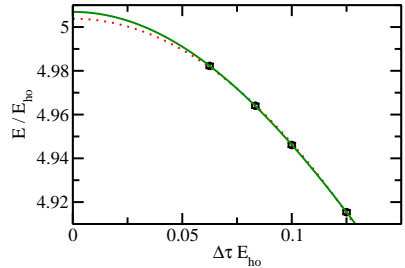}
\caption{Illustration of the dependence of the PIGS energy on
$\Delta \tau$ for the $(2,2)$ system.
The symbols with error bar show the PIGS energy
for $\tau = (E_{\rm{ho}})^{-1}$, obtained using 
the trial 
function given in Eqs.~(\ref{eq_psitrial_ff})-(\ref{eq_psitrial_correlations})
with $\alpha=0.505$.
  The solid and dotted lines show second- and fourth-order
fits to the energies (see text for details).
 }\label{figconv}
\end{figure}

Using $\alpha=0.505$ and $\tau=(E_{\rm{ho}})^{-1}$,
the symbols with error bars in Fig.~\ref{figconv}
show the PIGS energies for the
$(N/2,N/2)=(2,2)$ system for four different $\Delta \tau$.
The solid line shows a second-order fit of the form
$a+b\Delta\tau^2$.
The extrapolated $\Delta \tau=0$ energy 
is
$5.0038(12)E_{\rm{ho}}$, where the error bar in brackets 
represents the fit uncertainty, which takes
the error bars of the finite $\Delta \tau$
PIGS energies into account.
For comparison, the dotted line shows a fourth-order fit of the form
$a+b\Delta\tau^2+c \Delta \tau^4$.
The extrapolated $\Delta \tau=0$ energy 
is
$5.0069(29)  E_{\rm{ho}}$, where the error bar in brackets 
represents---as before---the fit uncertainty, which takes
the error bars of the finite $\Delta \tau$
PIGS energies into account.
The fact that the extrapolated second- and fourth-order energies agree
within error bars suggests that the extrapolated energies are reasonably good.
This is confirmed by comparing with 
the highly-accurate energy 
$E_{\rm{ECG}}$ obtained via a basis set expansion approach, which
employs explicitly correlated Gaussians (see column~6 of Table~\ref{tab1}). 
Assuming, for a moment, that the $\tau=(E_{\rm{ho}})^{-1}$
result is identical to that for the $\tau \rightarrow \infty$ limit, 
we can estimate the systematic uncertainties of the 
second- and fourth-order extrapolations.
The fourth-order energy agrees with $E_{\rm{ECG}}$ 
(see Table~\ref{tab1}) within error bars
while the second-order energy deviates by about four standard deviations.
We thus estimate that the systematic error that
originates from the second-order fit 
is of the order of
0.1\%.
This suggests that one needs to use the fourth- or 
an even higher-order extrapolation
scheme or perform additional calculations for smaller $\Delta \tau$
if the statistical error is of the order of 0.1\% or smaller.

\begin{figure}
\centering
\includegraphics[angle=0,width=0.4\textwidth]{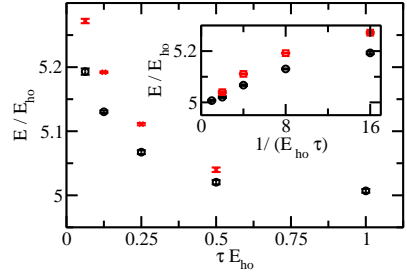}
\caption{Illustration of the dependence of the PIGS energy on 
the propagation time $\tau$
for the $(2,2)$ system.
  The circles and crosses 
show the extrapolated $\Delta \tau=0$ energies with error bars
(the fourth-order scheme is used), obtained using 
the trial function given in Eqs.~(\ref{eq_psitrial_ff})-(\ref{eq_psitrial_correlations})
with $\alpha=0.505$ and $1$, respectively.
The inset replots the energies
as a function of $1/\tau$.
 }\label{figtrial}
\end{figure}

To analyze how close the $\tau=(E_{\rm{ho}})^{-1}$ 
energy is to the $\tau \rightarrow \infty$ limit,
circles with error bars in Fig.~\ref{figtrial}
show the extrapolated energy,
using the fourth-order scheme, of the $(2,2)$ system
for various $\tau$; as before, $\alpha$ is set to $0.505$.
As expected, the energy decreases with increasing $\tau$
and flattens out for large $\tau$.
To better show how the energy behaves with increasing $\tau$,
the inset replots the extrapolated $\Delta \tau=0$ energies
with error bars 
as a function of $1/ \tau$.
It can be seen that the energies for the 
two smallest $1/\tau$ (two largest $\tau$)
do not agree within error bars. This means that,
strictly speaking, the large $\tau$ limit has not yet been reached.
However, as discussed further below, going to larger $\tau$ is 
rather challenging because of the Fermi sign problem. 
This implies that, ultimately,
the accuracy of the PIGS energy is limited, as already alluded to above,
by the systematic error that originates from not going to the
$\tau \to \infty$ limit and not by the statistical error bars.
For the $N \ge 6$ systems, the computational time is chosen such
that the statistical error is of the order of the estimated
systematic error; the reasoning behind this is
that a smaller statistical error would not allow one to gain more
insight into the exact value of the energy.

For comparison, the crosses with error bars in the main part of 
Fig.~\ref{figtrial}
show the extrapolated $(2,2)$ PIGS energies for $\alpha=1$.
These energies lie above those for $\alpha=0.505$
for all $\tau$, reflecting the fact that the
trial function with $\alpha=0.505$ 
provides a better description
of the $(2,2)$ system than the 
trial function with $\alpha=1$.
The difference between the PIGS energies for the calculations
with the two different $\alpha$ values
decreases with increasing $\tau$, reflecting the fact that,
in principle, any trial function that has finite
overlap with the exact ground state wave function could be used.
However, the better $\psi_T$, the smaller the resulting 
error bars.

\begin{figure}
\centering
\includegraphics[angle=0,width=0.4\textwidth]{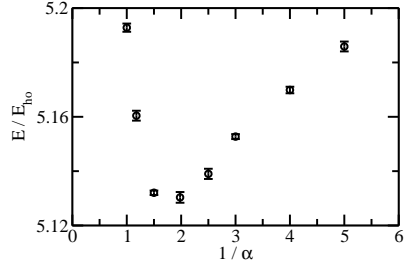}
\caption{
Illustration
of the dependence of the PIGS energy 
for the $(2,2)$ system on the trial function parameter $\alpha$.
  The symbols
show the extrapolated $\Delta \tau=0$ energies
with error bars 
(the fourth-order scheme is used) for $\tau=0.125 (E_{\rm{ho}})^{-1}$
as a function of $1/\alpha$, obtained using 
the trial function given in Eqs.~(\ref{eq_psitrial_ff})-(\ref{eq_psitrial_correlations}).
 }\label{figvaryalpha}
\end{figure}

To more systematically investigate the dependence of the 
PIGS energy, and correspondingly the speed of the convergence with
increasing $\tau$, on $\alpha$, we fix $\tau$ at $0.125(E_{\rm{ho}})^{-1}$.
These small $\tau$ calculations are computationally comparatively 
inexpensive and hence allow one to
survey the $\alpha$ dependence more exhaustively. 
Ultimately, one needs, of course, to go to larger $\tau$.
However, to find the best $\alpha$ (or more generally, the best
trial function), it is often times sufficient to
consider a relatively small $\tau$.
Figure~\ref{figvaryalpha} shows the PIGS energy with error
bars as a function of $\alpha^{-1}$ for $\tau=0.125 (E_{\rm{ho}})^{-1}$.
The lowest energy is obtained for $\alpha$ around $0.5$,
confirming our choice of $\alpha$ based on the hyperspherical
coordinate approach.
This suggests that the optimal 
$\alpha$ could alternatively be determined iteratively.
To this end,
let us assume that the ground state energy is unknown.
One would then chose an initial value of $\alpha$
to obtain a first PIGS energy estimate for small $\tau$.
Using this (non-converged)
PIGS energy, one would obtain an improved 
$\alpha$ value using Eq.~(\ref{eq_alpha2}) and perform another 
PIGS calculation. After a few iterations, the optimal
$\alpha$ value would be found.

\begin{figure}
\centering
\includegraphics[angle=0,width=0.4\textwidth]{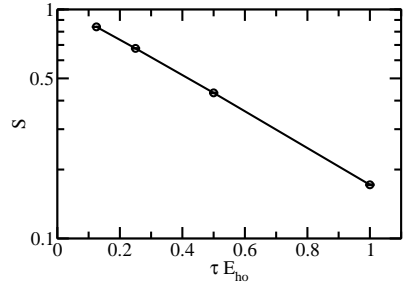}
\caption{Symbols show the quantity
   $S$,
Eq.~(\ref{eq_sfactor}), as a function of $\tau$ for the $(2,2)$ system.
The error bars are smaller than the symbol size.
The 
   time step $\Delta\tau$ is fixed at $0.125 (E_{\rm{ho}})^{-1}$.
As a guide to the eye, the solid line connects consecutive
data points.
 }\label{figsign}
\end{figure}

As already alluded to, the Fermi sign problem limits
the maximum propagation time $\tau$ that can be reached with
a finite amount of computational resources.
The symbols in Fig.~\ref{figsign} show the quantity $S$
[see Eq.~(\ref{eq_sfactor})],
which appears in the denominator of
the expression for all observables,
as a function of $\tau$ for the $(2,2)$ 
system for the trial function with $\alpha=0.505$.
For this series of calculations, $\Delta \tau$ is
fixed at $\Delta \tau = 0.125 (E_{\rm{ho}})^{-1}$,
i.e., the number of time slices increases with increasing
$\tau$.
The solid line shows a fit to the data, demonstrating that
$S$ decreases exponentially with increasing $\tau$ or,
equivalently, increasing number of time slices.
An $S$ value close to 1 indicates that the sign problem is irrelevant.
The smaller $S$, the more severe the sign problem becomes.
As a consequence, for a given $\tau$, there exists a maximum
$\Delta \tau$ for which the calculation is feasible.
For smaller $\Delta \tau$, the errors that originate from the
sign problem are too large to be useful.
For large $\tau$, the smallest $\Delta \tau$
that can be treated reliably might not be sufficiently small to
allow for a reliable extrapolation to $\Delta \tau=0$.
For the $(2,2)$ system, e.g., $\tau=(E_{\rm{ho}})^{-1}$
is a good compromise. The excited state contributions have,
essentially, decayed and the extrapolation to the $\Delta \tau=0$ limit 
is reliable.

In addition to the $(2,2)$ system, we treat the $(3,3)$, $(4,4)$,
and $(5,5)$ systems 
using the
$\alpha$ values determined from the known ground state
energies via Eq.~(\ref{eq_alpha2}).
The value of $\tau$ (see column~3 of 
Table~\ref{tab1}) is chosen such that the estimated systematic error,
due to the use of a finite $\tau$, is
comparable to or smaller than the error of the extrapolated $\Delta \tau=0$
energy for this $\tau$ (see columns~4-6 of Table~\ref{tab1}).
As $N$ goes up, the propagation time $\tau$ is chosen to be smaller and smaller
(see Table~\ref{tab1}).
The reason is that the simulations for larger $N$
are more computationally demanding since the Fermi sign problem
becomes more severe with increasing $N$. This means that we are 
limited by the number of time slices and, correspondingly, 
the largest $\tau$ we can use. 
The number of time slices $2n+1$ considered in the $\Delta \tau \to 0$
extrapolation are chosen based on our
detailed analysis of the $(2,2)$ system.
Because of the relatively small $\tau$ considered, the 
energies reported for the $N/2=4$ and 5 systems should be regarded,
within error bars, as variational upper bounds.
Our $(3,3)$ energy agrees, within error bars, with the basis set expansion energy
$E_{\rm{ECG}}$ but lies slightly below the diffusion Monte Carlo energy 
$E_{\rm{DMC}}$.
For the $(4,4)$ and $(5,5)$ systems,
the PIGS energies agree, within error bars, with $E_{\rm{ECG}}$
and $E_{\rm{DMC}}$.
For the $(5,5)$ system, we performed an additional calculation 
using---as before---$\tau=0.125 (E_{\rm{ho}})^{-1}$
but using an $\alpha$ value that is
larger than that listed in Table~\ref{tab1},
namely $\alpha=0.335$. 
The extrapolated $\Delta \tau=0$ PIGS energy is $16.22(14)E_{\rm{ho}}$,
which agrees within error bars with our result listed in Table~\ref{tab1}. 
The larger error bar reflects the fact that the larger $\alpha$ 
value provides a less good trial function.

The trial function used so far 
[see Eqs.~(\ref{eq_psitrial_ff})-(\ref{eq_psitrial_correlations})]
contains a single adjustable parameter, namely $\alpha$, that 
primarily determines the correlations in the hyperradial degree of freedom.
Our goal is now to design a trial function that provides
an improved description of the hyperangular degrees of freedom.
In doing so, we are guided by the analytically known wave function of the
harmonically-trapped $(2,1)$ system 
with $(L,\Pi)=(0,+1)$ symmetry at 
unitarity~\cite{castin06l}.
The hyperangular part of the
wave function that yields the lowest energy with $(0,+1)$
symmetry is proportional to 
\begin{equation}
(1-\hat{P}_{12})\sin(\bar{s}_0(\theta_{1}-\pi/2))/\sin(2\theta_1),
  \label{eq_21exacteq}
\end{equation}
where $\theta_{1}=\arcsin(r_{13}/(2\sqrt{\bar{R}}))$, 
$\bar{s}_0=2.166$,
and
$\bar{R}^2=\sum_{j<k}r_{jk}^2/N$.
Application of $\hat{P}_{12}$ changes $\theta_1$ into $\theta_2$, where
$\theta_{2}=\arcsin(r_{23}/(2\sqrt{\bar{R}}))$.
The factor $\sin(\bar{s}_0(\theta_{1}-\pi/2))$ enhances the probability to find
two particles at vanishing hyperangle $\theta_{1}$, i.e., at
vanishing distance between the unlike particles 1 and 3.
In the non-interacting limit,
$\bar{s}_0$ is equal to $4$, which implies that
the probability to find two 
particles at vanishing hyperangle $\theta_{1}$ vanishes.
The quantities $\bar{R}$ and $\bar{s}_0$, which are defined 
by excluding the
center-of-mass degrees of freedom, are related to
$R$ and $s_0$, $\bar{R}^2=R^2 - N(R_{\rm{cm}})^2$
and 
$\bar{s}_0=s_0-3/2$.

Motivated by the hyperangular wave function of the $(2,1)$ system
with $(0,+1)$ symmetry
at unitarity, we consider the following alternative form of the trial function,
\begin{equation}
  \psi_T=
f_{\rm{trap}}(\mathbf{R}) \Phi_{\alpha,\beta,\gamma}(\mathbf{R}),
\end{equation}
where
\begin{equation}
\Phi_{\alpha,\beta,\gamma}(\mathbf{R})=
\frac{\bar{R}^\alpha}{(a_{\rm{ho}})^{\alpha+3N/2}}\hat{\mathcal{P}}
  \prod_{j=1}^{N/2}\frac{\sin[\beta(\theta_j-\pi/2)](\cos\theta_j)^{\gamma+1}}
{\sin(2 \theta_j)}
  \label{eq_psitrial_correlations2}
\end{equation}
and
$\theta_j=\arcsin(r_{j,N/2+j}/(2 \sqrt{\bar{R}}))$.
The factor $(\cos \theta_j)^{\gamma+1}$ is introduced to increase the tunability
of the trial function.
As before, the optimal value of $\alpha$ is obtained by matching to the known hyperradial
solution. This yields $\alpha=0.166$. 
The values of $\beta$ and $\gamma$, in contrast, are determined
by performing PIGS simulations for small $\tau$.

Considering the $(2,2)$ system
and using $\alpha=2.5091$, $\beta=1.1$, and $\gamma=2$, we obtain 
the extrapolated $\Delta \tau=0$ PIGS energy of $5.022(3)E_{\rm{ho}}$ 
for $\tau=0.0625 (E_{\rm{ho}})^{-1}$.
This energy is significantly lower than the 
PIGS energy of $5.193(4)E_{\rm{ho}}$ that we obtained using the correlation
factor $\Phi_{\alpha}(\mathbf{R})$ for the same $\tau$.
For $\tau=0.25 (E_{\rm{ho}})^{-1}$, we obtain 
$5.011(8)E_{\rm{ho}}$, which agrees  within error bars with the
ground state energy $E_{\rm{ECG}}$ (see Table~\ref{tab1}).
Considering the $(3,3)$ system
and using the trial parameters $\alpha=5.837$, $\beta=1.5$, and $\gamma=8$,
we obtain the extrapolated $\Delta \tau=0$ PIGS energy of $8.357(8)E_{\rm{ho}}$ 
for $\tau=0.25 (E_{\rm{ho}})^{-1}$.
For comparison, the $(3,3)$ PIGS energy reported in Table~\ref{tab1}
is for a larger $\tau$, namely $\tau = 0.5 (E_{\rm{ho}})^{-1}$.
For the $(4,4)$ and $(5,5)$ systems, the trial function with
the correlation factor $\Phi_{\alpha,\beta,\gamma}$ did not
yield an improved energy compared to that for $\Phi_{\alpha}$.  
The reason could be that the trial parameters were not fully
optimized or that the degrees of freedom of
the larger systems are less
well described by the trial function
(e.g., that three- and higher-body correlations are needed).

In addition to the energies, we use the PIGS approach to calculate structural 
properties.
The scaled pair distribution functions reported below are obtained for
a finite $\Delta \tau$; no extrapolation to the $\Delta \tau=0$ limit was performed.
To determine a suitable $\Delta \tau$, we consider the $(2,2)$
system
and perform calculations for $\tau =0.25 (E_{\rm{ho}})^{-1}$
using three different $\Delta \tau$, i.e., 
$\Delta\tau = 0.25(E_{\rm{ho}})^{-1}$, $0.125(E_{\rm{ho}})^{-1}$, 
and $0.0625(E_{\rm{ho}})^{-1}$.
We find that 
the scaled
pair distribution function for $\Delta\tau =0.25(E_{\rm{ho}})^{-1}$ 
differs slightly from those for
$\Delta\tau =0.125 (E_{\rm{ho}})^{-1}$ and $\Delta\tau=0.0625(E_{\rm{ho}})^{-1}$.
However, no visual difference is observed between the 
scaled pair distribution
functions for 
$\Delta\tau =0.125(E_{\rm{ho}})^{-1}$ and $\Delta\tau =0.0625(E_{\rm{ho}})^{-1}$.
Motivated by this observation, we calculate the 
scaled pair distribution functions
for the spin-balanced systems with $N/2 \le 5$ using 
$\Delta\tau =0.125(E_{\rm{ho}})^{-1}$.
The propagation times and trial functions are the same as
those used to obtain the energies reported in Table~\ref{tab1}.

\begin{figure}
\centering
\includegraphics[angle=0,width=0.4\textwidth]{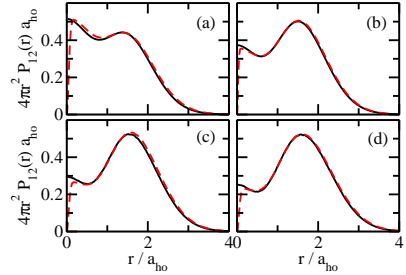}
\caption{
Scaled pair distribution functions $4 \pi r^2 P_{12}(r)$ for
the
(a) $(2,2)$,
(b) $(3,3)$,
(c) $(4,4)$, and
(d) $(5,5)$ systems at unitarity.
  The solid lines show our PIGS results for the model
Hamiltonian with two-body zero-range interactions.
The results are obtained using the propagation times and trial function
parameters listed in Table~\ref{tab1}.
The imaginary time step is set to $\Delta \tau = 0.125 (E_{\rm{ho}})^{-1}$.
  For comparison, the dashed lines show basis set expansion
based results~\cite{yin15} for an attractive two-body Gaussian interaction
with finite effective range (see text for details).
 }\label{figpair}
\end{figure}

Solid lines in Figs.~\ref{figpair}(a)-\ref{figpair}(d)
show the resulting scaled pair distribution function for the
spin-balanced systems with $N=4-10$. 
For comparison, the dashed lines show the scaled pair distribution function
obtained from basis set
calculations for an attractive two-body Gaussian potential with
infinitely large $s$-wave scattering length and
effective range of approximately $0.12a_{\rm{ho}}$
(the range is $0.06 a_{\rm{ho}}$)~\cite{yin15}. 
The Gaussian potential used supports exactly one zero-energy 
two-body bound state in free space.
If the effective range were taken to zero, the two different approaches
should yield the same result.
As can be seen from Fig.~\ref{figpair},
the zero-range results deviate a bit from the finite-range results
at small interparticle distances; in particular, the scaled pair distribution
functions take a finite value for vanishing interparticle distance
if the zero-range interaction model is used and
go to zero if the finite-range interaction model is used.
At larger interparticle distances, the agreement between the dashed and solid lines 
is quite good, suggesting that the PIGS 
scaled pair distribution functions are,
indeed, quite well converged for $r \gtrsim 0.5a_{\rm{ho}}$.

To check the convergence at small interparticle distances, we 
report the contact $C$ obtained from the $r=0$ value of the scaled pair
distribution function~\cite{tan1},
\begin{eqnarray}
C=4\pi n_1 n_2 \lim_{r\to 0} 4\pi r^2 P_{12}(r).
\label{eq_contact_pair}
\end{eqnarray}
The resulting contacts are summarized in Table~\ref{tabcontact}.
The contact calculated by the PIGS approach is, for all $N$
considered, larger than the contact calculated by extrapolating
basis set expansion results 
(namely, the slope of the energy with respect
to the inverse of the $s$-wave
scattering length) for finite-range interactions
to the zero-range limit.
As $N/2$ increases from $2$ to $5$,
the difference between the two sets of results
increases from about two to
about six standard deviations.
The numerical determination of the contact through the
scaled pair distribution function is, in general, quite
challenging since the contact probes a small portion
of the Hilbert space. As a consequence, the convergence of the contact
can be slow.
It is presently unclear which set of results is more reliable 
and if, possibly,
the errorbars of the results obtained by
either of the two methods was underestimated: 
the basis set expansion results
may be ``contaminated'' by 
basis set extrapolation and zero-range extrapolation
errors while the PIGS results may be ``contaminated''
by finite propagation time and finite time step errors.
\begin{table}[!tbp]
  \centering
  \caption{
    Contact for spin-balanced two-component Fermi gas ($N/2\le5$) with
    zero-range interactions at unitarity.
    Column 2 reports the value of the 
contact $C_{\rm{PIGS}}$ obtained from our
PIGS simulations with $\Delta\tau =0.125(E_{\rm{ho}})^{-1}$.
    The error reported does not include the systematic error introduced by
    not extrapolating to zero $\Delta\tau$.
    Column 3 shows the contact $C_{\rm{ECG}}$ from the literature, 
obtained by extrapolating the contact obtained by the ECG
    approach for finite-range
interactions to the zero-range limit~\cite{yin15}.
  }
\begin{tabular} 
{c c c}
    \hline
    \hline
    $(n_1,n_2)$ & $C_{\rm{PIGS}} a_{\rm{ho}}$& $C_{\rm{ECG}}a_{\rm{ho}}$ \\
    \hline
    (2,2) & 25.91(6) & 25.74(1) \\
    (3,3) & 42.26(5) & 40.39(8) \\
    (4,4) & 59.5(2) & 55.4(5) \\
    (5,5) & 79.3(3) & 72.3(8) \\
    \hline
    \hline
  \end{tabular}
  \label{tabcontact}
\end{table}

\subsection{Non-interacting Fermi gas with a single impurity}
\label{sec_application_fermiimbalanced}
This section considers a non-interacting Fermi
gas with a single impurity, i.e., the $(n_1,n_2)=(N-1,1)$ system
with $N=3-5$.
In a zeroth-order approximation, one can think of this system as consisting 
of one up-down pair and $N-2$ unpaired spin-up atoms.
Of course, this picture needs to be refined to account for the
fact that the system contains $N-1$ identical fermions
and for the fact that the presence of additional spin-up fermions
``disturbs'' the single pair.
Nevertheless, the simple picture correctly
suggests that the ground state of 
the $(N-1,1)$ system does not have
$(L,\Pi)=(0,+1)$ symmetry.
Rather,
the ground state of the $(2,1)$, $(3,1)$ and $(4,1)$
systems has $(L,\Pi)=(1,-1)$, $(1,+1)$ and $(0,-1)$
symmetry~\cite{castin06l,kevinenergy_spectrum_p7_10,rakshit12}.
Roughly, this can be understood by realizing that the 
$(2,1)$, $(3,1)$ and $(4,1)$
systems contain one, two and three unpaired
spin-up atoms, each of which carry one quantum of angular momentum
(the $p$-shell is being filled).
Thus, due to the angular momentum
carried by the single unpaired spin-up atom
the $(2,1)$ system has $(1,-1)$ symmetry. 
In the $(3,1)$ system, the angular momenta of each of
the two unpaired atoms can couple to an angular momentum
$0$, $1$, or $2$, with the parity being even. The $(1,+1)$
channel turns out to have the lowest energy~\cite{rakshit12}.
Last, in the $(4,1)$ system, the angular momenta of each of
the three unpaired atoms can couple to an angular momentum
$0$, $1$, $2$, or $3$, with the parity being odd. Consistent with the idea
of a closed shell configuration, the $(0,-1)$
channel turns out to have the lowest energy.

As our first application of the PIGS approach
to spin-imbalanced systems,
we treat the $(2,1)$ system with $(L,\Pi)=(0,+1)$ symmetry.
This application illustrates that the PIGS approach can be used 
to describe the energetically lowest-lying state
(not the ``absolute ground state'')
of a given symmetry.
Following the logic that motivated the trial function given in
Eqs.~(\ref{eq_psitrial_ff})-(\ref{eq_psitrial_correlations}),
we write
\begin{eqnarray}
  \psi_T(\mathbf{R})= f_{\rm{trap}}(\mathbf{R})
\Phi_{\alpha'}(\mathbf{R}),
\label{eq_trialspinimba}
\end{eqnarray}
where
\begin{eqnarray}
\Phi_{\alpha'}(\mathbf{R})=
(a_{\rm{ho}})^{-\alpha'(N-2)-3N/2+1}
\hat{\mathcal{P}}
\left( 
\frac{\prod_{j=2}^{N-1}(r_{j,N})^{\alpha'}}
{r_{1,N}}
\right).
\end{eqnarray}
Using a relatively small $\tau$, namely $\tau=0.25 (E_{\rm{ho}})^{-1}$,
Table~\ref{tab21} shows the extrapolated $\Delta \tau=0$ energy. 
It lies about six sigma above the exact energy
obtained within the hyperspherical coordinate 
approach~\cite{castin06l}.
If we repeated the calculation for larger $\tau$,
we would expect to obtain a
PIGS energy closer to the exact energy.
To prove that the lowest $(0,+1)$ energy can be obtained 
exactly within the
PIGS approach, i.e., to prove that the PIGS approach does, indeed, preserve the
symmetry of the trial function, we use the exact (analytically known)
eigen state~\cite{castin06l} as the trial function.
The resulting extrapolated $\Delta \tau=0$ energy
(see Table~\ref{tab21})
agrees to within error bars with the exact energy.

As a proof-of-principle, we apply the PIGS
approach to the 
$(2,1)$ system with $(L,\Pi)=(1,-1)$ symmetry at unitarity.
Knowing that the orbital angular momentum is carried by
the Jacobi vector $\mathbf{r}_{13,2}$ when
particles 1 and 3 form a pair and by the
Jacobi vector $\mathbf{r}_{23,1}$ when particles 2 and 3 form a 
pair~\cite{castin06l,kevinenergy_spectrum_p7_10},
we write
\begin{eqnarray}
\psi_T(\mathbf{R}) = (a_{\rm{ho}})^{-9/2} 
f_{\rm{trap}}(\mathbf{R})
\left( \frac{\bar{R}}{a_{\rm{ho}}} \right)^{\alpha}
{\hat{\cal{P}}}
\left(
\frac{x_{13,2}}{r_{13,2}}
\right).
\label{eq_trial_21system_oneminus}
\end{eqnarray}
Owing to the three-fold degeneracy of $L=1$ states,
alternatively one can use $y_{13,2}$ or
$z_{13,2}$ instead of
$x_{13,2}$.
In an equivalent formulation,
the term $x_{13,2}/r_{13,2}$
in Eq.~(\ref{eq_trial_21system_oneminus}) is
replaced by $Y_{1,m_l}(\hat{\mathbf{r}}_{13,2})$,
where $Y_{1,m_l}$ denotes the spherical harmonic and $m_l$ can take the values
$\pm1$ and $0$. Since the spherical harmonics with $m_l \ne 0$
are complex, spherical harmonics
are less convenient from a numerical/implementation
perspective than the real version
used in Eq.~(\ref{eq_trial_21system_oneminus}).
Using Eq.~(\ref{eq_trial_21system_oneminus})
with $\tau=0.5 (E_{\rm{ho}})^{-1}$,
we find the extrapolated $\Delta \tau=0$ energy
$E_{\rm{PIGS}}=4.276(9)E_{\rm{ho}}$, which agrees,
within error bars, with the exact zero-range energy from Ref.~\cite{castin06l}
(see row 3 of Table~\ref{tab21}).

The above trial function can be extended to the $(3,1)$
system. Assuming the formation of a pair consisting
of atoms 1 and 4, one quantum of orbital
angular momentum each is assumed to be carried by
the vectors $\mathbf{r}_{14,2}$ and $\mathbf{r}_{14,3}$.
Coupling $Y_{1,m_1}(\mathbf{r}_{14,2})$
and $Y_{1,m_2}(\mathbf{r}_{14,3})$ such that the resulting function
has $(L,\Pi)=(1,-1)$ symmetry, we obtain the desired
correlation factor.
Since we prefer to work with real quantities,
we write
\begin{eqnarray}
  \psi_T(\mathbf{R})=
  (a_{\rm{ho}})^{-6} f_{\rm{trap}}(\mathbf{R})
  \left( \frac{\bar{R}}{a_{\rm{ho}}} \right)^{\alpha}
  \hat{\mathcal{P}} 
\left(
  \frac{( \mathbf{r}_{14,2} \times \mathbf{r}_{14,3}) \cdot {\hat{\mathbf{z}}}}
       {r_{14,2} r_{14,3}}
  \right),
  \label{eq_trial31}
  \end{eqnarray}
where the dot product serves to select the $z$-component of the vector
that results when taking the cross product.
Instead of the $z$-component, the $x$- or $y$-components can
be used.
For the $(4,1)$ system, we
use
\begin{eqnarray}
  \fl
  \psi_T(\mathbf{R})=
  (a_{\rm{ho}})^{-15/2} f_{\rm{trap}}(\mathbf{R})
  \left( \frac{\bar{R}}{a_{\rm{ho}}} \right)^{\alpha}
  \hat{\mathcal{P}} 
\left(
  \frac{( \mathbf{r}_{15,2} \times \mathbf{r}_{15,3}) \cdot {\mathbf{r}_{15,4}}}
       {r_{15,2} r_{15,3} r_{15,4}}
  \right),
  \label{eq_trial411}
\end{eqnarray}
which has the desired $(L,\Pi)=(0,-1)$ symmetry.
Alternatively, one could use 
\begin{eqnarray}
  \psi_T(\mathbf{R})=
  (a_{\rm{ho}})^{-21/2} f_{\rm{trap}}(\mathbf{R})
  \left( \frac{\bar{R}}{a_{\rm{ho}}} \right)^{\alpha}
  \hat{\mathcal{P}} 
\left(
  ( \mathbf{r}_{15,2} \times \mathbf{r}_{15,3}) \cdot {\mathbf{r}_{15,4}}
  \right).
  \label{eq_trial412}
\end{eqnarray}
The resulting extrapolated $\Delta \tau=0$ energies
for the $(3,1)$ and $(4,1)$ systems are
reported in Table~\ref{tabn1}.
The $(3,1)$ PIGS energy deviates by two sigma from the highly-accurate
basis set expansion energy $E_{\rm{ECG}}$.
The small disagreement may be due to the fact that $\tau$ is not quite
large enough or that the error bar of the extrapolated
energy is, in fact, slightly larger than
what is reported in Table~\ref{tabn1}.
The $(4,1)$ PIGS energies for $\tau=0.5 (E_{\rm{ho}})^{-1}$
and $\tau= (E_{\rm{ho}})^{-1}$ agree, within error bars, with the energy
$E_{\rm{DMC}}$.
Since the diffusion Monte Carlo energy was not extrapolated to the zero-range limit, the
true zero-range energy is probably somewhat smaller than $E_{\rm{DMC}}$.

\begin{table}[!tbp]
  \centering
  \caption{Spin-imbalanced fermionic $(2,1)$ system with
zero-range interactions at unitarity.
Columns 2, 3, and 4 report the propagation
time, the $n$ used (the number of time slices is 2$n$+1),
and the trial function
employed for two different symmetries (see column 1).
Column 5 shows the resulting PIGS
energy with error bars. For the $(0,+1)$ channel, 
no extrapolation to the $\Delta \tau =0$ limit was performed.
For the $(1,-1)$ channel,
a 
second-order extrapolation was used. 
For comparison, column 6 shows the exact 
zero-range energies obtained using the formalism
developed in Ref.~\cite{castin06l}.
  }
\begin{tabular} 
{c c c c c c c}
    \hline
    \hline
    $(L,\Pi)$ & $\tau E_{\rm{ho}}$ & $n$ used & trial function &
$E_{\rm{PIGS}}/E_{\rm{ho}}$  &
    $E_{\rm{exact}}/E_{\rm{ho}}$\\
    \hline   
     $(0,+1)$& 0.25&4 & Eq.~(\ref{eq_trialspinimba}), $\alpha=1.66622$  &  4.687(4) & 4.66622  \\ 
     $(0,+1)$& 0.25&4 & Eq.~(\ref{eq_21exacteq})&   4.676(10)  & 4.66622\\ 
    $(1,-1)$& 0.5&2,4 & Eq.~(\ref{eq_trial_21system_oneminus}), $\alpha=0.772724$ & 4.276(9)  & 4.27272\\
    \hline
    \hline
  \end{tabular}
  \label{tab21}
\end{table}

\begin{table}[!tbp]
  \centering
  \caption{Spin-imbalanced fermionic $(N-1,1)$ system ($N=4$ and $5$)
with zero-range
interactions at unitarity.
Column~1 lists the $(N-1,1)$ system considered.
The symmetry of the ground state is reported in column~2.
Columns 3, 4, 5, and 6 report the propagation
time,  the $n$ used
(2$n$+1 is the number of time slices included in the 
extrapolation of the energy to $\Delta \tau=0$), the
value of $\alpha$, and the equation number of the trial function
used.
Column 7 shows the resulting extrapolated $\Delta \tau=0$ PIGS
energy with error bars, obtained using a
second-order extrapolation. 
For comparison,
column 8 reports energies
obtained using the explicitly 
correlated Gaussian approach~\cite{blume10unequal,rakshit12};
the energies, which are denoted by $E_{\rm{ECG}}$, 
are obtained by extrapolating a series of finite-range
energies to the zero-range limit.
Column 9 reports the $(4,1)$ energy $E_{\rm{DMC}}$ 
for a square-well potential with range $r_0=0.01 a_{\rm{ho}}$
obtained
using
the diffusion Monte Carlo
method~\cite{blume08trappedgas}; no 
extrapolation to the zero-range limit was done.
The uncertainties for the ECG and DMC calculations are,
according to Refs.~\cite{blume10unequal,rakshit12,blume08trappedgas},
in
the
last digit reported.
  }
\begin{tabular} 
{c c c c c c c c c}
    \hline
    \hline
     & $(L,\Pi)$ & $\tau E_{\rm{ho}}$ & $n$  &  $\alpha$ & $\psi_T$ &$E_{\rm{PIGS}}/E_{\rm{ho}}$ & $E_{\rm{ECG}}/E_{\rm{ho}}$ & $E_{\rm{DMC}}/E_{\rm{ho}}$        \\
    \hline   
    $(3,1)$ &$(1,+1)$& 0.5 &4, 8&  0.791& (\ref{eq_trial31})& 6.60(1) &  6.5819 & \\
    $(4,1)$ &$(0,-1)$& 0.5 &2, 4&  0.667& (\ref{eq_trial411})& 8.93(7) & 8.95 & 8.93 \\
    $(4,1)$ &$(0,-1)$& 1   &4, 8 & 0& (\ref{eq_trial412})& 8.92(4) & 8.95 & 8.93 \\
    \hline
    \hline
  \end{tabular}
  \label{tabn1}
\end{table}

\begin{figure}
\centering
\includegraphics[angle=0,width=0.4\textwidth]{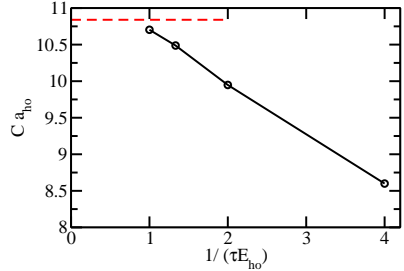}
\caption{Symbols show the
  contact $C$ of the $(3,1)$
  system as a function of $\tau^{-1}$. The 
  time step $\Delta\tau$ is fixed at $0.125(E_{\rm{ho}})^{-1}$.
As a guide to the eye, the symbols are connected by a solid line.
  For comparison, the dashed horizontal line shows the
  contact $C=10.84(2)(a_{\rm{ho}})^{-1}$, obtained by extrapolating
basis set expansion
  results for finite-range interactions to the zero-range limit.
 }\label{figcontactimpurity}
\end{figure}

In addition to the energy, we determine the contact
$C$ from the $r=0$ behavior of the scaled
pair distribution function [see Eq.~(\ref{eq_contact_pair})].
In general, the convergence
rate of the energy and that of other observables can be different.
For the case at hand, namely the contact,
this can be understood by realizing that only a small
fraction of the wave function amplitude is located at small
$r$. Thus, while the energies shown in Table~\ref{tabn1}  appear to be
converged, the contact may not be.
Indeed, this is what we find.
Figure~\ref{figcontactimpurity}
shows the contact of the $(3,1)$
system as a function of $\tau^{-1}$.
The calculations are performed for 
$\Delta \tau =0.125(E_{\rm{ho}})^{-1}$.
The contacts for the two largest $\tau$ values considered
differ by roughly 2\%. For even larger $\tau$, the curve
should flatten out. Thus, we can interpret our calculations as providing
a lower
bound on $C$, $C \ge 10.7(a_{\rm{ho}})^{-1}$.
Indeed, calculations that employ a correlated Gaussian basis
set yield a contact of $C=10.84(2) (a_{\rm{ho}})^{-1}$
(see the dashed horizontal line in Fig.~\ref{figcontactimpurity}),
which is close to the PIGS contact for the largest $\tau$ considered.
For $\tau =0.5 (E_{\rm{ho}})^{-1}$, 
we performed 
an additional calculation for a smaller $\Delta \tau$, i.e., for 
$\Delta \tau =0.0625 (E_{\rm{ho}})^{-1}$.
The results
for $\Delta\tau =0.125 (E_{\rm{ho}})^{-1}$
and $\Delta \tau = 0.0625 (E_{\rm{ho}})^{-1}$
differ by 0.4\%, which is small compared to the error introduced
by not extrapolating to the 
$\tau = \infty$ limit.

\section{Summary and outlook}
\label{sec_summary}
This article provided a detailed introduction to
the path integral ground state Monte Carlo
(PIGS) algorithm. Implementation details
and convergence properties were discussed using
general arguments and subsequently illustrated
for selected systems and observables.
While the primary focus was on the PIGS 
approach, several aspects are applicable more
broadly. 
For example, the discussion of the error analysis 
is relevant to essentially all Monte Carlo 
algorithms and many features of the path
generation apply also to the finite-temperature
path integral Monte Carlo approach.

The PIGS algorithm takes a trial function, which
is selected by the simulator, and propagates it in imaginary time. For
sufficiently large imaginary time $\tau$, the 
lowest eigen state of the system Hamiltonian, which
has finite overlap with the trial function,
is being projected out. A crucial question is how to assess
whether the resulting eigen energy
and other observables are truly converged.
This question is particularly pressing for fermionic
systems, for which the largest $\tau$ considered is
restricted by numerical instabilities due to the Fermi sign
problem. In the absence of identical fermions, the convergence analysis is
relatively simple since there are essentially no
restrictions on the $\tau$ that can be considered.
The applications to fermions considered in this tutorial
employed a multi-faceted approach to the convergence analysis.
For small systems, comparisons with established
literature results were used as a benchmark.
For larger systems, an analysis of the error
bars was used to establish where the Fermi sign problem
sets in. For $\tau$ not noticably impacted by the Fermi
sign problem, the resulting energies, extrapolated to the infinite
time slice limit, provided variational upper bounds. 
Additionally, the calculations were performed for
different trial functions and the runs were
checked for consistency. Ultimately, there is no guarantee 
that the resulting observables are not biased by the
trial function. However, the various checks provide one with tools for
(roughly) estimating and minimizing the variational bias.

The sample applications presented concern
strongly-correlated Fermi gases.
In cold atom experiments, the two-body
van der Waals length is typically much smaller than the
average interparticle spacing and the two-body
$s$-wave scattering length (for two-component Fermi
systems, this is the interspecies (and not the
intraspecies) scattering length). Thus,
cold atom systems realize, to a very good approximation, 
idealized systems in which the two-body interaction
range
is zero. From a theoretical point of view,
a vanishing two-body range is particularly interesting as this
implies that the range drops out of the problem.
For infinitely large two-body $s$-wave scattering length, e.g., the
system exhibits a scale invariance,
reflecting underlying symmetries of the Hamiltonian.
While scale-invariance based arguments and formulations have
led to a great deal of insight into these paradigmatic,
strongly-correlated systems, few analytical or
numerical techniques 
exist that can reliably predict the  energy, Tan contact,
superfluid fraction, or other observables.
The PIGS approach 
treats two-body zero-range interactions by building the exact 
two-body Bethe-Peierls
boundary condition into the propagator using the pair
product approximation. 
For fermions, the treatment is limited to small number
of particles since the Fermi sign problem becomes
exponentially more severe with increasing number of 
identical fermions. The system sizes considered in this article
are the same as those that have been treated by the explicitly
correlated Gaussian basis set expansion approach~\cite{yin15}.
For both (the PIGS and basis set expansion approach), the
computational effort increases tremendously as $N$ is
increased beyond what is considered in this article.

Future applications of the PIGS approach to systems
with zero-range interactions may include unequal-mass
two-component Fermi gases, Fermi gases in non-spherically
symmetric external traps (including effectively
low-dimensional systems), or Bose droplets without and
with an impurity.
These applications can be tackled with the technology 
already developed.
An interesting and challenging future development 
is the treatment of spin-orbit coupled systems, where
the spatial degrees of freedom are coupled to the
spin degrees of freedom. It will be interesting to
marry the treatment of spin degrees of freedom with the use
of two-body zero-range interactions.

\section{Acknowledgments}
\label{sec_acknow}
We are grateful to Xiangyu Yin
for providing the scaled pair distribution functions
from Ref.~\cite{yin15}
(dashed lines in Fig.~\ref{figpair}) in tabular form.
Support by the National
Science Foundation (NSF) through Grant No.
PHY-1415112
is gratefully acknowledged.
This work used the Extreme Science and Engineering
Discovery Environment (XSEDE), which is supported by
NSF Grant No. OCI-1053575, and the
WSU HPC.
\newcommand*{\newblock}{Bibliography}

\end{document}